\documentclass[reprint,twocolumn]{revtex4}
\usepackage{graphicx}%
\usepackage{dcolumn}%
\usepackage{bm}%
\usepackage{epsfig,colordvi}
\usepackage{color} 
\usepackage{here}
\usepackage{soul}

 \def\be{\begin{equation}}
 \def\ee{\end{equation}}
 \def\bea{\begin{eqnarray}}
 \def\eea{\end{eqnarray}}
 \def\bean{\begin{eqnarray*}}
 \def\eean{\end{eqnarray*}}
 \def\gsim{\mathrel{\rlap{\lower0.2em\hbox{$\sim$}}\raise0.2em\hbox{$>$}}}
 \def\ksim{\mathrel{\rlap{\lower0.2em\hbox{$\sim$}}\raise0.2em\hbox{$<$}}}
 \def\kg{\mathrel{\rlap{\lower0.25em\hbox{$>$}}\raise0.25em\hbox{$<$}}}

\begin{document}

\title{Parton-Hadron-Quantum-Molecular Dynamics (PHQMD) - A Novel Microscopic  
N-Body Transport Approach for Heavy-Ion Collisions, Dynamical Cluster Formation
and Hypernuclei Production}

\author{J. Aichelin$^{1,2}$, E. Bratkovskaya$^{3,4}$, A. Le F\`evre$^3$, V. Kireyeu$^5$, V. Kolesnikov$^5$, 
 Y. Leifels$^3$, V. Voronyuk$^5$ and G. Coci$^3$}
\affiliation{$^1$ SUBATECH, Universit\'e de Nantes, IMT Atlantique, IN2P3/CNRS
4 rue Alfred Kastler, 44307 Nantes cedex 3, France}
\affiliation{$^2$ Frankfurt Institute for Advanced Studies, Ruth Moufang Str. 1, 60438 Frankfurt, Germany}
\affiliation{$^3$ GSI Helmholtzzentrum f\"ur Schwerionenforschung GmbH,
  Planckstr. 1, 64291 Darmstadt, Germany} 
\affiliation{$^4$ Institut f\"ur Theoretische Physik, Johann Wolfgang Goethe-Universit\"at,
Max-von-Laue-Str. 1, 60438 Frankfurt am Main, Germany}
\affiliation{$^{5}$ Joint Institute for Nuclear Research, Joliot-Curie 6, 141980 Dubna, Moscow region, Russia}

\date{\today}

\begin{abstract} \noindent

Cluster and hypernuclei production in heavy-ion collisions 
is presently under active experimental and theoretical investigation.
Since clusters are weekly bound objects, their production is very sensitive 
to the dynamical evolution of the system and its interactions. The theoretical description
of cluster formation is related to the n-body problem.
Here we present the novel n-body dynamical transport approach PHQMD (Parton-Hadron-Quantum-Molecular-Dynamics) which is designed to provide a microscopic description of nuclear cluster and hypernucleus formation as well as of general particle production in heavy-ion reactions at relativistic energies. In difference to the coalescence or statistical models, often 
used for the cluster formation, in PHQMD clusters are formed dynamically due to the interactions between baryons described on a basis of Quantum Molecular Dynamics (QMD)
which allows to propagate the n-body Wigner density and n-body correlations in phase-space, essential for the cluster formation.
The clusters are identified by the  MST (Minimum Spanning Tree) or the SACA (‘Simulated Annealing Cluster Algorithm’) algorithm which  finds the most bound configuration of nucleons and clusters.
Collisions among hadrons as well as Quark-Gluon-Plasma formation and parton dynamics 
in PHQMD are treated in the same way as in the established PHSD (Parton-Hadron-String-Dynamics)
transport approach.
In order to verify our approach with respect to the general dynamics 
we present here the first PHQMD results for general 'bulk' observables 
such as rapidity distributions and transverse mass spectra for hadrons ($\pi, K, \bar K$,
$p$, $\bar p$, $\Lambda$, $\bar \Lambda$) from SIS to RHIC energies.
We find a good description of the "bulk" dynamics which allows us to
proceed with the results on  cluster production, including hypernuclei.
\end{abstract}


\maketitle

\section{Introduction}
There are a variety of evidences that a new state of matter, a quark-gluon plasma (QGP), 
has been created in the experiments at the Relativistic Heavy Ion Collider (RHIC) at Brookhaven 
and at the Large Hadron Collider (LHC) at CERN \cite{QM2017}. 
The QGP has been predicted by lattice gauge
calculations (lQCD) \cite{Borsanyi:2013bia,Bazavov:2014pvz}, in which the Lagrangian of 
Quantum Chromo-Dynamics (QCD),  describing strongly-interacting matter, is calculated on the computer. 
One of the unsolved questions is how the fraction of the matter in QGP phase changes when lowering the beam energy 
and at which beam energy a QGP ceases to be created. At low beam energies, around 
a few $A$GeV, heavy-ion collisions (HIC) are successfully 
described by models which are based on hadronic degrees-of-freedom only.
From experimental data at RHIC and LHC we know that at ultrarelativistic energies the baryon
chemical potential in the midrapidity region is close to zero. 
By decreasing the beam energies one tests higher baryonic chemical potentials.
 However, for a large baryon chemical potential lQCD
calculations cannot guide us because of the sign problem. Phenomenological models, like those 
based on the Nambu-Jona-Lasinio Lagrangian, predict that the smooth
transition (crossover) between the hadronic world and the QGP at vanishing baryon chemical potential 
\cite{Borsanyi:2013bia,Bazavov:2014pvz} becomes a first order phase transition
for finite chemical potentials \cite{ref1storder1,ref1storder2}.

In order to study nuclear matter at high baryon densities
 presently two accelerators are under construction, 
the Facility for Antiproton and Ion Research (FAIR) in Darmstadt and 
the Nuclotron-based Ion Collider fAcility (NICA) in Dubna. They
 will become operational in the next years. 
Moreover, the presently running BES-II (Beam Energy Scan) at RHIC, 
which includes a fixed target program, provides  experimental data in this energy regime.
The scientific goal of all these experimental efforts is to study those observables which may carry information on the existence 
of the QGP and the nature of its phase transition to the hadronic world. These observables include the particle yields, rapidity and transverse
momentum spectra of produced hadrons, their fluctuations and correlations with  particular focus on  the fluctuations of baryons, 
production of strange and multi-strange baryons as well as 
cluster and hypernuclei production.

The study of cluster and hypernucleus production, which reflects the phase space density 
during the expansion phase,  is of particular interest from experimental as well as from  theoretical side.
Experimentally clusters have been observed at all energies: from low energies - measured by  ALADIN \cite{Schuttauf:1996ci,Sfienti:2006zb}, INDRA \cite{Nebauer:1998fy}, FOPI \cite{Reisdorf:2010aa}, HypHI \cite{Rappold:2015una} Collaborations, 
to (ultra-) relativistic energies - measured by NA49 \cite{Anticic:2016ckv}, 
STAR \cite{Abelev:2010rv,Agakishiev:2011ib}, 
ALICE \cite{Adam:2015yta,Adam:2015vda,Acharya:2017bso} Collaborations).

The multiplicity of the produced clusters at midrapidity 
is related  to the phase space distribution of baryons at their creation point 
and therefore a change of the fluctuations - like expected in the neighborhood of a first order
phase transition - will be directly reflected in the cluster multiplicity \cite{Shuryak:2018lgd}. 
On the other hand, without identifying clusters, single particle observables such as the baryon spectra
cannot be correctly interpreted. This is especially important at low collision energies.
For example, in central Au+Au collisions at 1.5 $A$GeV only  65\% of the total baryon charge is observed 
as protons as has been measured by FOPI Collaboration \cite{Reisdorf:2010aa}, 
the rest is  bound predominantly in light clusters.
Composite clusters show different rapidity distributions, 
in-plane flows and  $p_T$ spectra than free protons.
Therefore, for the theoretical understanding of single baryon spectra measured at those energies, one has to take into account
the formation of clusters, otherwise predictions of observables  are not precise, especially at low energies. 
 
Among the clusters,  hypernuclei which contain at least one hyperon (strange baryon) are the most interesting observables. 
The formation of hypernuclei  in heavy-ion reactions has been a subject of many theoretical studies - cf. \cite{Shuryak:2018lgd,Wakai88,Rudy95,Gaitanos09,Topor10,Botvina,Ko15,Andronic:2010qu}. Recent experimental results 
\cite{Rappold:2015una,Agakishiev:2011ib,Adam:2015vda} have shown that hypernuclei and anti-hypernuclei can be formed in heavy-ion collisions from SIS to LHC energies. Detailed theoretical calculations have identified two sources of hypernuclei in these reactions:
In the overlap region of target and projectile, hyperons are produced in energetic first chance  NN collisions. They a) may migrate into the cold spectator matter being there absorbed to form heavy hypernuclei or 
b) may stay
in the participant region, which expands, and their interaction with the surrounding nucleons allows them to form  light clusters and hence light hypernuclei.
In view of their small binding energy and their hot environment this is like the creation of 'ice in a fire'. Nevertheless, such hypernuclei have been found around 
midrapidity in RHIC and LHC experiments \cite{Rappold:2015una,Adam:2015vda}.

The two production mechanisms of hypernuclei may shed light on the theoretical understanding of the  dynamical evolution of heavy-ion reactions which cannot be addressed by other probes. In particular, 
the formation of heavy projectile/target like hypernuclei elucidates the physics at the transition region
between spectator and participant matter. 
Since hyperons are produced in the overlap region, multiplicity as well as rapidity distributions of hypernuclei formed in the target/projectile region depend crucially on the interactions of the hyperons with the hadronic matter, e.g. cross sections and potentials.
On the other hand, midrapidity hypernuclei test the phase space distribution of baryons in the expanding participant matter,
especially whether the phase space distributions of strange and non-strange baryons are similar  and whether they are in thermal equilibrium.
The present data \cite{David:1998qu,Ritman:1995tn} does not allow for an conclusive answer.
The description of cluster and hypernuclei formation is a challenging theoretical task
which requires \\
I) the  microscopic dynamical description of the time evolution of heavy-ion collisions;\\
II) the modelling of the mechanisms for the cluster formation.\\

The existing transport approaches are either based on 
i) the Quantum Molecular Dynamics (QMD) algorithms for the propagation of particles 
with mutual density dependent 2-body potential interactions, e.g.  QMD  \cite{Aichelin:1991xy,Aichelin:1987ti,Aichelin:1988me,David:1998qu}, 
IQMD \cite{Hartnack:1997ez},  UrQMD \cite{Bass:1998ca,Bleicher:1999xi} etc.
or on
ii) the mean-field based approaches such as different types of semi-classical 
(Vlasov)Boltzmann-Uehling-Uhlenbeck ((V)BUU) models realized in terms of different numerical
codes known as 
BUU \cite{Kruse:1985pg,Aichelin:1985zz,BUU}, AMPT \cite{AMPT}, 
HSD \cite{Ehehalt:1996uq, Cassing:1999es}, PHSD \cite{Cassing:2008sv},
GiBUU \cite{GiBUU}, SMASH \cite{Weil:2016zrk} etc.
There are also models based on a cascade type propagation as the Quark-Gluon String Model (QGSM) \cite{QGSM}.
 
The mean-field models reproduce well the single particle observables, however,
they are not suited for describing cluster formation since they propagate the single-particle
distribution function (realized with the test particle method) in a mean-field potential calculated
by averaging over many parallel ensembles.  This approach smears out the initial n-body correlations 
as well as the dynamical correlations due to the interactions which develop during the whole time evolution of the system .  

For the production  of clusters, which are n-body correlations in phase space,  one needs to calculate
the time evolution of the n-body Wigner density \cite{Gossiaux:1994jq}.
Most of the presently available QMD approaches (QMD, IQMD) are limited to 
nonrelativistic energies. 
The only exception is the UrQMD approach, which has been used for study of deuteron and light nuclei production via coalescence  \cite{Feckova:2016kjx}.

Cluster formation has often been described either by a coalescence model
\cite{Zhu:2015voa,Feckova:2016kjx} or statistical methods  
\cite{Botvina,Botvina_HSD} assuming that during the heavy-ion reaction at least a subsystem achieves thermal equilibration. 
Both of these models have serious drawbacks. The most essential is that they are not able 
to address the question of  how the clusters are formed and what we can learn from the 
cluster formation about the reaction dynamics. 

In the coalescence model the multiplicity of clusters  depends 
crucially on external parameters and the time $t_C$, when instantaneously the coalescence is calculated,  as well as on the coalescence parameters. 
It neglects that energy and momentum conservation require the presence of another hadron during the cluster formation process
and assumes that, after the clusters are identified at $t_C$, no further interactions of the cluster nucleons take place.

Such a sudden freeze-out is not in line with other observables like the resonance production. 
Decay products of resonances can interact with the surrounding medium -- being absorbed or rescattered, therefore, the resonances cannot be identified anymore by the invariant mass method. Consequently, one observes experimentally a decrease of the multiplicity of 
resonances in comparison to the statistical model prediction.
Such an effect is not properly treated within coalescence models.

There are some efforts made to improve the coalescence picture by extending
it to the Wigner density approach. In this case the cluster formation at
 $t_C$ is calculated by projecting the n-body Wigner density, which 
is propagated in the transport model, on the Wigner density of the ground states of the 
2, 3 or 4-body clusters. One uses a simple parametrization of the ground state 
wave function of the clusters which reproduces their {\it rms} (root-mean-square) radius. 
The Wigner density method allows to predict the momentum distribution of these 
clusters and has been applied for the deuteron formation in  heavy-ion reactions \cite{Zhu:2015voa}.
The drawbacks, however, remain that the origin of the cluster formation cannot be studied and that the dynamical cluster formation is
reduced to a projection on the cluster Wigner density at a given time point $t_C$ during the reaction. 

Statistical fragmentation models are based on the strong assumption that a thermal equilibrium 
is obtained  in the heavy-ion reactions, at least in a limited rapidity interval.  The single particle spectra of protons and produced 
hadrons do not support such an assumption \cite{Hartnack:2011cn}, at least not at the intermediate energies 
($1 A$GeV $\le E_{beam} \le 30 A$GeV) on which we focus in this study. The statistical fragmentation model
assumes, furthermore, that equilibrium is maintained during the expansion of the system up to 
very low densities where cluster formation sets in. The ingredients of the model - like the treatment 
of free and bound neutrons, the initial temperature and the baryon chemical potential - are fitted to the experimental observations.
The multiplicity of clusters observed with the high energy beams at RHIC and LHC experiments 
can be quantitatively described by a statistical model calculations 
using  the same parameters as for description of hadron multiplicities.
The light cluster production can be described as well  by a coalescence model  \cite{Adam:2015vda}.
Moreover, in  Ref. \cite{Danielewicz:1991dh} deuterons are produced and propagated by Green function techniques.
In  Ref. \cite{Oliinychenko:2018ugs} the deuteron production in Pb+Pb central collisions at the LHC energies 
is assumed to be a final state interaction simulated by a two step process  $p+n \to d^\prime$ and  $d^\prime +\pi \to d+\pi$   
including a fictitious  resonance $d^\prime$.

In order to overcome these limitations we advance the novel Parton-Hadron-Quantum-Molecular 
Dynamics (PHQMD) transport approach. The goal of this approach is to provide a dynamical description for the formation of light and heavy clusters and hypernuclei in relativistic heavy-ion collisions based on a microscopic origin, i.e. on the interaction between nucleons and hyperons which leads to the binding of clusters. Since clusters are weakly bound objects, they are very sensitive to the general dynamics of the system
and to the interactions of the constituents, i.e. to the propagation and collisional  
interaction described by the kinetic equations-of-motions:\\
i) The PHQMD is based on the QMD dynamics for the propagation of the baryons,
realized by density dependent 2-body potential interactions  
\cite{Aichelin:1991xy,Marty:2012vs,Marty:2014zka},
 which allow (contrary to the mean-field approaches) to propagate the n-body phase-space correlations between baryons.\\
ii) At high energy heavy-ion collisions the dynamics is dominated by
the multiparticle production at the early stage with the formation of 
the QGP and partonic interactions. For the description of collisions and the QGP dynamics 
in PHQMD we adopt the collision integral of the Parton-Hadron-String Dynamics approach (PHSD) \cite{Cassing:2008sv,Cassing:2008nn,Cassing:2009vt,Bratkovskaya:2011wp,Linnyk:2015rco}
which was well tested in the reproduction of experimental data on "bulk" dynamics from SIS to LHC energies.
Moreover, the original PHSD mean-field propagation  (realized within the parallel 
ensemble method) for baryons is kept as an option, too, which will allow to investigate the 
differences between both approaches - i.e. the influence of the QMD versus 
mean-field based propagation on "bulk" observables.

Thus, PHQMD provides a fully microscopic description of the time evolution
of the system and the interactions between particles - on the hadronic and partonic levels.  
Due to that in PHQMD the clusters are formed dynamically. 
This means that at the end 
of the heavy-ion reaction the same  potential interaction, which is present during 
the whole time evolution, forms bound clusters of nucleons which are well distinct 
in phase space from other clusters and free nucleons. 
This differentiates our approach from coalescence models where at a given time point a coalescence radius in phase space is employed without considering whether the coalescing nucleons are still strongly interacting
with nucleons which do not belong to the cluster. 

These clusters can be identified by two methods:
either by the minimum spanning tree 
(MST) procedure \cite{Aichelin:1991xy} or by a cluster finding algorithm based on 
the simulated annealing technique, 
the Simulated Annealing Clusterization Algorithm (SACA) \cite{Puri:1996qv,Puri:1998te}. Presently 
an extended version -- the  ‘Fragment Recognition In General Application’ (FRIGA) 
\cite{FRIGA2019} is under development
which includes symmetry and pairing energy as well as hyperon-nucleon interactions.

The MST algorithm is based on spatial correlations and it is effective in finding 
the clusters at the end of the reaction. 
In order to identify the cluster formation  already at early times of the reaction, 
when the collisions between the nucleons are still on-going  and the nuclear density is high, 
the SACA approach is used. It is based on the idea of 
Dorso and Randrup \cite{Dorso:1992ch} that the most bound configuration of nuclei and nucleons 
evolves in time towards the final cluster distribution. 
The validity of this idea has been confirmed in numerical studies \cite{LeFevre:2009er,Gossiaux:1997hp,Fevre:2007pr}.

First results from the combined  PHSD/SACA approach have been reported in \cite{Fevre:2015fua}. 
There we have applied SACA at some fixed time using the nucleon distribution from the PHSD at 11.45 GeV 
for semi-peripheral Au+Au collisions. Moreover, the first attempt to identify hypernuclei with FRIGA
has been reported in Refs. \cite{Fevre:2015fua,FRIGA2019}.

In this study we present the first results from the PHQMD approach.
In order to validate the general dynamics in the PHQMD we start with presenting the results 
on 'bulk' observables, covering the energy range from SIS to RHIC, and compare the 
PHQMD results with the PHSD results in order to identify the difference between the 
QMD and mean-field propagation of baryons, it's influence on stopping of protons
and, correspondingly, on the "chemistry" production and pressure redistribution
in the interacting system  by looking on rapidity distributions and transverse mass $m_T$ 
or transverse momentum $p_T$ spectra for hadrons ($\pi, K, \bar K$,
$p$, $\bar p$, $\Lambda$, $\bar \Lambda$) from SIS to RHIC energies.
Then we proceed with the first PHQMD results on dynamical cluster formation, 
including hypernuclei, based on the MST and SACA models.
Furthermore, we verify our model for the cluster production in comparison to the available experimental data at SIS energies, show the capacity of PHQMD
to create clusters and hypernuclei at higher energies and make  predictions 
for future FAIR and NICA experiments.

Our paper is organized as follows:
We describe in Section II  the basic ideas of the PHQMD model. 
In Section III we detail the algorithms (SACA and MST) which allow to identify clusters 
in a dynamical model.
In Section IV we present the results from the PHQMD for the 'bulk' observables
such as rapidity distributions and transverse mass or momentum spectra and compare them 
to available data from $E_{beam}=1.5 A$GeV  up to 21.3 $A$TeV. 
Section V is devoted to the study of clusters. We confront our results with the 
presently existing data for heavy clusters and explore the formation
of light clusters at midrapidity. 
Finally we present in Section VI our conclusions.


\section{Model description: the PHQMD approach}

In this section we describe the basic ideas of the PHQMD approach.
The PHQMD is a n-body microscopic transport approach which describes the time evolution of an interacting system by  solving the kinetic equations-of-motion 
which contains i) the propagation of degrees-of-freedom with their potential
interaction as well as ii) their scattering described by "collision integrals".
iii) Moreover, the dynamically formed clusters are identified by the SACA and MST algorithms.

i) The propagation of baryons in PHQMD follows the Quantum Molecular Dynamics (QMD) 
approach where baryons are described by Gaussian wave functions. 
In QMD the particles propagate under the influence of mutual 2-body forces which
are density dependent in order to approximate $n$-body forces $(n>2)$. 
The density is defined by the sum of the squares of the wave functions of all other
nucleons. Both, density independent and density dependent two-body forces are 
necessary to obtain a maximum of the binding energy at normal nuclear matter density. 
In such an approach 'actio' is equal to 'reactio' and, therefore, energy and momentum are 
strictly conserved. The strength of the interaction is chosen in a way
that in infinite matter a given nuclear EoS is reproduced. 
The time evolution of the wave functions is determined according to a variational principle
\cite{Feldmeier:1989st}. This approach conserves the phase-space correlations 
in the system and does not suppress fluctuations as mean-field based kinetic approaches.  
Since clusters are n-body correlations this approach is well suited to address the
creation and time evolution of clusters.

ii) The PHQMD incorporates the collision integrals of the Parton-Hadron-String Dynamics 
(PHSD) approach \cite{Cassing:2008sv,Cassing:2008nn,Cassing:2009vt,Bratkovskaya:2011wp,Linnyk:2015rco}
which describes all interactions in the system - from primary 
hadron collisions to the formation of the QGP in terms of strongly interacting quasiparticles - massive quarks and gluons, partonic interactions with further
dynamical hadronization up to hadronic interactions during the final stage of 
the expansion. Moreover, the propagation of partonic degrees-of-freedom 
is adopted from the PHSD, too, and based on the Kadanoff-Baym equations for the
dynamics of strongly interacting systems \cite{Cassing:2008nn,Cassing:2008sv}.

iii) At the later stage of the reaction, after hadronization and
resonance decays, the cluster recognition is performed by the SACA or MST algorithms
which determine whether baryons are bound in clusters or not.

In the next sub-sections we will discuss the ideas of the PHQMD approach 
in more detail, however, for the readers who are familiar with the PHSD
and QMD approaches one can summarize by saying that PHQMD combines the description of the QGP and hadronic interactions of the PHSD approach with the n-body dynamics and initial distributions of baryons from QMD. 
Additionally,  cluster recognition algorithms MST and SACA, which is based on the finding of configurations with minimal binding (negative)energy calculated by the Weizs\"acker mass formula, are applied.

\subsection{Stages of the Nucleus-Nucleus Collisions; the Collision Integral}

Nucleus-nucleus collisions in the PHQMD (similar to the PHSD) follow the 
following steps:

$\bullet$  In the beginning of the nucleus-nucleus collision (an initialization of 
the nuclei as well as QMD propagation of nucleons  will be discussed in the next section) two nuclei are approaching each other until they start to overlap such that
individual nucleon-nucleon primary collisions take place. At relativistic energies
the description of such primary collisions with multi-particle production
is based on the LUND string model \cite{LUND} which describes the energetic 
hadron-hadron collisions by the creation of excited color-singlet states, 
denoted by "strings" which are realized within the FRITIOF \cite{FRITIOF} 
and PYTHIA models \cite{Sjostrand:2006za}
(cf. the HSD review \cite{Cassing:1999es} for the description of string dynamics
in HICs). 
A string is composed of two string ends corresponding to the leading constituent quarks 
(antiquarks) of the colliding hadrons and a color flux tube (color-electric field) in between. 
As the string ends recede, virtual $q\bar q$ or $qq\bar q\bar q$ pairs are produced in the
uniform color field by a tunneling process (described by the Schwinger formula \cite{Schwinger}), causing the breaking of the string.
The produced quarks and antiquarks recombine with neighbouring partons 
to "prehadronic" states which will approach hadronic quantum states 
(mesons or baryon-antibaryon pairs )
after a formation time $\tau_f\sim 0.8\,$fm/c (in the rest-frame of the string).
In the calculational frame of heavy-ion reaction (which is chosen to be the initial $NN$ center-of-mass frame) the formation time then is $t_F=\tau_F \cdot \gamma$, 
where $\gamma=1/\sqrt{1-v^2}$ and  $v$ is the velocity of the particle in the calculational frame. 

In the string decay, the flavor of the produced quarks is determined
via the Schwinger formula \cite{Schwinger,Sjostrand:2006za}, which
defines the production probability of massive $s\bar s$ pairs with
respect to light flavor production $(u\bar u,d\bar d)$ pairs:
\begin{equation}
\label{Schwinger-formula} \frac{P(s\bar s)}{P(u\bar
u)}=\frac{P(s\bar s)}{P(d\bar
d)}=\gamma_s=\exp\Bigl(-\pi\frac{m_s^2-m_{u,d}^2}{2\kappa}\Bigr)\,,
\end{equation}
with $\kappa\approx 0.176\,$GeV$^2$ denoting the string tension
and $m_{u,d,s}$ -- the constituent quark masses for strange
and light quarks. For the constituent quark masses
$m_u\approx0.35\,$GeV and $m_s\approx0.5\,$GeV are adopted in the vacuum; the
production of strange quarks thus is suppressed by a factor of
$\gamma_s\approx 0.3$ with respect to the light quarks, which is the
default setting in the FRITIOF routines.
We note that in Refs. \cite{PHSD_CSR,Alessia} the PHSD has been extended to
include the chiral symmetry restoration effect (CSR) in the string decay 
by changing of constituent quark masses due to the dropping of 
the scalar quark condensate in a hadronic environment of a finite baryon and meson density.
After string decay the "leading hadrons", which originate from the string ends, 
can re-interact with hadrons almost instantly with reduced cross-sections 
(according to the constituent quark number) \cite{Cassing:1999es}.

$\bullet$ In case the local energy density in the cell is above the critical value of $\epsilon_c\sim 0.5\,$GeV/fm$^3$, the "deconfinement" (i.e. a transition of hadronic
to partonic degrees-of-freedom) is implemented by dissolving the  
"pre-hadrons" (the string decay products which are in this cell) into the massive colored quarks/antiquarks and mean-field energy, keeping "leading hadrons" out of dissolution
(cf. Refs. \cite{Cassing:2008sv,Cassing:2009vt} for the details). This procedure allows
to keep the microscopic description of changing degrees-of-freedom by conserving the
energy-momentum, charge, flavour etc.

In PHQMD (as well as in PHSD) the partonic (or the QGP phase) is based on the Dynamical Quasi-Particle Model (DQPM) \cite{Cassing:2007yg,Cassing:2007nb} which describes the properties of QCD (in equilibrium) in terms of resummed single-particle Green's functions. Instead of massless partons of pQCD, in PHSD the gluons and quarks are massive 
strongly-interacting quasi-particles which reflects the non-perturbative nature of the strong interaction. The properties of off-shell quasi-particles are described by spectral functions (imaginary parts of the complex propagators) with temperature-dependent masses and widths.  The widths and pole positions of the spectral functions are defined by the real and imaginary parts of the parton  self-energies and the effective coupling strength which is fixed by adjusting lQCD results for the entropy density \cite{Aoki:2009sc,Cheng:2007jq} 
(using in total three parameters).
The details of the DQPM model, adopted in the PHQMD, can be found in Appendix A.

$\bullet$  Within the QGP phase, the partons (quarks, antiquarks and gluons) scatter and propagate in a self-generated scalar mean-field potential \cite{Cassing:2009vt}. 
On the partonic side the following elastic and inelastic
interactions are included $qq \leftrightarrow qq$, $\bar{q}
\bar{q} \leftrightarrow \bar{q}\bar{q}$, $gg \leftrightarrow gg$,
$gg \leftrightarrow g$, $q\bar{q} \leftrightarrow g$  exploiting
'detailed-balance' with temperature dependent cross sections
(as in the PHSD 4.0) (cf. \cite{Ozvenchuk:2012fn,Linnyk:2015rco}).
The propagation of off-shell partons in PHQMD (as well as in PHSD) is fully determined by
the Cassing-Juchem off-shell transport equations based on the Kadanoff-Baym equations (cf. the review \cite{Cassing:2008nn}).

$\bullet$ The expansion of the system leads to a decrease of the local energy density and, once the local energy density becomes close to or lower than $\epsilon_c$, the massive colored off-shell quarks and antiquarks hadronize to colorless off-shell mesons and baryons. The hadronization process is defined by covariant transition rates 
(see Appendix B) and fulfills the energy-momentum and quantum number conservation in each event \cite{Cassing:2009vt}.

$\bullet$ In the hadronic "corona" (i.e. the region with low energy density) 
as well as in the late hadronic phase after hadronization, or even the whole 
dynamics at low bombarding energies (without the formation of the QGP), the hadrons are 
interacting and propagating. 
The hadronic degrees-of-freedom in the PHQMD are the baryon octet and decouplet, the ${0}^{-}$ and ${1}^{-}$ meson nonets and higher resonances.
The hadronic interactions include elastic and inelastic collisions between baryons,
mesons and resonances (including the backward reactions through detailed balance)  in line with the HSD approach \cite{Ehehalt:1996uq,Cassing:1999es}. 
We note that in the PHQMD (as well as in the PHSD) the multi-meson
fusion reactions to baryon-antibaryon pairs and backward reactions ($ n \ mesons \leftrightarrow B+\bar B$) 
are included, too \cite{Cassing:2001ds,Seifert:2017oyb}. 
The PHQMD (as well as PHSD) incorporates also the in-medium effects, i.e. the changes
of hadronic properties in the dense and hot environment, such as a collisional broadening 
of spectral functions of vector mesons ($\rho, \omega, \phi, a_1$)
\cite{Bratkovskaya:2007jk}, strange mesons $K, \bar K$ \cite{HSDK} 
and strange vector mesons $K^*, \bar K^*$ \cite{Ilner:2016xqr}. 
The propagation of the off-shell mesonic states is described also by the 
Cassing off-shell transport equations \cite{Cassing:2008nn,HSDK}.
Contrary to the PHSD the propagation of baryonic states in the PHQMD 
follow the QMD equations (see Section II.C).

In the PHQMD approach the full evolution of a relativistic heavy-ion collision, 
from the initial hard NN collisions out-of-equilibrium, the formation of a partonic state 
up to the hadronisation as well as final interactions of the resulting hadronic particles, is described 
on the same footing.

\subsection{Initialization of the nuclei}

As mentioned above, we adopt the parallel ensemble method for the PHQMD
approach for both dynamical options: QMD (where the parallel ensembles are independent) 
and mean-field (MF) used in the PHSD.
In the  MF (i.e. PHSD) mode the initialization in coordinate space is realized 
by point-like test particles, randomly distributed according to the Wood-Saxon 
density distribution and in momentum space according to the local Thomas-Fermi 
distribution in the rest frame of the nucleus. 
\begin{figure}[t!]%
\centering
\includegraphics[scale=0.425]{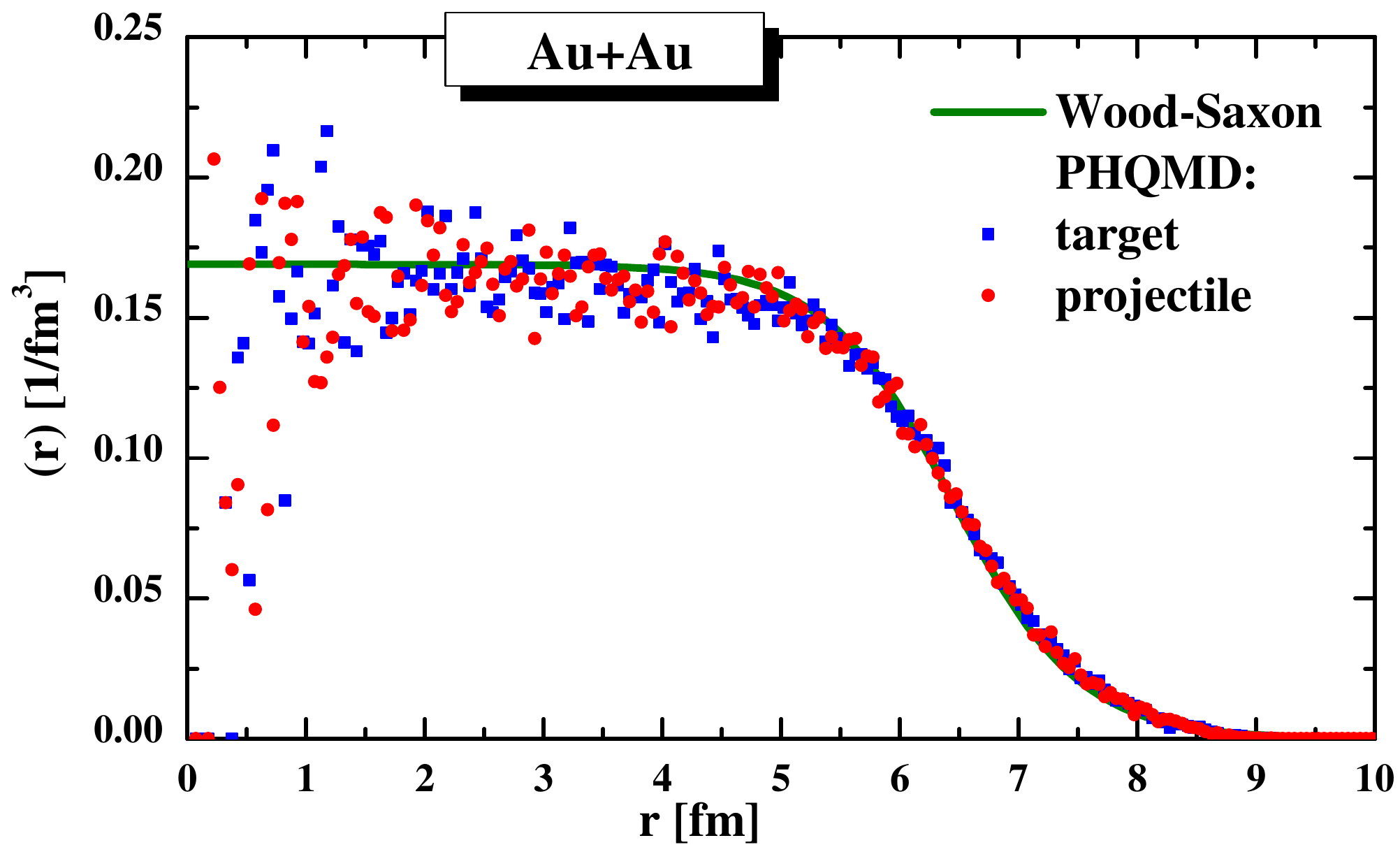}
\caption{\label{fig:WS} (Color online) 
The averaged (over 250 events) density distribution of target (blue squared) and projectile (red dots) nucleons 
in Au+Au collisions obtained from the QMD initialization in comparison to the Wood-Saxon distribution, 
eq. (\ref{eq:WS}), (solid line).}  
\end{figure} 

In the QMD mode we use the single-particle Wigner density of the nucleon $i$,
which is given by  
\begin{equation} \label{fdefinition}
 f ({\bf r_i, p_i,r_{i0},p_{i0}},t) = \frac{1}{\pi^3 \hbar^3 }
 {\rm e}^{-\frac{2}{L} ({\bf r_i} - {\bf r_{i0}} (t) )^2   }
 {\rm e}^{-\frac{L}{2\hbar^2} ({\bf p_i - p_{i0}} (t) )^2},
\end{equation}
where the Gaussian width $L$ is taken as  $L=2.16$ fm$^2$.
We will use the $\hbar = c = 1$  convention for further consideration. The corresponding single particle density is obtained by an integration of the single-particle Wigner density over the momentum of nucleon $i$:
\bea
\rho_{sp}({\bf r_i,r_{i0}},t)=
\int d{\bf p_i}  f ({\bf r_i, p_i,r_{i0},p_{i0}},t) \nonumber\\
= \Big(\frac{2}{\pi L}  \Big)^{3/2}{\rm e}^{-\frac{2}{L} ({\bf r_i} - {\bf r_{i0}} (t) )^2}.
\label{rhosp}
\eea 

The total one-body Wigner density is the sum of the Wigner densities of all nucleons.
To initialize the nuclei we choose randomly the position of nucleons
${\bf r_{i0}}(t=0)$ according to the Wood-Saxon density distribution.
We take care that the distribution is smooth by requiring a minimal phase space 
distance between the nucleons. 
Figure \ref{fig:WS} shows the nucleon density distribution (averaged over 250 QMD events) 
of target and projectile nucleons in Au+Au collisions in comparison to the Wood-Saxon distribution 
\begin{equation} \label{eq:WS}
\rho^{WS}(r)={\rho_0 \over 1+e^{r-R_A \over a} },
\end{equation}
where $R_A=r_0 A^{1/3}$ is the radius of nuclei with mass number $A$ with $r_0=1.125$ fm, $\rho_0=0.1695$ fm$^{-2}$, $a=0.535$~fm.

To initialize the nuclei in momentum space, we chose randomly the momenta  of nucleons
${\bf  p_{i0}} (t=0)$ according to the Thomas-Fermi distribution with the additional 
requirement that the nucleons are bound
\be
 0 \le \, \sqrt{m^2+{\bf p_{i0}}^2(t=0)}-m \, \le \, - <V({\bf r_{i0}})>,
\label{mommax}
\ee
where $m$ is the mass of a nucleon.
Here the expectation value of the potential energy $<V({\bf r_{i0}})>$ 
(which we discuss in the next subsection) is negative.
This procedure gives a lower momentum to those nucleons which are located close 
to the surface because there the density is lower. Finally we take care that 
$\sum_i {\bf p_{i0}}(t=0)=0$ by adding a common momentum to all nucleons. 

With such determined momenta and positions we calculate the average binding energy
of the nucleons and compare the result with the Bethe-Weizs\"acker mass formula. 
It turned out that we underestimate slightly the average binding energy independent of 
the size of the nucleus. To obtain the right binding energy we multiply 
finally all momenta by a common factor (close to one) and the same for all nucleons. It depends on the value of $L$. 
 Before the nuclei collide, target and projectile nucleons are boosted into the nucleus-nucleus center-of-mass frame and get Lorentz contracted.

\subsection{QMD Propagation}

The propagation of the Wigner density is determined by a variational principle \cite{Feldmeier:1989st}, which has been developed  for  the Time Dependent Hartree-Fock approach, 
\be
\delta \int_{t_1}^{t_2} dt <\psi(t)|i\frac{d}{dt}-H|\psi(t)> = 0.
\ee
In our approach we assume that the n-body Wigner density is  the direct product of the single particle Wigner densities.  There are also QMD versions which use a Slater determinant,  FMD  \cite{Feldmeier:1989st} and AMD \cite{Ono:1992uy}, but due to the difficulty to formulate collision terms these approaches have only been applied to low energy heavy-ion collisions. Assuming that the wave functions have a Gaussian form and that the width of the wave function is time independent one obtains for the time evolution of the centroids of the Gaussian single particle wave functions two equations which resemble the equation of motion of a classical particle with the phase space coordinates ${\bf r_{i0},p_{i0}}$ \cite{Aichelin:1991xy}.
The difference is that here the expectation value of the quantal Hamiltonian is used and not a classical Hamiltonian: 
\begin{equation}
\dot{r_{i0}}=\frac{\partial\langle H \rangle}{\partial p_{i0}} \qquad
\dot{p_{i0}}=-\frac{\partial \langle H \rangle}{\partial r_{i0}} \quad .
\label{prop}
\end{equation}
These time evolution equations are specific for Gaussian wave functions. 
For other choices of wave functions the time evolution equations would be different. 
The Hamiltonian of the nucleus is the sum of the Hamiltonians of the nucleons, composed of kinetic and two body potential energy.
\be
H = \sum_i H_i= \sum_i  (T_i+V_i)  = \sum_i  (T_i+\sum_{j\neq i}  V_{i,j}).
\ee
The interaction between the nucleons has two parts, a local Skyrme type interaction and a Coulomb interaction
\begin{eqnarray}
&&\phantom{a}\hspace*{-5mm}
V_{i,j}= V({\bf r_i, r_j,r_{i0},r_{j0}},t)  = V_{\rm Skyrme}+ V_{\rm Coul}  \label{EP} \\
&& \phantom{a}\hspace*{-5mm}
=\frac{1}{2} t_1 \delta ({\bf r_i} - {\bf r_j})  +  \frac{1}{\gamma+1}t_2 \delta ({\bf r_i} - {\bf r_j})  \,     
    \rho^{\gamma-1}({\bf r_i,r_j,r_{i0},r_{j0}},t) \nonumber \\  
&&\phantom{a}\hspace*{-5mm}
 +\frac{1}{2}  \frac{Z_i Z_j e^2}{|{\bf r_i}-{\bf r_j}|} , \nonumber
\end{eqnarray}
with the density $ \rho({\bf r_i,r_j,r_{i0},r_{j0}},t)$ defined as
\bea
& &\rho({\bf r_i,r_j,r_{i0},r_{j0}},t) =\nonumber \\
&=&C\frac{1}{2}\Big[ \sum_{j,i\neq j} \big(\frac{1}{\pi L}\big)^{3/2}e^{-\frac{1}{L}({\bf r_i-r_j-r_{i0}(t)+r_{j0}(t)})^2} \nonumber \\
&+&\sum_{i,i\neq j} \Big(\frac{1}{\pi L}\Big)^{3/2}e^{-\frac{1}{L}({\bf r_i-r_j-r_{i0}(t)+r_{j0}(t)})^2}\Big].
\label{dens1}
\eea
where $C$ is a correction factor explained below. 

We define the 'interaction' density $\rho_{int}({\bf r_{i0}},t)$,
which for nonrelativistic case can be written as
\be \label{rhoint}
\rho_{int} ({\bf r_{i0}},t) = C \sum_{j,j\neq i}(\frac{1}{\pi L })^{3/2}
{\rm e}^{-\frac{1}{L} ({\bf r_{i0}}(t) - {\bf r_{j0}} (t) )^2}.
\ee

The interaction density  has twice the width of the particle density, Eq. (\ref{rhosp}),
and is obtained by calculating the expectation value of  the local Skyrme potential which is $ \propto \delta ({\bf r_i-r_j})$.
The correction factor $C$ in Eq. (\ref{dens1}) depends on $L$. 
It is introduced because nuclear
densities are calculated differently in mean-field approaches -- for which the Skyrme parametrization
has been developed -- and QMD approaches.  In mean-field transport or hydrodynamical approaches the density, which enters the density dependent two body interaction, is obtained
by summing over all particles in the system $\rho_{int}^{MF} ({\bf r_{i0}},t) = \sum_{j}...$ . In QMD
type approaches we have to exclude self-interactions and therefore, the density which enters the density
dependent interaction is the sum over all nucleons with the exception of that nucleon on which this
density dependent potential acts,  $\rho_{int} ({\bf r_{i0}},t) = \sum_{j\neq i} ...$ .
Both differ by $(\frac{1}{\pi L })^{3/2}$. To compensate for the lower density in the QMD type approaches compared to the
mean-field approaches 
we introduce the correction factor $C$ which is adjusted numerically to achieve equality of both densities. With this correction factor we can use
also for the QMD approach the Skyrme potentials.

The expectation value of the potential energy $V_i$,
$\langle V_i\rangle=\langle V({\bf r_{i0}},t)\rangle $, of the nucleon i is given by
\bea
&&\langle V({\bf r_{i0}},t)\rangle=\sum_{j,j \neq i}\int d^3r_id^3r_jd^3p_id^3p_j
 V({\bf r_i, r_j,r_{i0},r_{j0}}) \nonumber \\ 
&&\times f ({\bf r_i, p_i,r_{i0},p_{i0}},t) f ({\bf r_j, p_j, r_{j0},p_{j0}},t).
\label{Vpot}
\eea
 Numerical test have shown that the time evolution of the system does not change if we replace
1/2($\rho_{int} ({\bf r_{i0}},t) + \rho_{int} ({\bf r_{j0}},t)) $
 by $\rho_{int} ({\bf r_{i0}},t) $ or by $\rho_{int} ({\bf r_{j0}},t)$. 
For the Skyrme  potential we can therefore use the analytical form 
\begin{equation} \label{eosinf} \langle V_{Skyrme}({\bf r_{i0}},t) \rangle \,=\, \alpha
\left(\frac{\rho_{int} ({\bf r_{i0}},t)}{\rho_0}\right) +
        \beta \left(\frac{\rho_{int} ({\bf r_{i0}},t)}{\rho_0}\right)^{\gamma}.
\label{Upot}
\end{equation}
The expectation value  of the Coulomb interaction can also be calculated analytically.

The expectation value of the  Hamiltonian which enters in Eq. (\ref{prop}) is finally given by
\begin{eqnarray}
\langle H \rangle &=& \langle T \rangle + \langle V \rangle
\label{ham} \\
&=& \sum_i (\sqrt{p_{i0}^2+m^2}-m)  +\sum_{i}  \langle V_{Skyrme}({\bf r_{i0}},t)\rangle .
\nonumber 
\end{eqnarray}

The nuclear equation of state (EoS) describes the variation of the energy
$E(T=0,\rho/\rho_0)$ when changing the nuclear density in infinite matter 
to values different from the saturation density $\rho_0$ for zero temperature. 
In infinite matter the density is position independent and we can use Eq. (\ref{Upot}) to connect our
Hamiltionian with nuclear matter properties because for a given value of  $\gamma$ the parameters $t_1, t_2$  in eq.~(\ref{EP})  are uniquely related to the coefficients $\alpha, \beta$ of the EoS, eq. (\ref{Upot}). 
Values of these parameters for the different model choices can be found in Tab.~\ref{eostab}.

\begin{table}[hbt]
\begin{tabular}{lcccc}
 &$\alpha$ (MeV)  &$\beta$ (MeV) & $\gamma$ &\quad K\ [MeV] \\
\hline
 S & -390 & 320 & 1.14 & 200 \\
 H & -130 & 59  & 2.09 & 380 \\
\end{tabular}
\caption{Parameter sets for the nuclear equation of state used in
the PHQMD model.} \label{eostab}
\end{table}

\begin{figure}[t!]%
\centering
\includegraphics[scale=.5]{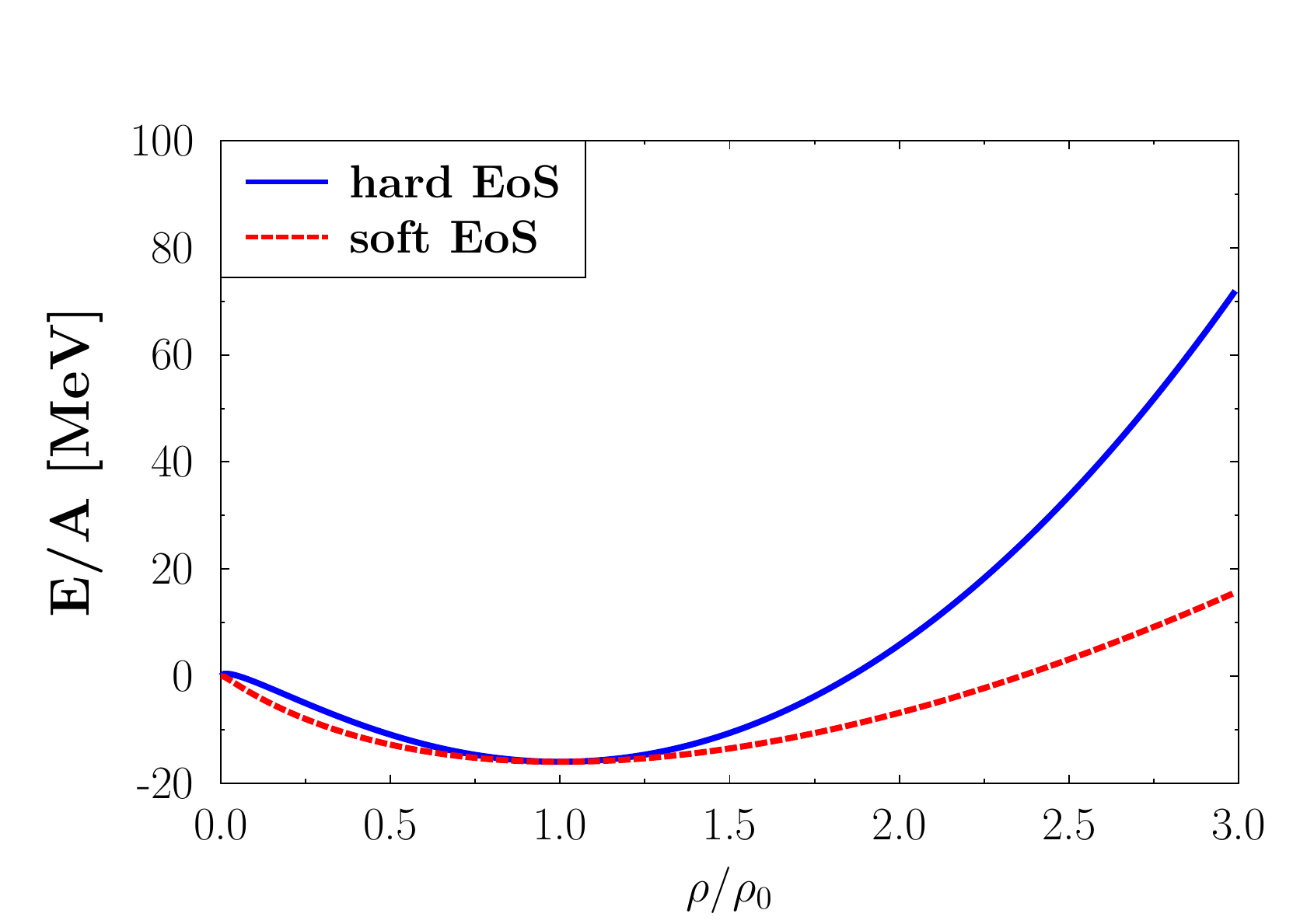}
\caption{\label{eos} (Color online) 
The energy per nucleon for the two EoS:
hard (solid blue line) and soft(dotted red line).} 
\label{EoS_HS}
\end{figure} 

Two of the 3 parameters of the Skyrme potential 
can be fixed by the condition that the energy per nucleon has 
a minimum of $\frac{E}{A}(\rho_0)=-16$ MeV at $\rho_0$. 

The third equation is historically provided by fixing  the compression modulus $K$
of nuclear matter,  the inverse of the compressibility $\chi = \frac{1}{V}\frac{d V}{d P}$,
which corresponds to the curvature of the 
Skyrme energy at $\rho=\rho_0$ (for $T=0$)  is also given in
Table I.
\be
K  = -V\frac{d P}{dV}= 9 \rho^2
\frac{{\rm \partial}^2(E/A(\rho))}{({\rm \partial}\rho)^2} \large|_{\rho=\rho_0} .\qquad
\ee
Here $P$ is the pressure in the system of volume $V$.
An equation-of-state with a rather low value
of the compression modulus $K$ yields a weak repulsion against
the compression of nuclear matter and thus describes "soft" matter
(denoted by "S"). A high value of $K$ causes a strong repulsion of
nuclear matter under compression (called a hard EoS, "H"). 
The hard and soft equations-of-state used in this study are illustrated in Fig. \ref{EoS_HS}. 

We stress again that for the present study we use
a  'static' form of Skyrme potential which depends only 
on the local density according to the Eq. (\ref{Vpot}). More realistic is a momentum dependent
Skyrme interaction. This will be the subject of future studies. Many observables show for
a soft momentum dependent interaction and a static hard interaction quite similar results \cite{Aichelin:1987ti}.
We also note that in the PHQMD we propagate non-strange baryon resonances (such as $\Delta$'s) 
in the same manner as nucleons assuming the same potential interaction 
as the nucleon-nucleon one while for strange baryon resonances 
(such as $\Lambda$'s, $\Sigma$'s) we assume 2/3 of the nucleon-nucleon potential.

The influence of the nucleon potential and hence of the EoS on hadronic observables
as well as on the cluster formation in heavy-ion collisions is well established at low energies 
(cf. e.g. \cite{Hartnack:2011cn}) where the nonrelativistic Hamiltonian formulation of QMD 
(presented in this section) is applicable. With increasing bombarding energies a relativistic
dynamics becomes more important. The relativistic formulation of molecular dynamics has 
been developed in  Ref.  \cite{Marty:2012vs}, however, the numerical realization of this method 
for realistic heavy-ion calculations is still not achievable with present 
computer power since it takes about two orders of magnitude longer time to simulate the reaction due to the inversion of high dimensional matrices.
Therefore, we are facing the problem of how to extend the nonrelativistic  
QMD approach to the high energy collisions, considered in this study, within a framework which can be numerically realized.

In order to extend our approach for relativistic energies, we introduce
the modified single-particle Wigner density $\tilde f$ of the the nucleon $i$
\bea
&& \tilde f ({\bf r_i, p_i,r_{i0},p_{i0}},t) =  \label{fGam} \\
&& =\frac{1}{\pi^3} {\rm e}^{-\frac{2}{L} ({\bf r_{i}^T}(t) - {\bf r_{i0}^T} (t) )^2} 
  {\rm e}^{-\frac{2\gamma_{cm}^2}{L} ({\bf r_{i}^L}(t) - {\bf r_{i0}^L} (t) )^2}.  \nonumber \\
&& \times {\rm e}^{-\frac{L}{2} ({\bf p_{i}^T}(t) - {\bf p_{i0}^T} (t) )^2} 
  {\rm e}^{-\frac{L}{2\gamma_{cm}^2} ({\bf p_{i}^L}(t) - {\bf p_{i0}^L} (t) )^2},  \nonumber 
\eea
which accounts for the Lorentz contraction of the nucleus in the beam $z$-direction,
in coordinate and momentum space by inclusion of $\gamma_{cm} =1/\sqrt{1-v_{cm}^2}$, where $v_{cm}$ is a velocity of the
bombarding nucleon in the initial $NN$ center-of-mass system.   
Accordingly, the interaction density (\ref{rhoint}) modifies as 
\bea
 \tilde \rho_{int} ({\bf r_{i0}},t) 
 &\to & C  \sum_j \Big(\frac{1}{\pi L }\Big)^{3/2} \gamma_{cm} \
 {\rm e}^{-\frac{1}{L} ({\bf r_{i0}^T}(t) - {\bf r_{j0}^T} (t) )^2} \nonumber \\ 
 &&\times  {\rm e}^{-\frac{\gamma_{cm}^2}{L} ({\bf r_{i0}^L}(t) - {\bf r_{j0}^L} (t) )^2}.  
 \label{densGam}
\eea
With these modifications we obtain
\begin{equation}
\langle \tilde H \rangle =
\sum_i (\sqrt{p_{i0}^2+m^2}-m)  +\sum_{i}  \langle \tilde V_{Skyrme}({\bf r_{i0}},t)\rangle .
\end{equation}
with
\begin{equation} \label{eosinfb} \langle \tilde V_{Skyrme}({\bf r_{i0}},t) \rangle \,=\, \alpha
\left(\frac{\tilde \rho_{int} ({\bf r_{i0}},t)}{\rho_0}\right) +
        \beta \left(\frac{\tilde \rho_{int} ({\bf r_{i0}},t)}{\rho_0}\right)^{\gamma}.
\label{Upot}
\end{equation}
with the time evolution equations (\ref{prop}). 

\begin{figure}[tp]%
\includegraphics[scale=.44]{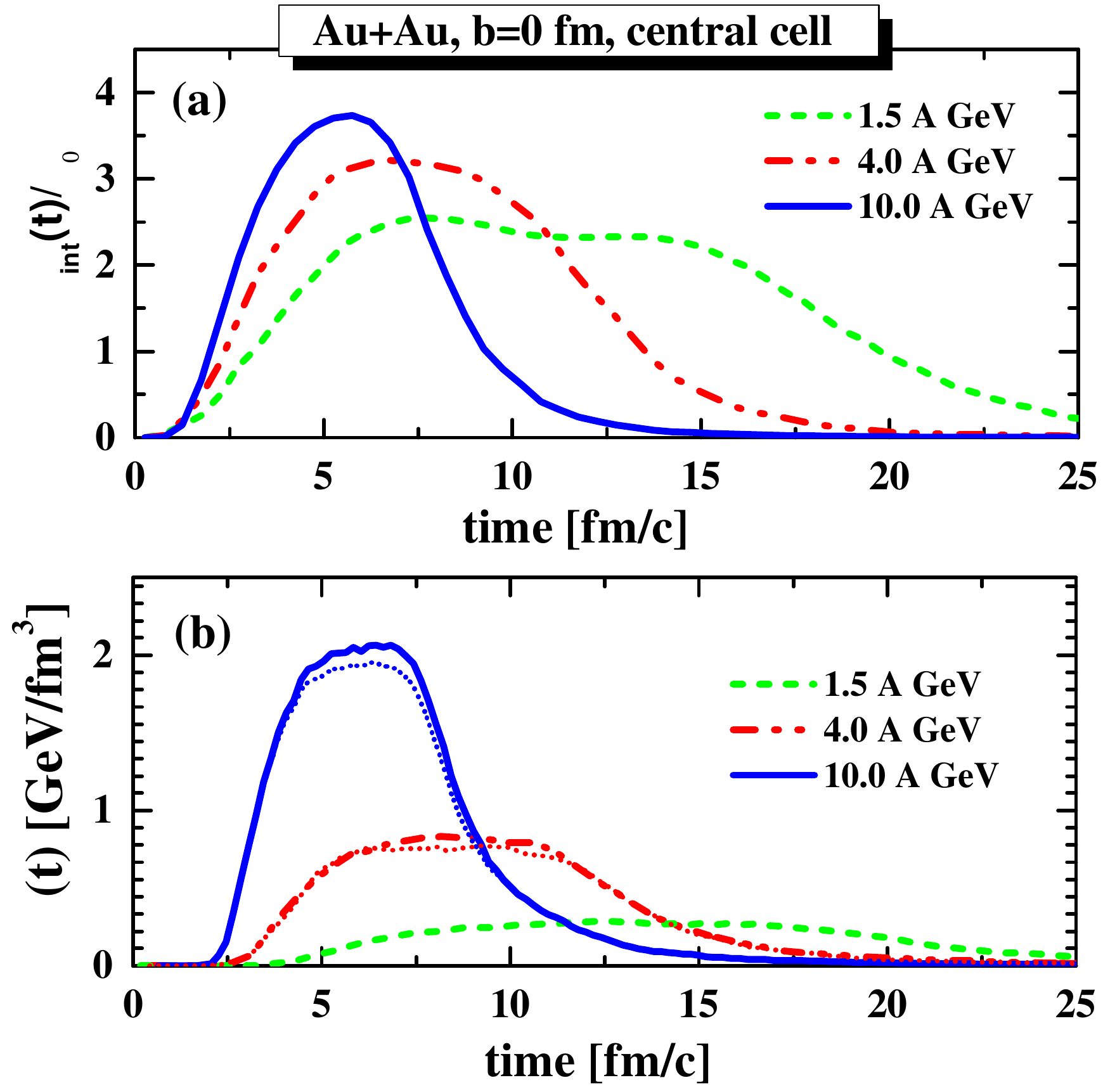}
\caption{\label{fig:DensEn} (Color online) 
Time evolution of the interaction density, scaled to the normal nuclear density
$\rho_0=0.168$ fm$^{-3}$, (upper plot) and the energy density 
(lower plot) in the central cell of volume ($27/\gamma_{cm}$) fm$^3$ of Au+Au collisions 
at $E_{beam}$ = 1.5 $A$GeV (dashed green lines), 4.0 $A$GeV (dash-dotted red lines) 
and 10.0 $A$GeV (solid blue lines), averaged over 200 events.
The dotted red and blue lines in the lower plot show the hadronic contribution to the energy
density for the corresponding beam energies  of 4.0 and 10.0 $A$GeV.
}
\end{figure}

To verify the applicability of our ansatz for the relativistic extension 
of the interaction density - eq. (\ref{fGam}),  
we have analyzed  the time evolution of the interaction density.
In the upper part of Fig. \ref{fig:DensEn}  we show the time evolution of the 
interaction density, scaled to the normal nuclear density $\rho_0=0.168$ fm$^{-3}$, 
of all baryons in the central cell of a volume ($27/\gamma_{cm}$) fm$^3$ of Au+Au collisions 
at $E_{beam}$ = 1.5 $A$GeV (dashed green lines), 4.0 $A$GeV (dash-dotted red lines) 
and 10.0 $A$GeV (solid blue lines), averaged over 200 events.
One can see that at 4 and 10 $A$GeV in the central cell a maximal density $\rho_{int}$
of 3$\div$3.5$\rho_0$ can be reached during the overlap of the nuclei.
The lower part of Fig. \ref{fig:DensEn} shows the time evolution of 
the local energy density $\varepsilon$ of all interacting particles (hadrons and partons), 
in the central cell (the color coding as for the upper plot). 
The dotted red and blue lines in the lower plot show the hadronic contribution to the energy
density for the corresponding beam energies  of 4.0 and 10.0 $A$GeV.
One can see that at the low bombarding energy of 1.5 $A$GeV the energy density in 
the center of the fireball is rather small and always below the critical one.
In spite that the energy density increases with bombarding energy, the matter
is still hadron dominated at 4 and 10 $A$GeV, the QGP is formed only in small droplets, i.e 
the fraction of the QGP in the total energy balance of the system is still very small at such energies and increases slowly with growing bombarding energy (cf. Fig. 4 in Ref. \cite{Konchakovski:2012yg}). This is related to the fact that in PHQMD (as in PHSD)
the leading hadrons -- the fastest ends of the decaying strings,
which are rather short (i.e. have little energy) at low energies -- are not dissolved 
(as explained in Section II.A) to partons and keep their hadronic identity.
However, with increasing bombarding energy the fraction of partonic degrees-of-freedom 
grows and at RHIC energies it dominates the hadronic one in the central cells.
Shortly after hadronization, the hadronic collisions are still frequent and the momentum transfer due to collisions is large relative  to the momentum transfer due to the 
potential interactions between hadrons. Only later during the expansion, 
when the mean-free path becomes large, the momentum change due to 
the potential interaction dominates again. 
However, the potential interaction in the QMD propagation is relevant 
for the spectators and baryons in the relatively cold 'corona' during
the whole time evolution of the system.

Summarizing, at higher beam energies the potential interactions are important in 
the following cases: \\
i) for midrapidity  baryons during the final hadronic phase of the expanding 
system when the mean free path of hadrons is long which might lead to the formation of 
light clusters.
In this expanding region the inverse slope parameters of the transverse energy spectra 
of the baryons are of the order of 100 MeV and therefore for all baryons we are in an
approximately nonrelativistic regime.   \\
ii) for spectator baryons (nucleons) at all energies and during the whole time evolution.
There the Pauli principle does not allow for collisions of nucleons because 
the phase space of the outgoing channel is already occupied by other nucleons. 
Thus, the rapidity distribution of spectators changes only little during the reaction 
and they are finally the source of heavy clusters. Here the relative momentum between two nucleons is of the order of the Fermi momentum and therefore we can as well apply nonrelativistic kinematics.

\subsection{Pauli blocking}
The collisions in the overlapping zone of projectile and target are rather energetic and therefore the phase space of their final state is empty. This is not the case for collisions in the spectator matter or for participants which enter the spectator matter. There, 
the final phase space is occupied in many cases,
 thus the collision is Pauli blocked. 
The evaluation of the Pauli blocking  is a nontrivial task in QMD calculations due to the problem to define a surface of the nucleus. 
For nucleons in the center of the reaction zone, where the phase space occupation is close to unity, one can calculate the phase space occupation and apply a Monte-Carlo approach to define whether the collision is allowed or not. 
At the surface it is more difficult because the initial nucleus has there a low phase space density.
For this case a special algorithm has been developed which blocks also the collisions close to the surface effectively. For a single Au nucleus, initialized with our initialization routine, where all collisions should be blocked, we obtain a blocking rate of 96\%.  More details of the Quantum Molecular Dynamics (QMD) approach can be found in \cite{Aichelin:1991xy,Hartnack:1997ez,Hartnack:2011cn}.

\section{Cluster formation: SACA and MST}   

\subsection{Algorithms for cluster formation}

Since the transport models propagate nucleons, one needs to define a consistent theoretical
approach to build clusters out of these nucleons. In our approach clusters are formed by the same nucleon-nucleon interactions 
which rule the time evolution of the system in the course of the heavy-ion collision. We call this {\it dynamical} cluster formation in contradistinction to models where fragments are created instantaneously at a given time like in coalescence models. As discussed in the introduction, we employ
here the following two procedures for the dynamical cluster identification: 
\begin{itemize}
\item MST ( Minimum spanning tree) \cite{Aichelin:1991xy}. \\
In this approach only the coordinate space information is used to define clusters. Therefore, this method can identify clusters only  when free nucleons and groups of nucleons, called clusters, are well separated in coordinate space at the end of the reaction.  Then two nucleons are considered as part of a cluster if their distance is less than $r_0 = 2.5 fm$. Nucleons which are connected by this condition form a cluster. Nucleons with a large relative momentum are no longer close to each other at late times. Consequently, additional cuts in momentum space change the cluster distribution only little.
\item SACA (Simulated Annealing Clusterization Algorithm) 
\cite{Puri:1996qv,Puri:1998te}. \\
To overcome the limitation that clusters can only  be identified at the end of the reaction we have developed the Simulated Annealing Cluster Algorithm (SACA) approach \cite{Puri:1996qv,Puri:1998te}. It is based on the idea of Dorso and Randrup \cite{Dorso:1992ch} that the most bound configuration of nucleons and clusters, identified during the reaction, has a large overlap with the final distribution of clusters and free nucleons.  This allows to study the clusterization pattern  early,  shortly after the passing time (the time the two nuclei need to pass each other) when the different final clusters still overlap in coordinate space.  Dorso and Randrup could demonstrate this for small systems and Puri et al. \cite{Puri:1996qv,Puri:1998te} found out that it is also true for large systems. To obtain the most bound configuration  one calculates for each possible configuration of clusters and free nucleons the total binding energy,  the sum of the binding energies of all clusters.  The potential interaction between clusters is neglected as well as that between free nucleons and clusters. The binding energy is calculated using the Skyrme interaction, eq. (\ref{Upot}).
This procedure allows to identify the clusters already early during the reaction and allows therefore for the study of the origin of physical processes which involve clusters.  To determine the most bound configuration, the simulated annealing technique has been employed \cite{Puri:1996qv,Puri:1998te},  a probabilistic numerical method (realized via a ‘Metropolis’ algorithm) for finding the global minimum of a given function under constraints. 
\end{itemize}

 For very late times the differences between a fully quantal and our semiclassical approach may influence the cluster distribution because the ground state of a cluster as a quantum system of fermions has to respect a minimal average kinetic energy of the nucleons (the Fermi energy if the nucleons are confined in a sphere)  whereas that of our semi-classical approach does not have to obey this condition. Therefore,  nucleons may still  be emitted even if in the corresponding quantum system this is not possible anymore. It takes, however, quite long, considerably more than 100 fm/c, until one of the cluster nucleons gains so much kinetic energy that it can overcome the potential barrier.

None of these approaches to determine clusters influences the time evolution of the heavy-ion reaction. The underlying PHQMD approach propagates in the QMD mode only baryons, but not clusters. If applied at different times during a heavy-ion reaction, the SACA approach allows to study the time evolution  of cluster formation. It has been shown that for large times SACA and MST yield very similar results \cite{Puri:1996qv,Puri:1998te} and that the results agree well with the experimental findings for clusters with $Z \ge 3$ \cite{Zbiri:2006ts}. 

We note that the clusterization algorithms (SACA and MST) find clusters in the rest frame 
of target/projectile spectators while the heavy-ion dynamics is realized in the initial $NN$ center-of-mass system in which spectators are
squeezed due to the Lorentz contraction of
initial nuclei at relativistic energies -- cf. Eq. (\ref{densGam}).
In order to obtain the right kinematical 'input' for finding the cluster in the 
spectator regions, we apply the inverse Lorentz  transformation  with $\gamma_{cm}$ containing the velocity between the $NN$ center-of-mass and  the respective rest system at
target/projectile region.
This approximation is justified even at high beam energies since with increasing $\gamma_{cm}$ the passing time of the heavy nuclei decreases as compared to $R/v_{Fermi}$ (where $R$ is the radius of the nucleus and $v_{Fermi}$ is the Fermi velocity). Thus, the spectators are practically frozen until the end of the violent part of the reaction.
Moreover, this approximation is applied for clusterization routines only and, thus, 
does not affect the general nucleon dynamics in the PHQMD.

 If one aims at a better quantitative description of lighter clusters or isotope yields additional efforts are necessary. The binding energy of those clusters cannot be described by the 
Weizs\"acker mass formula (which corresponds well to the cluster binding energies calculated by Skyrme type interactions  \cite{Aichelin:1991xy} -- as will be discussed later, but show shell effects and other quantum features. To study this, as well as the isotopic yields, the SACA algorithm is presently under improvement to include  shell effects, symmetry energy and pairing energy as well as the interaction between hyperons and nucleons \cite{FRIGA2019}).   Because the propagation of nucleons in PHQMD contains presently neither symmetry nor pairing energy terms we do not include these new features in this paper with the exception of the hyperon-nucleon interaction which is taken as 2/3 of the nucleon-nucleon interaction, assuming in this first study that the strange quark is inert. For the identification of the light clusters, $Z \le 2$, we use MST.

We note that the consistent description of cluster production
at high energies, where many resonances are excited,
is an open issue. Due to that we avoid in this study to apply
SACA/MST at very early times when the matter is
still resonance dominated, nevertheless even during the
later expansion the presence of resonances has to be accounted
for. To test their influence on the cluster yield we adopt the following procedure
in the present calculations:
1) at the selected time, before the
SACA/MST is called, we let "virtually" decay the baryon
resonances to nucleons and mesons. These decay nucleons
are then taken into consideration for cluster formation in the
SACA/MST algorithm (while the baryon resonances are propagated
further in the PHQMD code until their natural decay).
Under such an assumption we obtain a rather stable
pattern of clusters in time with the SACA/MST algorithm. \\
2) This we compare with the cluster yield obtained if we do not include
the nucleons from resonance "virtual" decays in SACA/MST, where
we find less clusters (by 5-10\%) at the early times of this study
since less nucleons are available for clusterization.
At the later time the results are similar in both scenarios. \\
Further inside in to the cluster formation and the role of resonances
can be expected by employing the persistent coefficient which measures
to which degree a cluster, measured at different times, contains the
same nucleons. This will be the subject of an upcoming study.

\subsection{QMD dynamics and cluster formation}

\begin{figure}[t]%
\centering
\includegraphics[scale=.5]{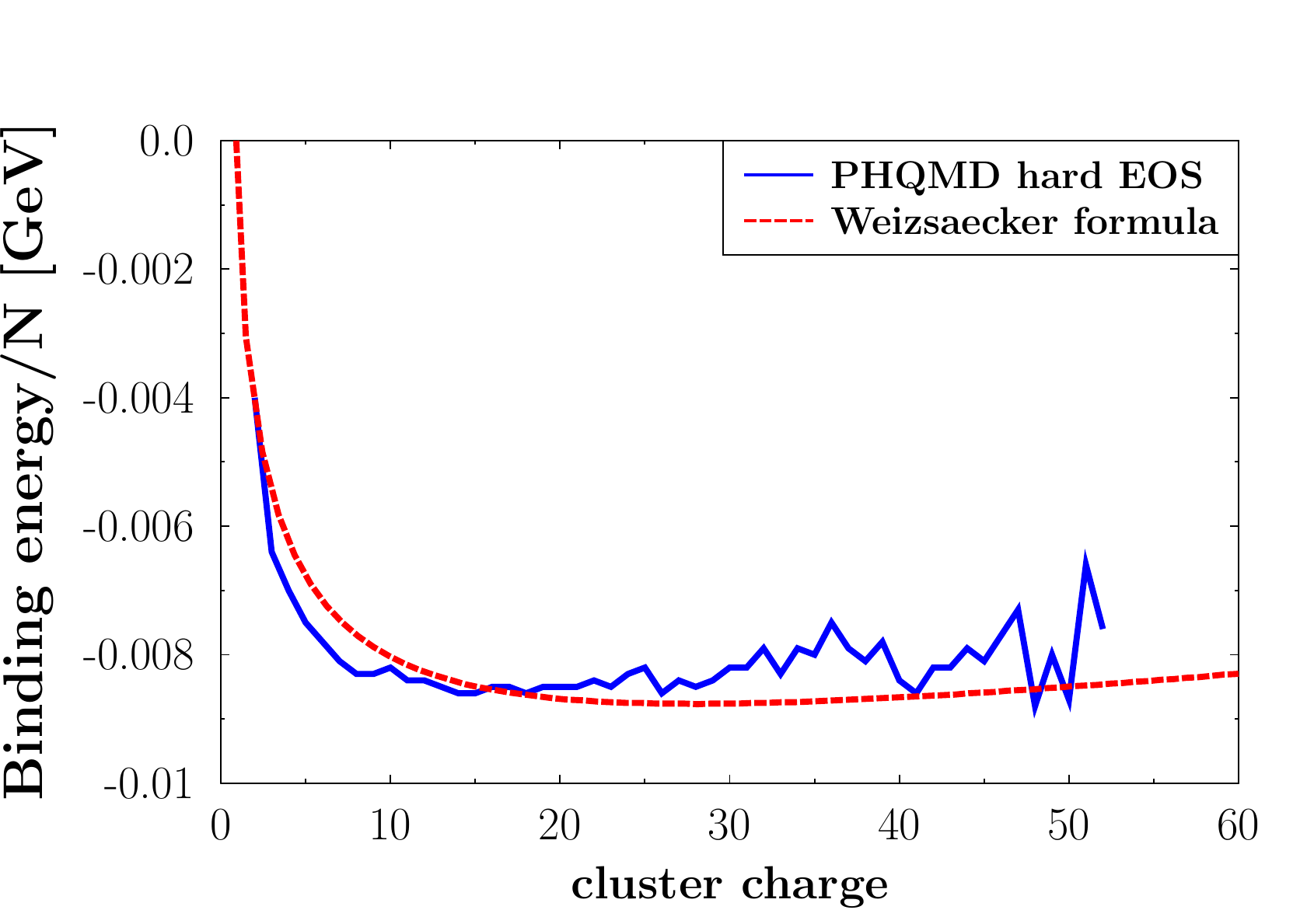}
\caption{\label{b8} (Color online) 
Average binding energy of the clusters identified by SACA algorithm from
the Au+Au collisions at 600 $A$GeV calculated within the PHQMD with the hard EoS 
as a function of the cluster charge calculated at late times (150 fm/c). }  
\end{figure} 

One of the conditions for any reasonable approach to cluster formation is the requirement that the binding energy of clusters is reproduced. A too small binding energy means that the clusters are excited and emit further nucleons or $\alpha$'s.  Fig. \ref{b8} shows the average binding energy of clusters at the end of a heavy-ion reaction of Au+Au at 600 $A$GeV as compared to the Weizs\"acker mass formula. The clusters have been determined by the SACA algorithm.
The binding energies do not vary for different beam energies and are stable from 75 fm/c on. We see that for clusters with $Z \ge 5$ the binding energy is close to that expected from the Weizs\"acker mass formula. This is all but self-evident. In PHQMD the density inside the clusters is given by the superposition of Gaussians and there is no well defined surface. The  binding energy is given by the expectation value of the Skyrme and Coulomb interaction for this spatial configuration supplemented by the total kinetic energy in the cluster rest system.

\begin{figure}[th!]%
\centering
\includegraphics[scale=.5]{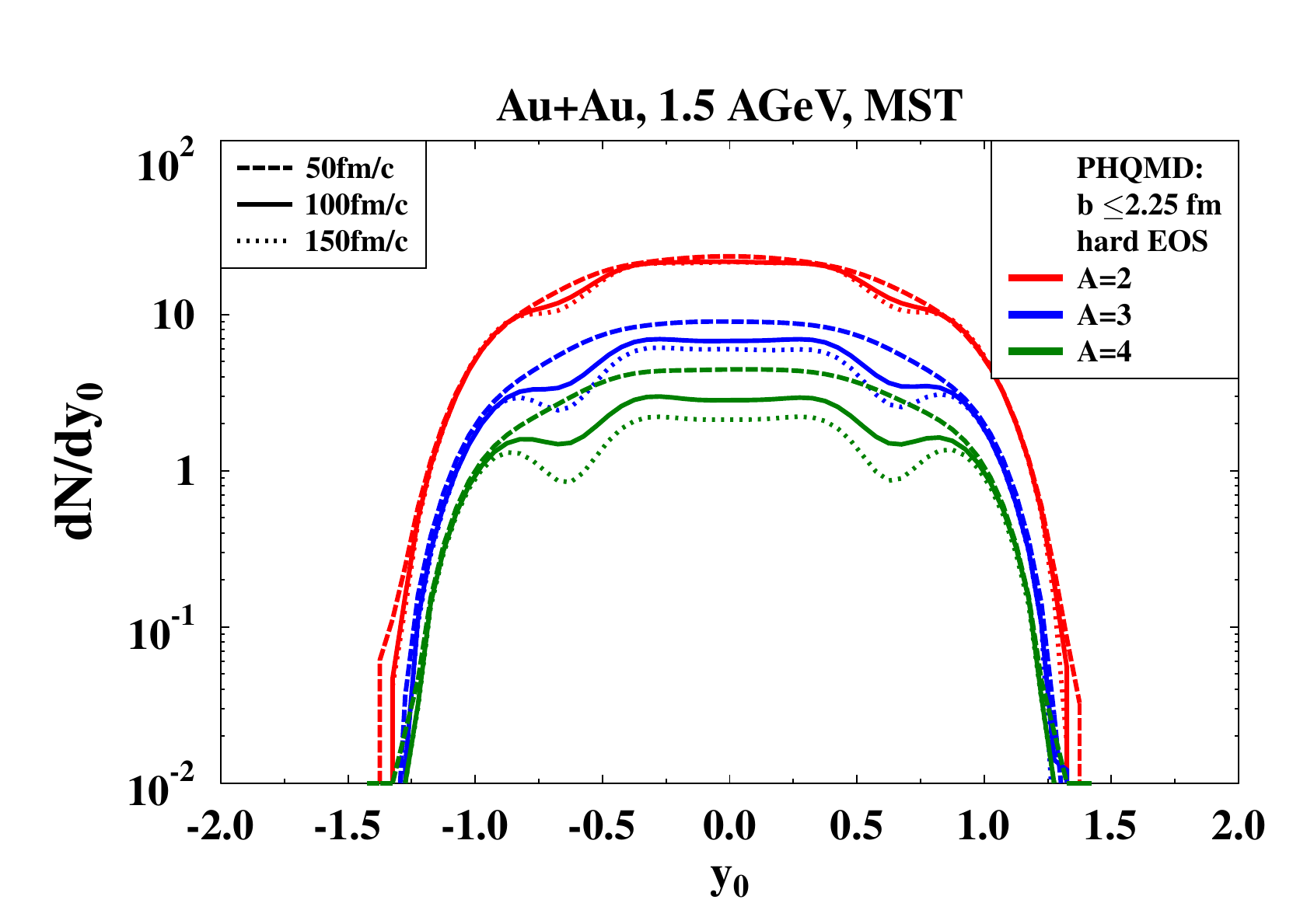}
\caption{\label{FOPI3} (Color online) 
Scaled rapidity distribution, $y_0=y/y_{proj}$, of clusters of  mass number
$A=2,3,4$ for central Au+Au collisions at 1.5  $A$GeV calculated
within the PHQMD approach. 
The clusters are determined by the MST algorithm at  $t=50$ fm/c, $t=100$ fm/c and $t=150$ fm/c.}
\end{figure}

The nucleon and cluster rapidity distribution is another key observable which characterizes an heavy-ion collision.  
In Fig. \ref{FOPI3} we display the scaled rapidity distribution $y_0=y/y_{proj}$ (where $y_{proj}$ is a projectile rapidity in the center-of-mass system)  of light clusters  of mass numbers $A=2,3,4$ for central Au+Au reactions at $E_{beam}=1.5$ 
$A$GeV. 
The clusters are determined by the MST algorithm at  $t=50$ fm/c, $t=100$ fm/c and $t=150$ fm/c.
 We see that the cluster yields are rather stable versus time.
We note that we find $\sim$ 10\% less clusters at 50 fm/c without accounting 
for nucleons from the 'virtual' decay of resonances for the cluster formation
- as discussed in Section III.A.
 
\begin{figure}[h!]%
\centering
\includegraphics[scale=.5]{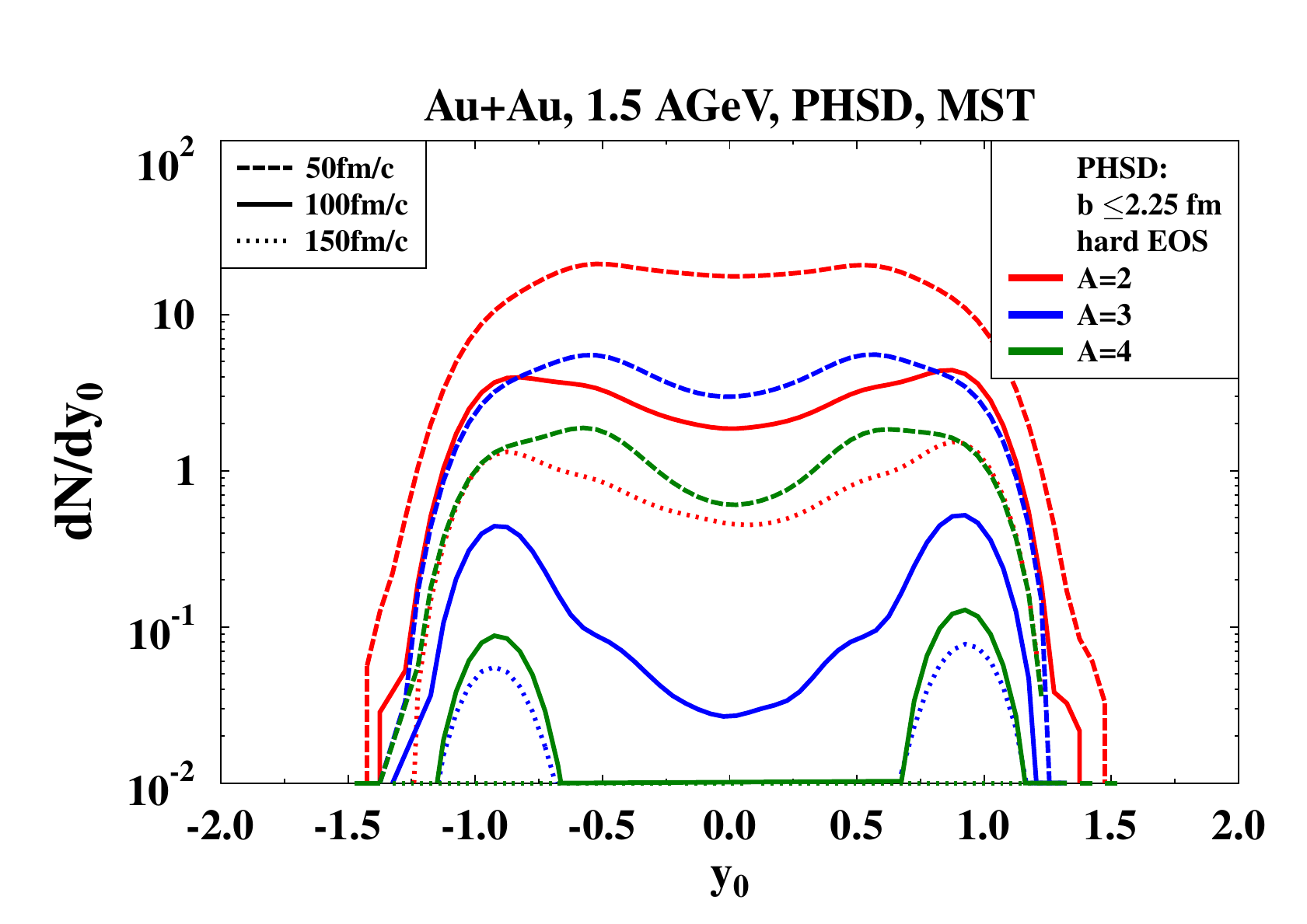}
\caption{\label{FOPI3yPHSD} (Color online) 
Scaled rapidity distribution, $y_0=y/y_{proj}$, of clusters of different mass number,
$A=2,3,4$, for central Au+Au reactions at 1.5 $A$GeV using 
the mean-field dynamics of the PHSD approach. 
The clusters are determined with the MST algorithm at  $t=50$ fm/c, $t=100$ fm/c and $t=150$ fm/c.}
\end{figure}

Fig. \ref{FOPI3yPHSD} presents the same scaled rapidity distribution of light clusters 
as in Fig. \ref{FOPI3}, however, calculated within the mean-field dynamics of PHSD.
One can see that the shape of the MF cluster distribution is rather different from that 
of QMD. Moreover, the MF cluster yield is not stable in time. 
This illustrates the limitation of the applicability of the mean-field  dynamics for the 
cluster identifications. 
We observe furthermore that in the mean-field approach the clusters at midrapidity disappear early
whereas those around projectile and target rapidity are longer present. This is expected because clusters 
at midrapdity are created by density fluctuations whereas those at projectile/target rapidity are mainly made
of spectators which disintegrate slowly in mean-field approaches. The disappearance of fragments and, even more,
the different times of disappearance questions the applicability of coalescence models to mean-field calculations. 


\begin{figure}[h!]%
\centering
\includegraphics[scale=.5]{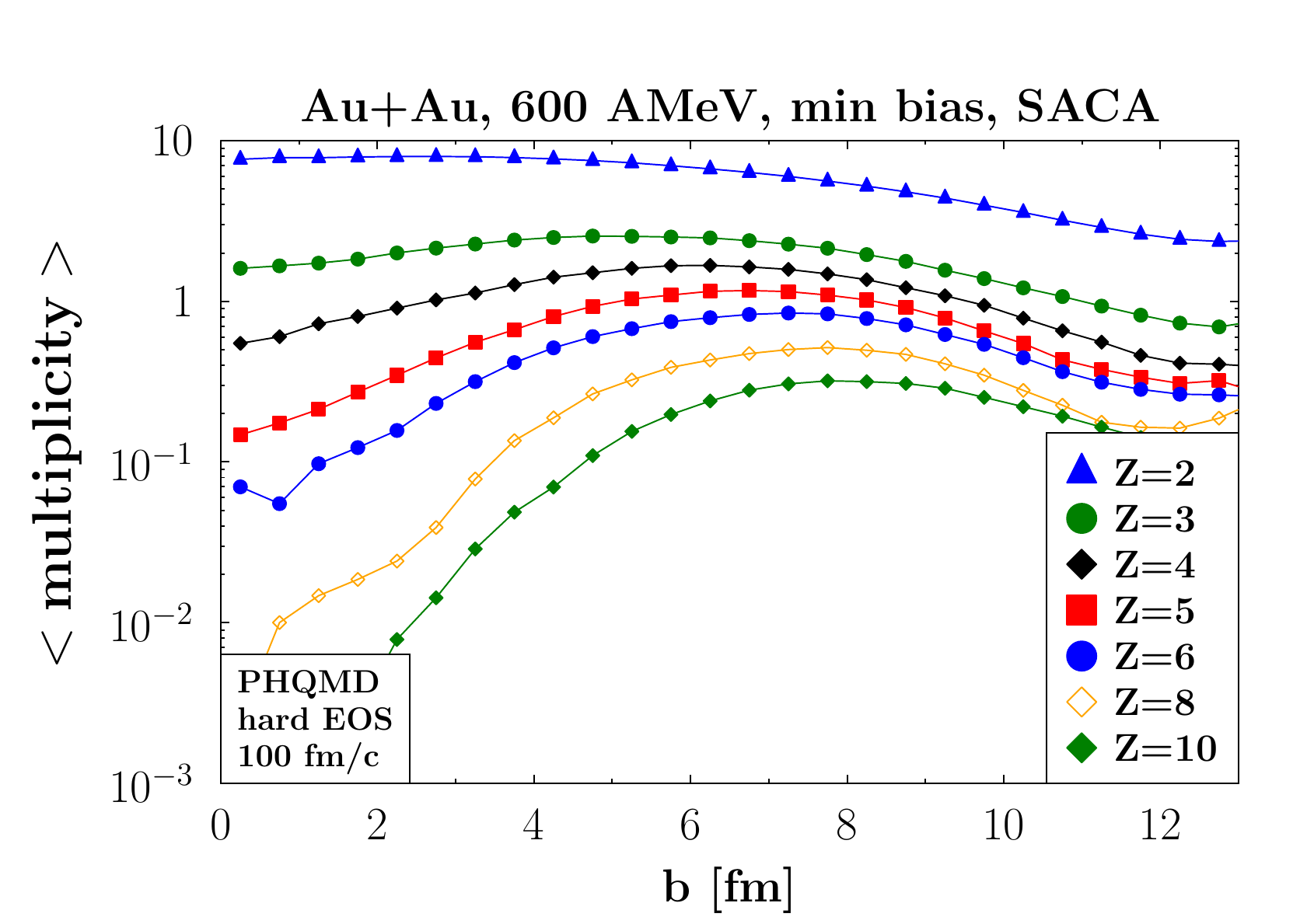}
\includegraphics[scale=.5]{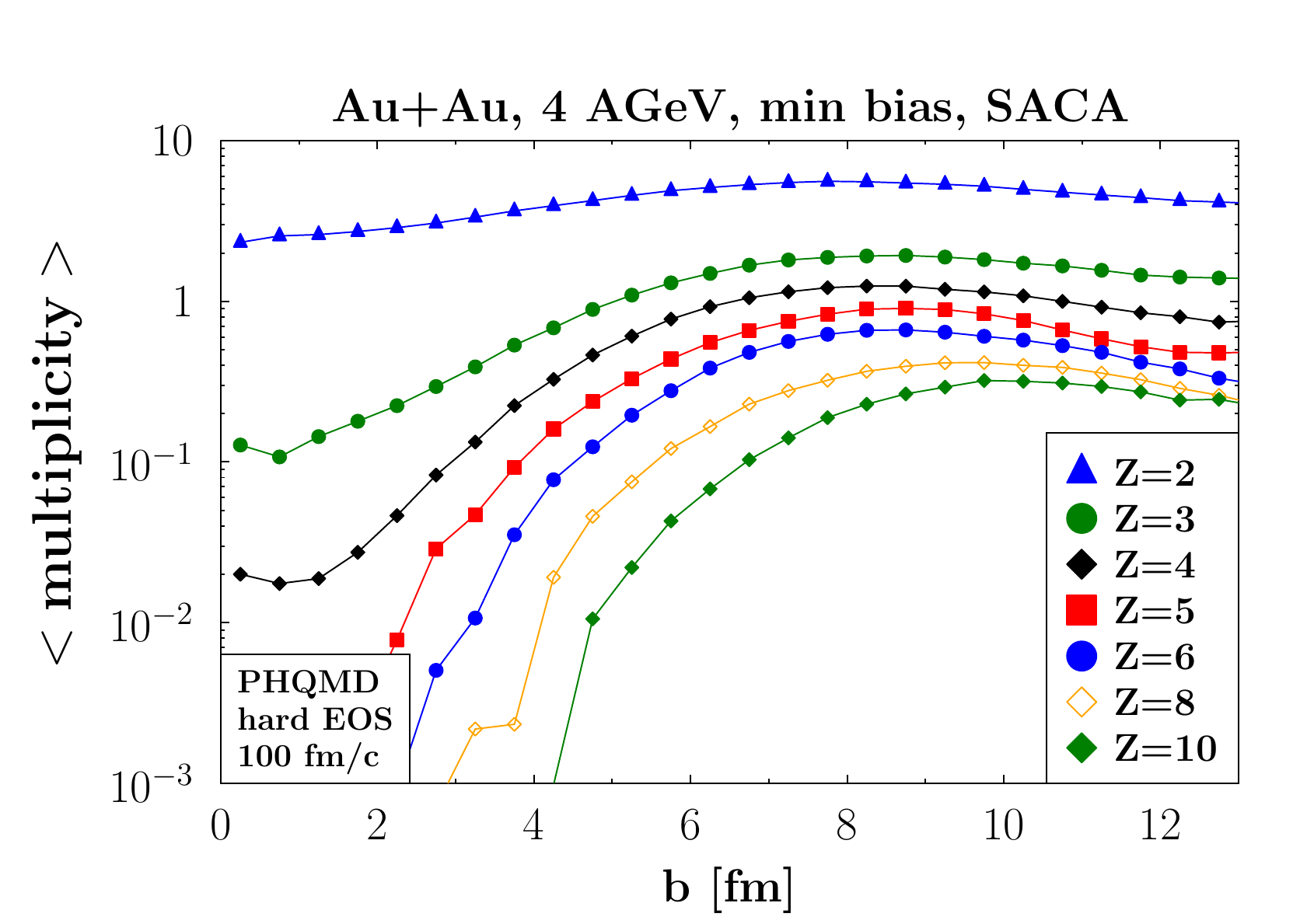}
\caption{\label{b2} (Color online) 
 Multiplicity of clusters  with $Z=2 - 10$  in Au+Au collisions 
as a function of the impact parameter for two different beam energies, 
$E_{beam}$ = 600 $A$MeV (upper plot) and 4 $A$GeV (lower plot)
calculated within the PHQMD (hard EoS). 
The SACA algorithm is used to identify the clusters at time 100 fm/c.}  
\end{figure}

Fig. \ref{b2}  displays the multiplicity of clusters with $Z=2 - 10$ for  Au+Au collisions as a function of centrality, represented by the impact parameter, for two different energies, 
$E_{beam}$= 600 $A$MeV (upper plot) and = 4 $A$GeV (lower plot).   
At very central collisions most of the nucleons are unbound, however, 
even if there some light clusters are produced whose number decreases with increasing 
the beam energy.  For larger impact parameters the intermediate mass clusters become important, they are mostly produced from the spectator matter.  The general trend is similar for both energies but the multiplicities differ in detail. The origin for this difference is that the number of participant nucleons, which enter the spectator matter and cause its instability, as well as the momenta of those nucleons, depend on the beam energy. 
\begin{figure}[h!]%
\centering
\includegraphics[scale=.5]{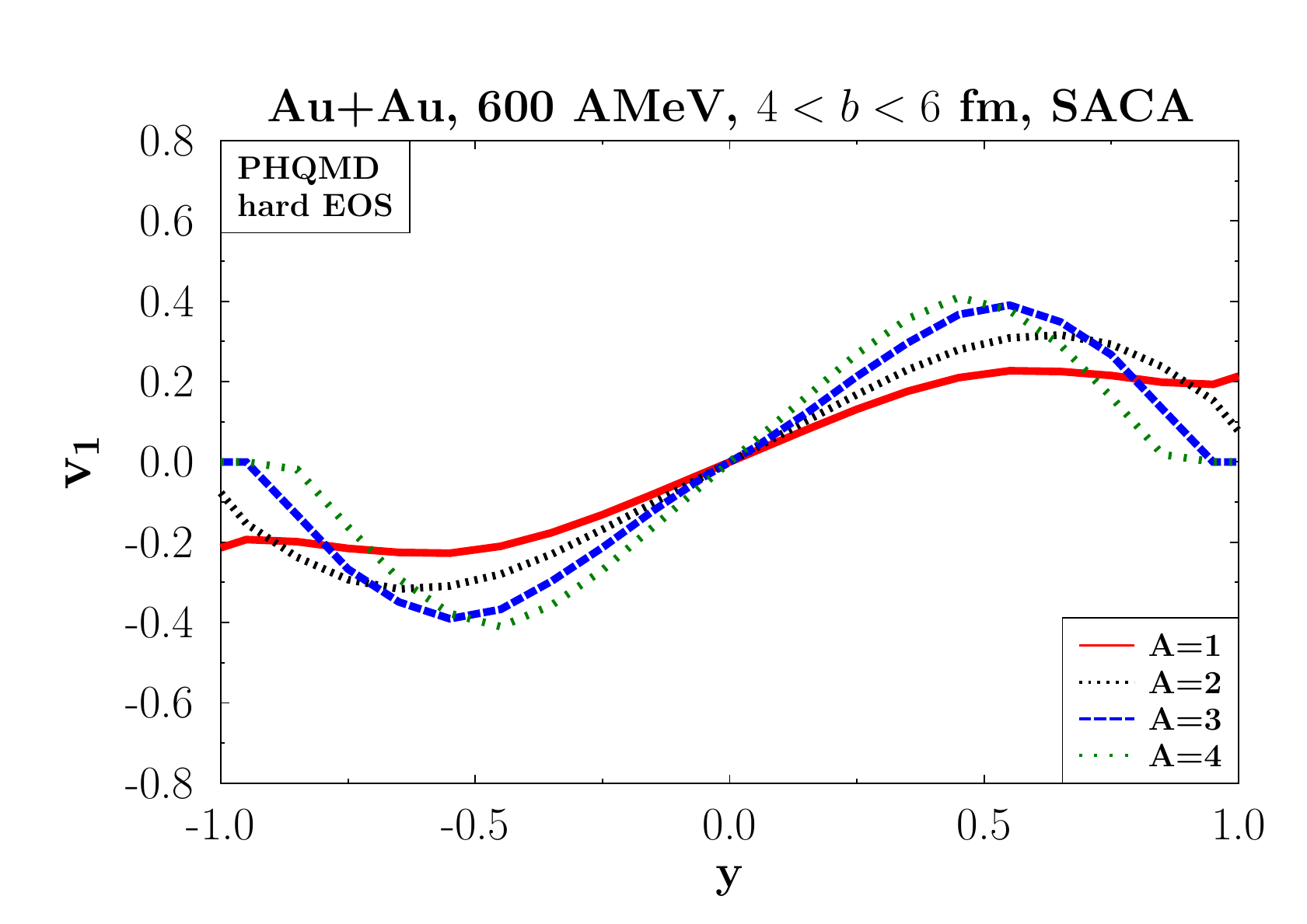}
\includegraphics[scale=.5]{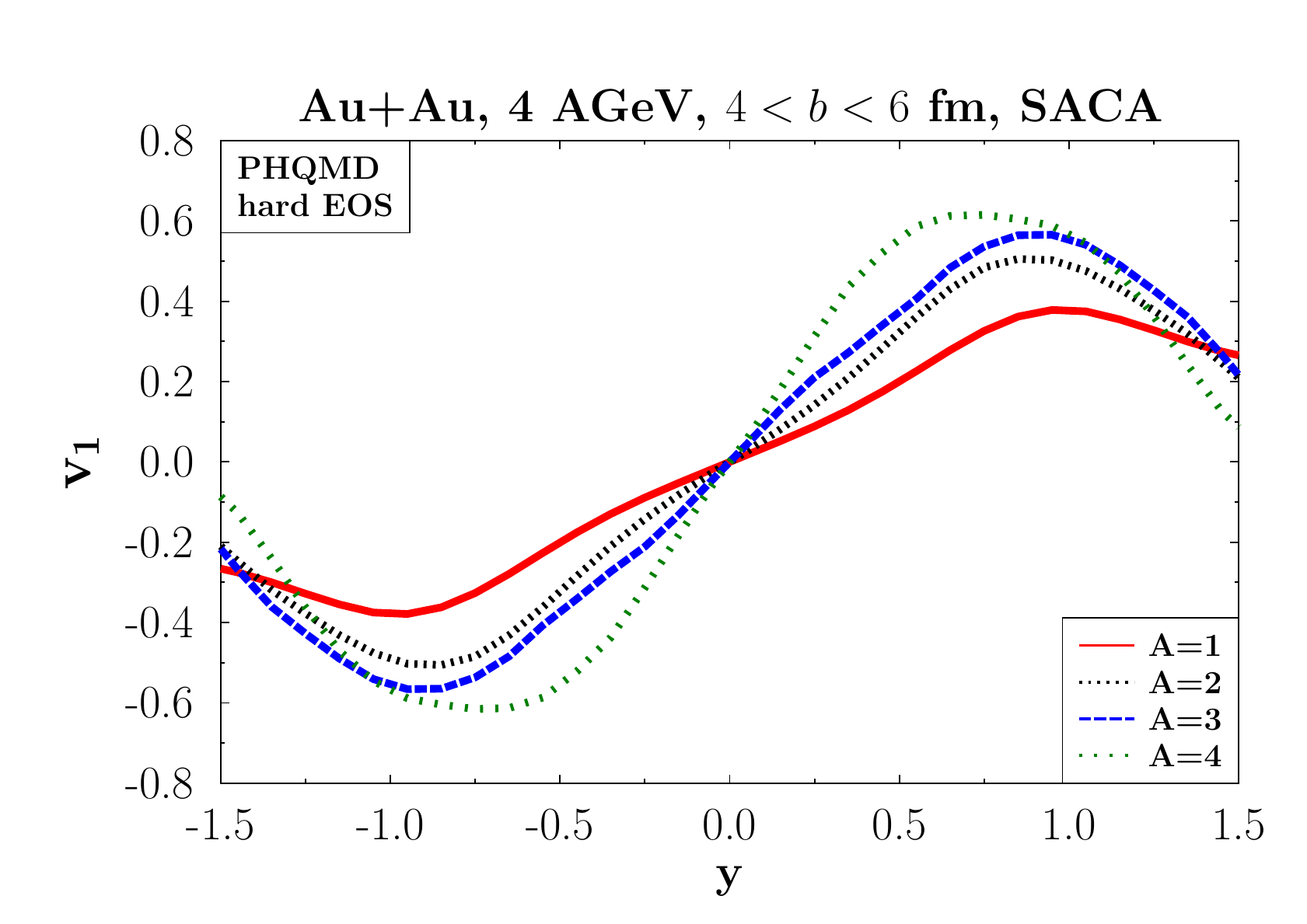}
\caption{\label{b4} (Color online) 
The in-plane flow, $v_1$, as a function of the rapidity $y$ 
for nucleons ($A=1$) and clusters of mass numbers $A=2,3,4$
for Au+Au collisions at 
beam energies of 600 $A$MeV ($y_{proj}=0.539$) -- upper plot, and of 4 $A$GeV 
($y_{proj}=1.17$) -- lower plot, for  an impact parameter range 
of $4\le b\le 6$ fm calculated within the PHQMD (hard EoS) using SACA algorithm
for the cluster recognition. } 
\end{figure} 

Another observable of interest is the in-plane flow, $v_1$, described by the first coefficient of the  Fourier expansion of the azimuthal distribution of nucleons or clusters 
\be
\frac{dN}{d\phi}= N_0(1+v_1 \cos{\phi} +2v_2 \cos{2\phi}....).
\ee
The in-plane flow is created, on the one side,  by the geometry of the reaction zone which allows hadrons with outward momentum to escape from reaction zone (and therefore even in cascade calculations a finite $v_1$ is obtained)  and, on the other side, by
the transverse force, $F_T$. This force is proportional to  the density gradient in transverse direction and is large at the interface between participant and spectator region. The relative importance of both sources of $v_1$ (geometrical and interaction) depends on the cluster size. 
Light clusters come predominantly from the transition region between spectators and participants and show a larger $v_1$ around projectile rapidity than single nucleons which come also from the high density participant region where the density gradient and therefore $v_1$ is smaller 
\cite{Reisdorf:2010aa}.
With increasing energy the passing time $t_{pass}$  decreases but on the other side the density gradient, and hence the force $F_T$  becomes steeper. Both effects almost compensate each other such that only a mild increase of  $\Delta p_T= F_T t_{pass}$ 
occurs. 

In Fig. \ref{b4} we show $v_1$ as a function of center-of-mass rapidity $y$ for nucleons ($A=1$) and clusters of different sizes ($A=2,3,4$), 
created in Au+Au collisions at two beam energies, $E_{beam}$ = 600 $A$MeV (upper plot)
and 4 $A$GeV (lower plot), for  an impact parameter range of
 $4\le b\le 6$ fm. 
One sees that $v_1$ increases with the mass number of the cluster.
 Even for light clusters  $v_1$ differs significantly from that of protons and neutrons $(A=1)$, 
in particular the slope at midrapidity (which is often used to characterize the in-plane flow for the cases where only a limited rapidity interval can been measured) 
differs significantly for different $A$. The tendency that the large clusters (which have a higher probability to come from 
the spectator matter) show a large $v_1$, is found to be the same for
both energies considered here, also the value of $v_1$ is similar. This mass dependence of the dynamical variables has also been found experimentally \cite{FOPI:2011aa}.

\begin{figure}[t!]%
\centering
\includegraphics[scale=0.45]{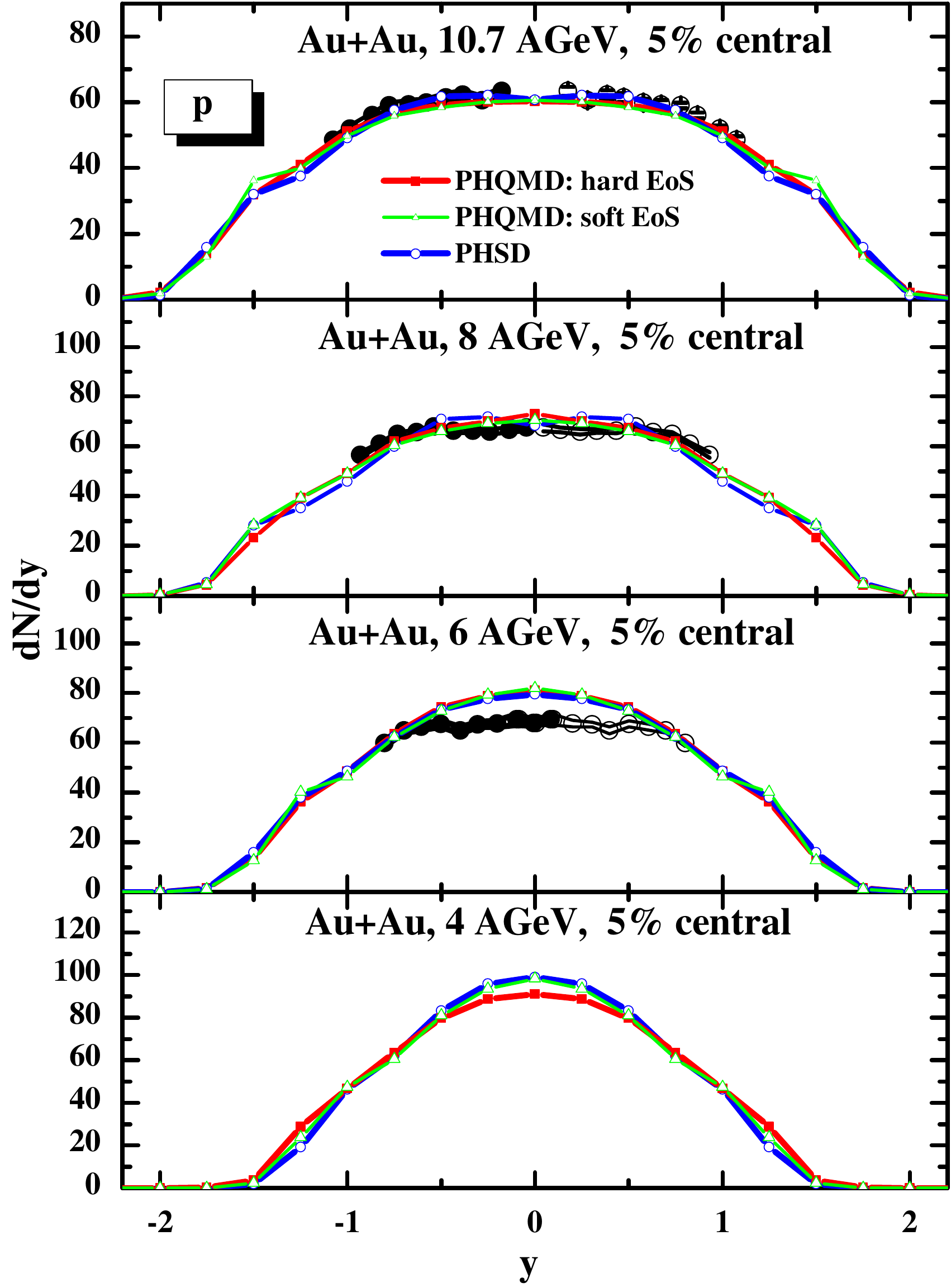}
\caption{\label{AGS2} (Color online) 
The rapidity distributions of protons for 5\% central Au+Au collisions
at 4, 6, 8, 10.7 $A$GeV (plots from lower to upper). 
The experimental data have been taken from Ref.~\cite{E917p02}. 
The full symbols correspond to the measured data, whereas
the open symbols are the data reflected at midrapidity.
Solid red lines with open squares refer to the PHQMD results with a hard EoS, 
the green line with open triangles for PHQMD results with a soft EoS, the blue lines with open circles  for the PHSD results.}
\end{figure}

\begin{figure}[t!]%
\centering
\includegraphics[scale=.45]{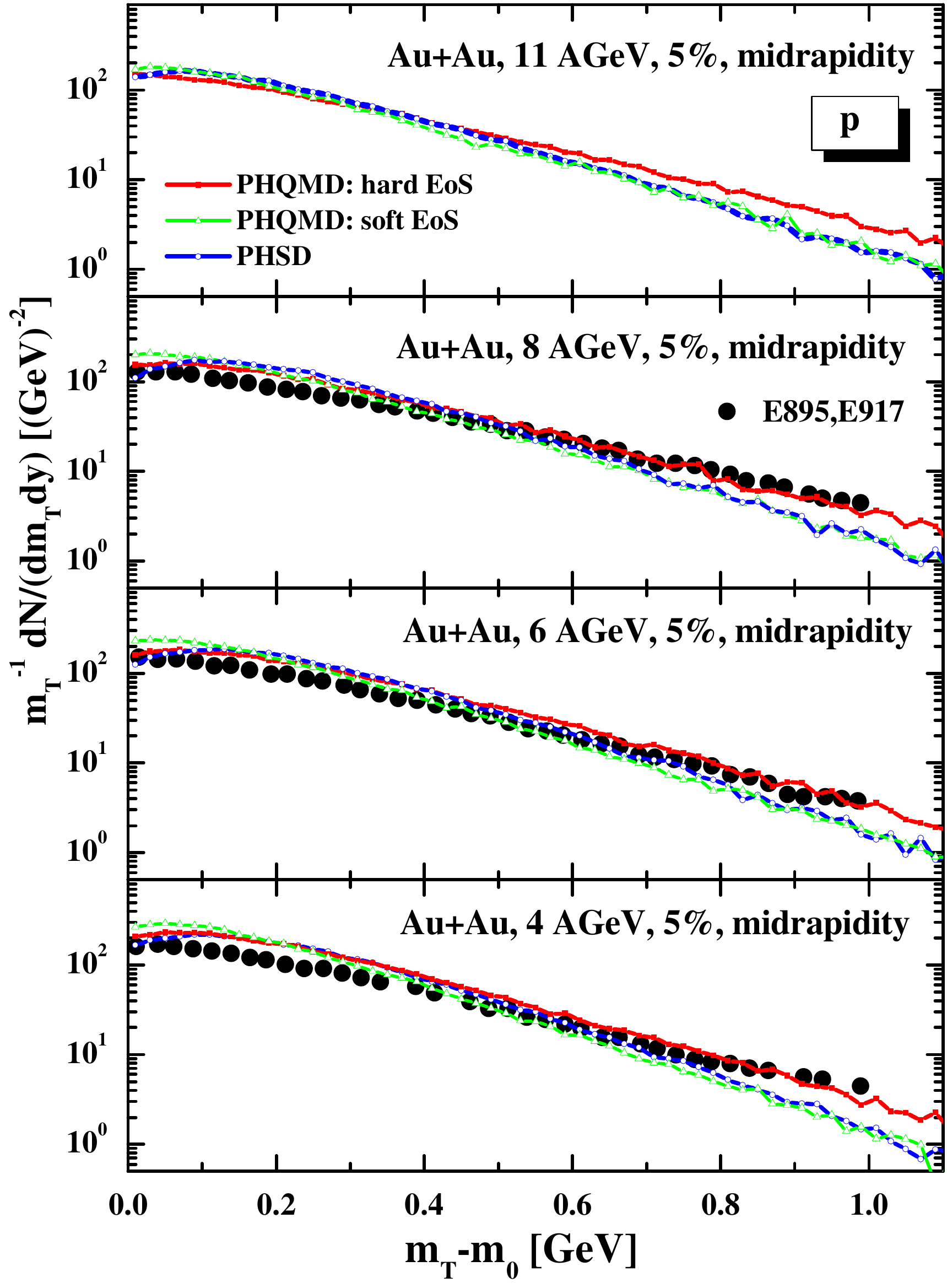}
\caption{\label{AGS2mt} (Color online) 
The transverse mass $m_T$-spectra of protons at midrapidity
for 5\% central Au+Au collisions
at 4, 6, 8, 10.7 $A$GeV (plots from lower to upper). 
The experimental data  have been taken from Ref.~\cite{E917p02}.
Solid red lines with open squares refer to the PHQMD results with a hard EoS, 
the green line with open triangles for the PHQMD results with a soft EoS, 
the blue lines with open circles  for the PHSD results.}
\end{figure}
\section{Results for hadronic spectra}

In this section we present the results of the PHQMD approach for the basic 'bulk'
observables like the rapidity $y$-distribution and the transverse mass $m_T$ spectra
of hadrons -- protons, anti-protons, pions, (anti-)kaons and (anti-)Lambdas
at a variety of energies -- from SIS to top RHIC energies - and confront
our results to the experimental data. All rapidities are measured in the center of mass of the nucleus-nucleus system.
We recall that the "bulk" observables have been extensively investigated in many
PHSD studies and a good agreement for a variety of 'bulk' as well as for the collective 
flows $v_n$, electromagnetic, heavy flavour etc. observables have been reported - 
cf. \cite{Cassing:2008sv,Cassing:2008nn,Cassing:2009vt,Bratkovskaya:2011wp,Linnyk:2015rco}.
However, it is necessary to verify the 'bulk' dynamics within the novel PHQMD approach 
because the  initialization of the nucleus as well as the nucleon dynamics
are realized differently.
In this respect the PHQMD provides an unique possibility to explore the differences
between the mean-field and the quantum-molecular dynamics since both are realized in the framework
of the same PHQMD code, i.e. both propagations can be tested while implying the collision
integral of PHSD.
This allows  to investigate how a different realization of the potential interaction 
-- MF versus QMD, may modify the trajectories of the individual nucleons in phase space. 
Also the interacting Gaussian wave functions in QMD with a given width have a different 
time evolution as compared to point like nucleons in a mean-field.  
Also we explore the influence of the EoS - hard vs. soft - realized with a static
density dependent potential in the QMD mode as discussed in Section IIC.


\subsection{AGS energies }

\begin{figure}[t!]%
\centering
\includegraphics[scale=.42]{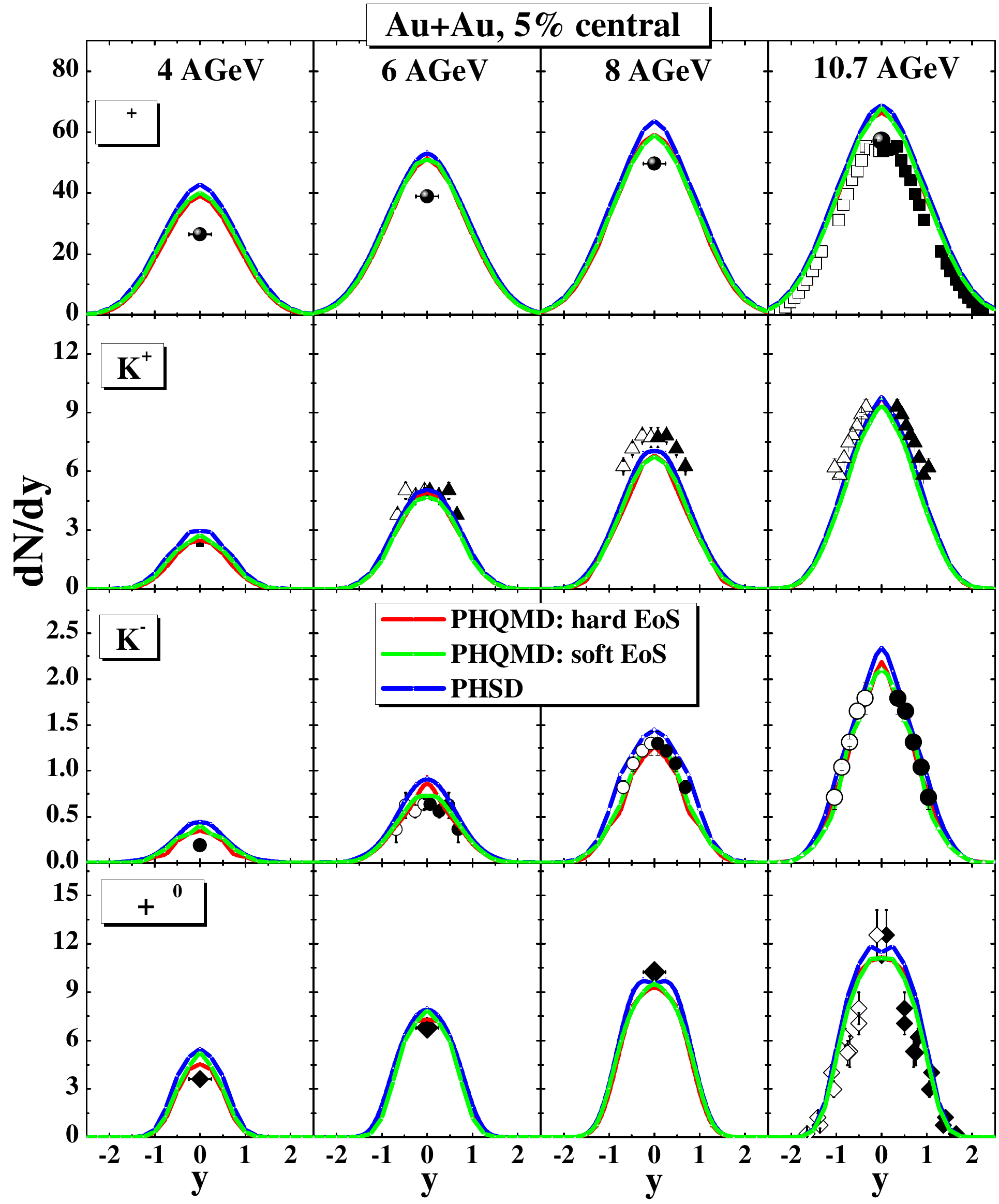}%
\caption{\label{yAGS} (Color online) 
The rapidity distributions of $\pi^+,  K^+, K^-$, and $\Lambda
+\Sigma^0$’s for 5\% central Au+Au collisions at 4, 6, 8 and 10.7 $A$GeV 
(plots from left to right) in comparison to the experimental data 
from Refs. \cite{E866E917,E917K,E866pi11,E877pi11,E891Lam,E896Lam}.
Solid red lines with open squares refer to PHQMD results with a hard EoS, 
the green line with open triangles for the PHQMD results with a soft EoS, 
the blue lines with open circles  for the PHSD results. }
\end{figure}

\begin{figure}[t!]%
\centering
\includegraphics[scale=0.4]{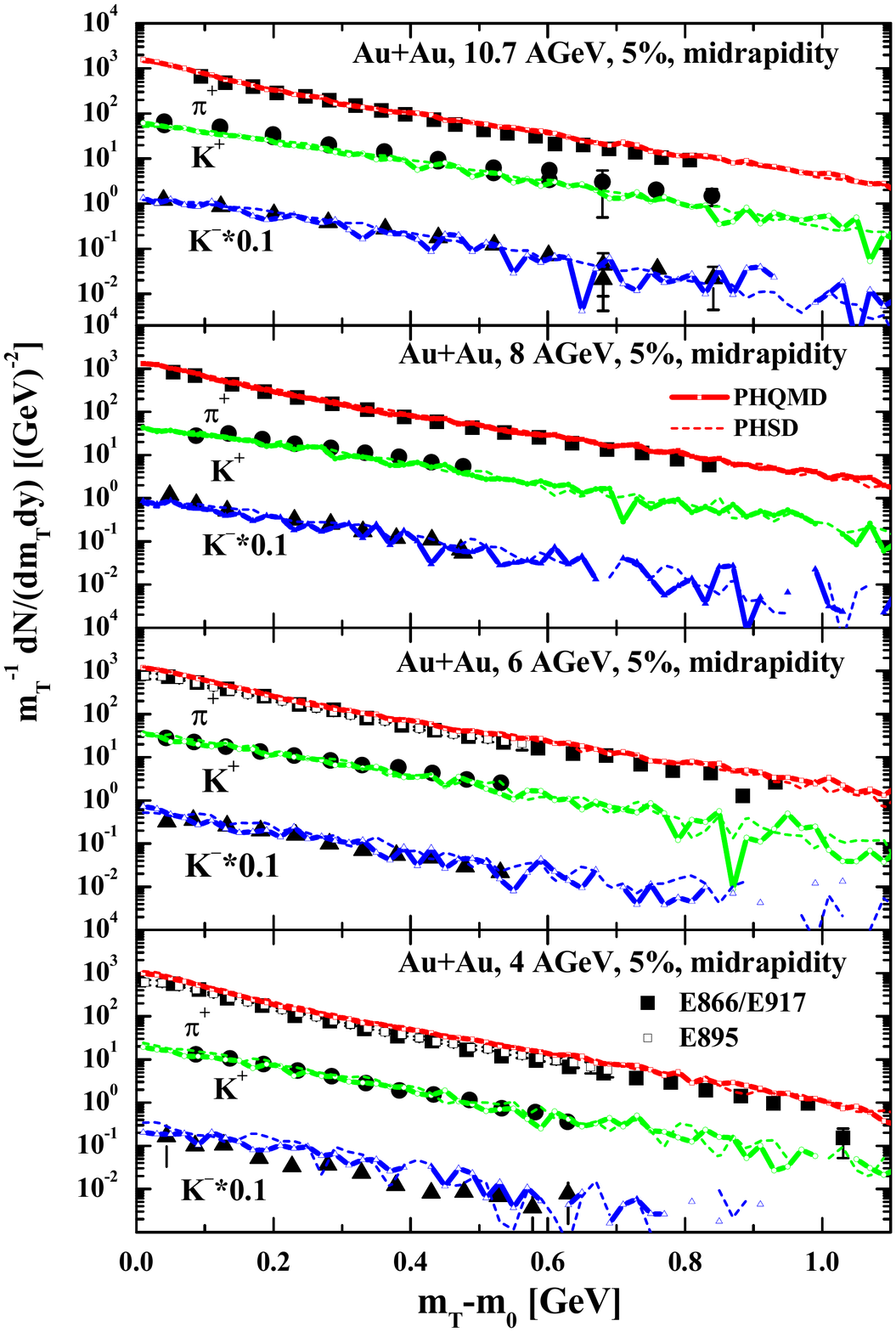}%
\caption{\label{AGS1mt}(Color online) 
The transverse mass $m_T$- spectra of $\pi^+,  K^+, K^-$, and $\Lambda
+\Sigma^0$’s at midrapidity for 5\% central Au+Au collisions at 4, 6, 8 and 10.7 $A$GeV 
(plots from lower to upper) in comparison to the experimental data 
from Refs. \cite{E866E917,E917K,E866pi11,E877pi11,E891Lam,E896Lam}. 
Solid lines with open symbols refer to PHQMD results with a hard EoS, 
the dashed line for the PHSD results.}
\end{figure}

We start our comparison by showing in Figs. \ref{AGS2} and \ref{AGS2mt}
the proton rapidity distributions and $m_T$ spectra  for central Au+Au collisions at 
beam energies of $4, \ 6,\ 8$ and 10.7 $A$GeV, calculated in  PHQMD 
with a hard and a soft EoS. The PHQMD results are compared with those from PHSD 
as well as with the AGS experimental data
\cite{E866E917,E917K,E866pi11,E877pi11,E891Lam,E896Lam,E917p02}.   
In the rapidity spectra the influence of the EOS becomes only slightly visible 
at the lowest beam energy but the transverse mass spectra show a sensitivity to 
the EOS at all energies. 
A hard EOS increases the slope of the spectra at large $m_T$ and lowers the yield 
at low $m_T$ as compared to a soft EoS. We find that the PHQMD with soft EoS agrees
very well with the PHSD result.
This agreement with experiment allows to conclude that the stopping of the nuclei in PHQMD 
is reasonably described. The latter is important for the interpretation of the results 
for the newly produced hadrons since their abundances are sensitive
to the energy loss of the initial colliding nucleons, i.e. to the fraction of
their kinetic energy which will be converted into mass production.

In Figs. \ref{yAGS}  we display the rapidity distribution and in Fig. \ref{AGS1mt} the
$m_T$- spectra of $\pi^+, K^+, K^-$  and $ \Lambda +\Sigma^0$, produced 
in central Au+Au collisions for different beam energies, $E_{lab} = 4,\ 6,\ 8$ and
10.7  $A$GeV. Again we compare here the PHQMD calculations with a soft and a hard EoS
with the PHSD results (we note that for the $m_T$ spectra we show only for hard PHQMD
and PHSD results for a more clear presentation). Contrary to the proton
$m_T$- spectra, which show a visible sensitivity to the EoS,
the spectra of newly produced hadrons indicate only a very mild dependence on the nucleon
potential -- all cases are rather similar to each other.

\subsection{SPS energies }

\begin{figure}[t!]%
\hspace*{-1cm}\includegraphics[scale=.45]{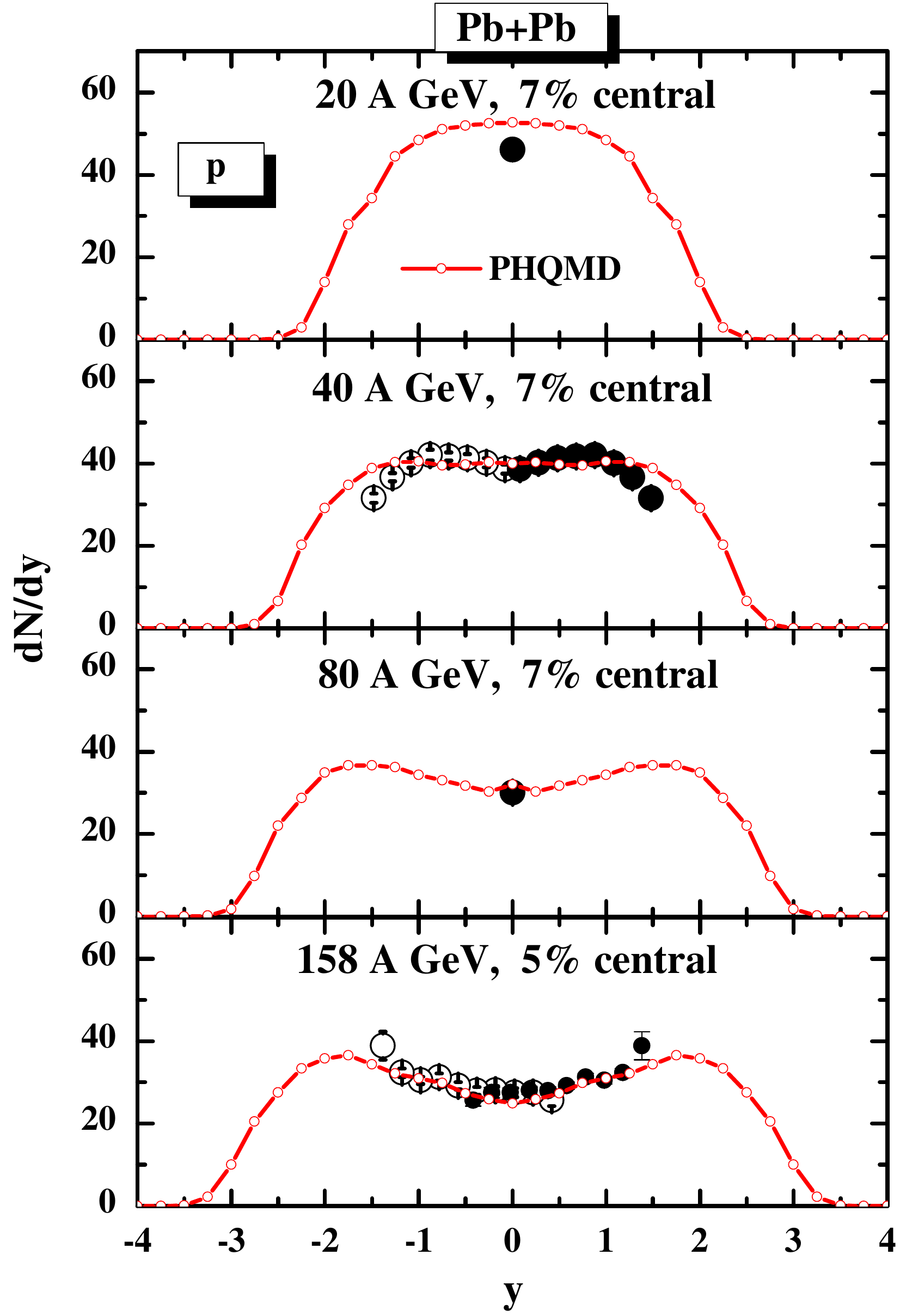}
\caption{\label{SPS} (Color online) 
The rapidity distributions of protons at midrapidity for 5\% central Pb+Pb collisions
at 20, 40, 80 and 158 $A$GeV (plots from upper to lower). 
The experimental data  have been taken from
Refs.~\cite{NA49pold,NA49_T,NA49_Lam}. The full symbols correspond to the measured data, whereas
the open symbols are the data reflected at midrapidity.
Solid red lines with open squares refer to PHQMD results with a hard EoS.}
\end{figure}

\begin{figure}[t!]%
\centering
\includegraphics[scale=.4]{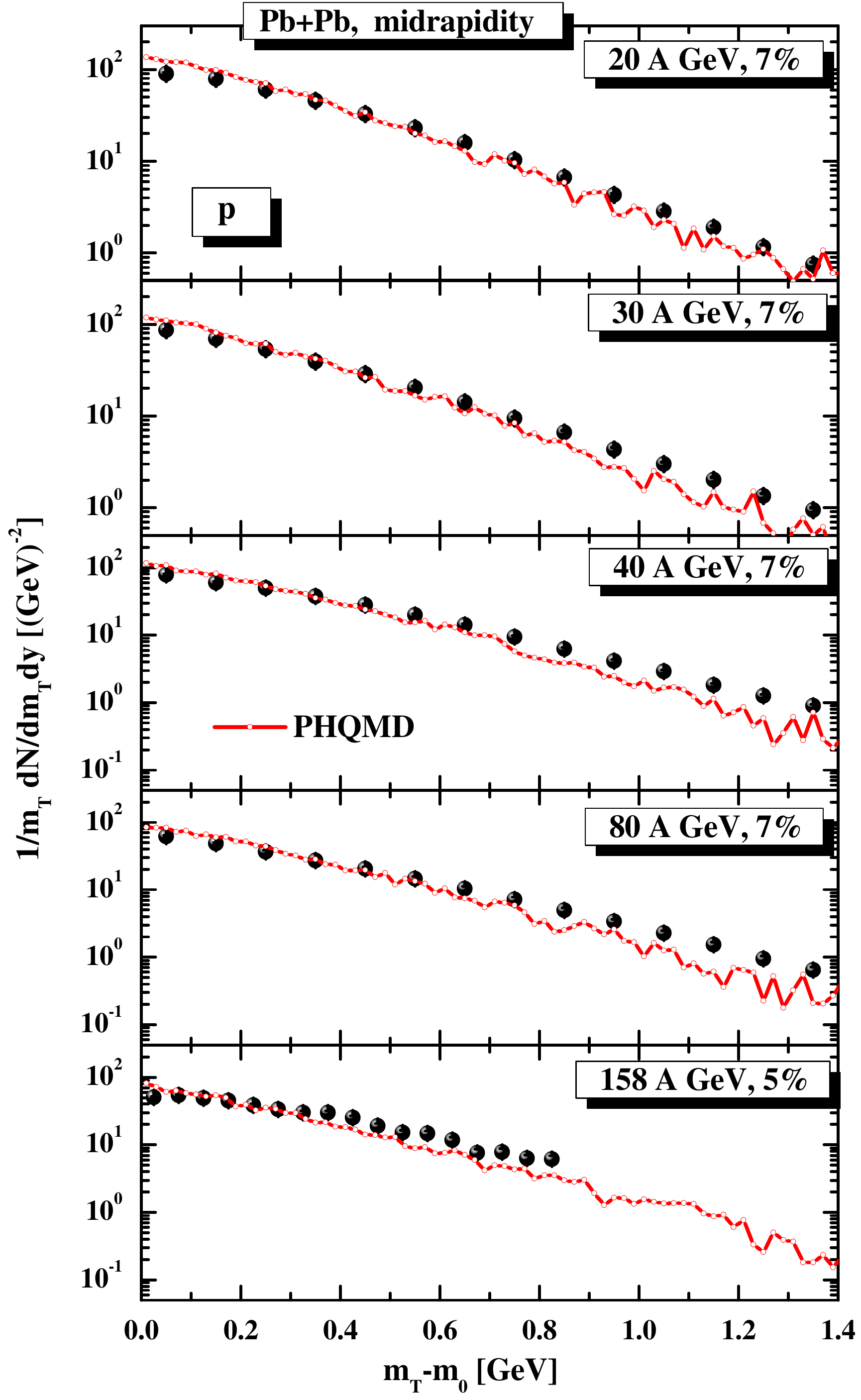}%
\caption{\label{mtSPSp} (Color online) 
The transverse mass $m_T$- spectra of protons for 5\% central Pb+Pb collisions at 20, 30, 40, 80 and 158 $A$GeV (plots from upper to lower), in comparison 
to the experimental data from NA49 Collaboration from Refs. \cite{NA49pold,NA49_T,NA49_Lam}.
Solid red lines with open squares refer to PHQMD results with a hard EoS. }
\end{figure}

Now we step up in energy and confront the PHQMD approach with the NA49 experimental
data at SPS energies. Again we start with checking the stopping of protons.
The proton rapidity spectra and $m_T$ spectra of PHQMD at $E_{beam} =20, 30, 40, 80$ and 158
$A$GeV, in comparison with the experimental data \cite{NA49pold,NA49_T,NA49_Lam}, 
are displayed in Figs. \ref{SPS} and \ref{mtSPSp}.  
Here  the solid red lines with open squares stay for the PHQMD results with a hard EoS. The PHQMD proton rapidity distribution and the $m_T$ spectra
show a reasonable agreement with experimental data, thus, the QMD
dynamics provide also a correct stopping at SPS energies similar 
to the AGS.

\begin{figure}[htp]%
\centering
\includegraphics[scale=0.42]{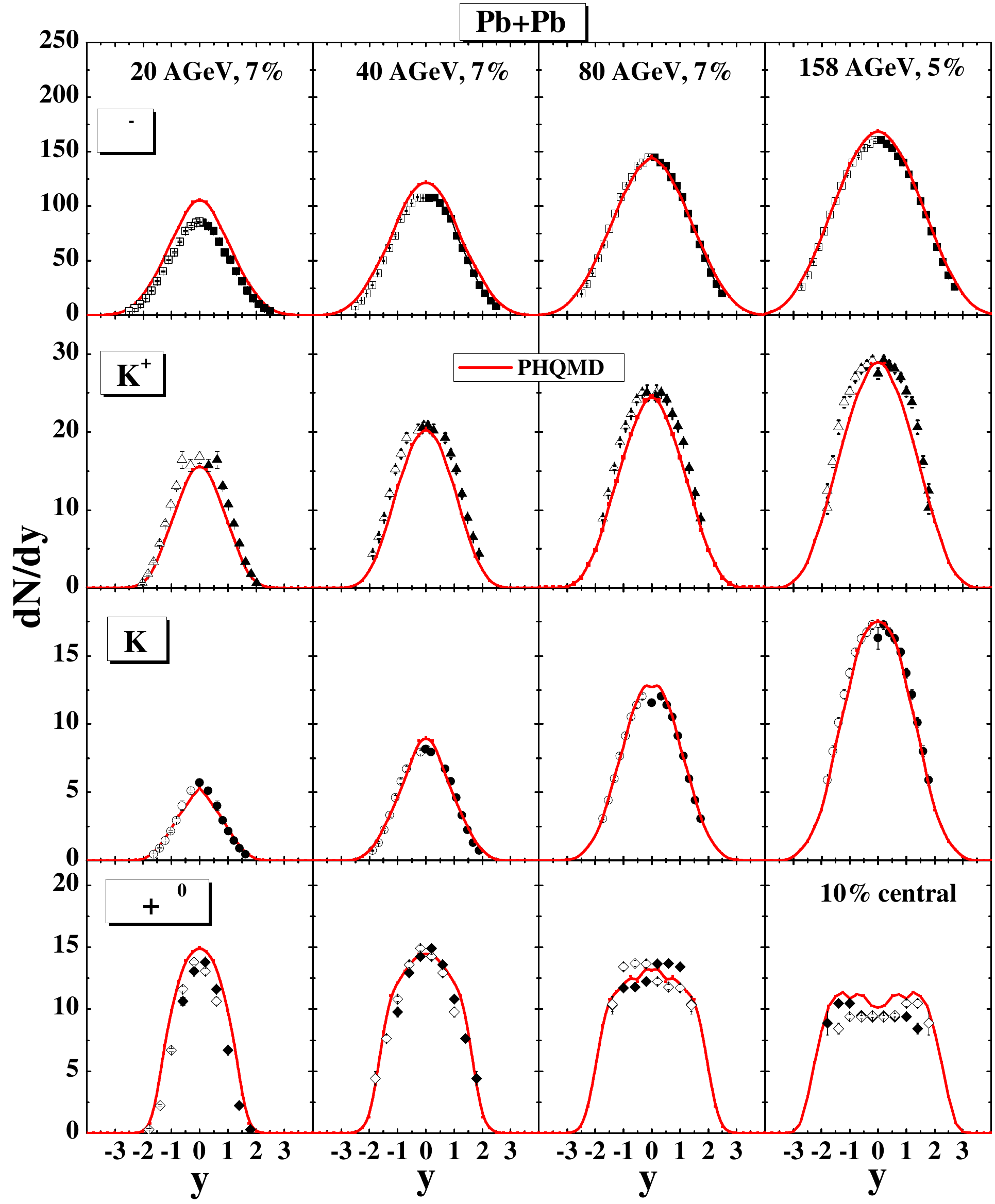}%
\caption{\label{ySPS} (Color online) 
The rapidity distributions of $\pi^+,  K^+, K^-$, and $\Lambda
+\Sigma^0$’s from PHQMD for 5\% central Au+Au collisions at 20, 40, 80 and 158 $A$GeV 
(plots from left to right) in comparison 
to the experimental data from NA49 Collaboration \cite{NA49pold,NA49_T,NA49_Lam}. }
\end{figure}

\begin{figure}[htp]%
\centering
\includegraphics[scale=0.42]{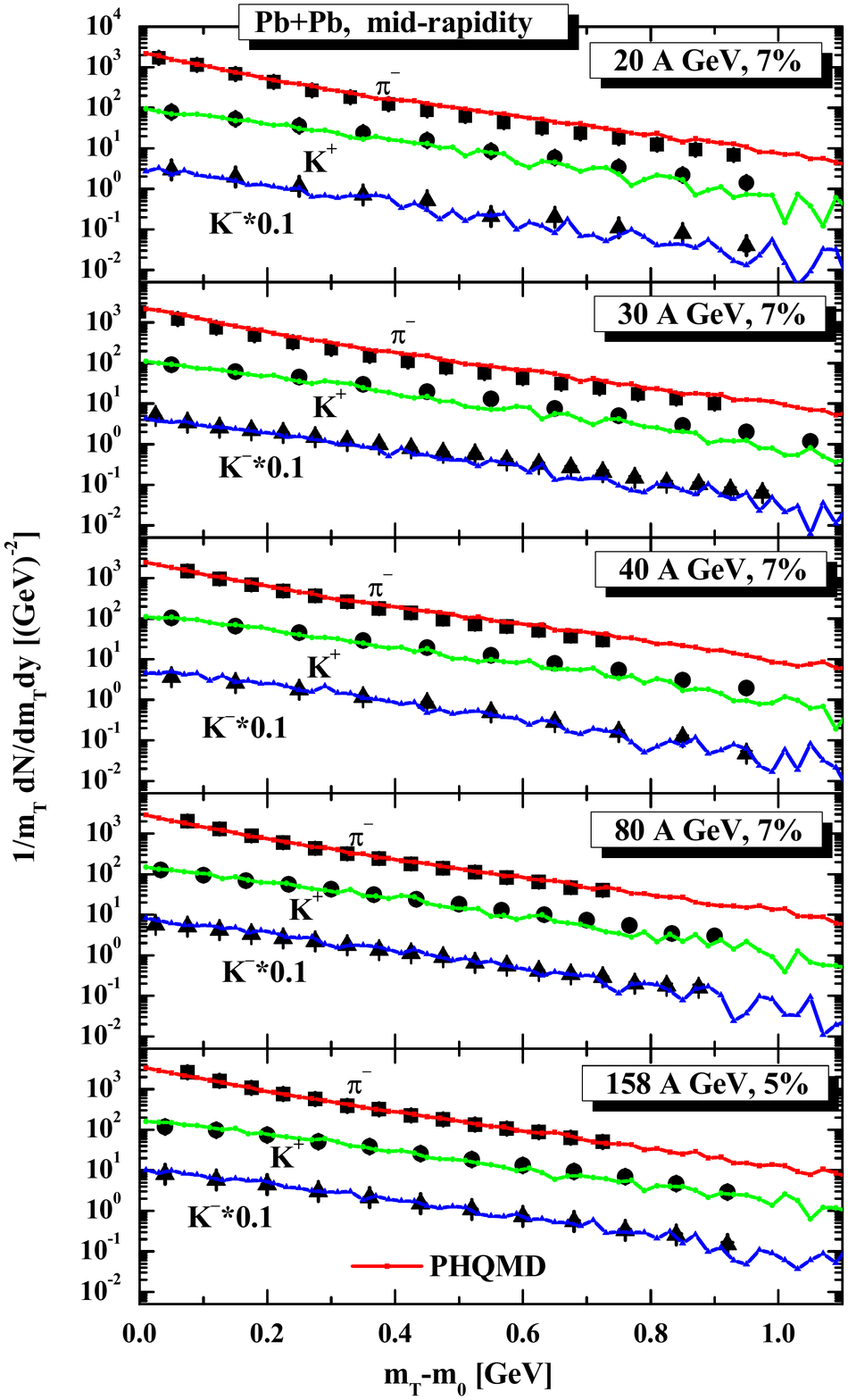}%
\caption{\label{mtSPS} (Color online) 
The  transverse mass $m_T$- spectra of $\pi^+,  K^+, K^-$, and $\Lambda
+\Sigma^0$’s at midrapidity from PHQMD for 5\% central Au+Au collisions at 20, 30, 40, 80 and 158 $A$GeV (plots from upper to lower) in comparison 
to the experimental data from the NA49 Collaboration \cite{NA49pold,NA49_T,NA49_Lam}. }
\end{figure}

In Figs. \ref{ySPS} and \ref{mtSPS} the y-distributions and $m_T$- spectra
of $\pi^+,  K^+, K^-$, and $\Lambda +\Sigma^0$’s for 5\% central Au+Au collisions 
at 20, 40, 80 and 158 $A$GeV are presented in comparison 
to the experimental data from the NA49 Collaboration \cite{NA49pold,NA49_T,NA49_Lam}. Here we find that the PHQMD 
agrees with the experimental data - similar to the PHSD -
since the dynamics of newly produced 
hadrons at high energies are dominated by collision integral
and is not very sensitive to the realization of nucleon dynamics -
via MF or QMD.

\subsection{RHIC BES energies }

Recent experimental measurements by the STAR Collaboration within the RHIC BES program
provide high precision experimental data at midrapidity.
Here we present selected results for the comparison of  PHQMD with RHIC BES
data. A more systematic study on this issue is in preparation. 

\begin{figure*}[htp]%
\centering
\includegraphics[scale=.8]{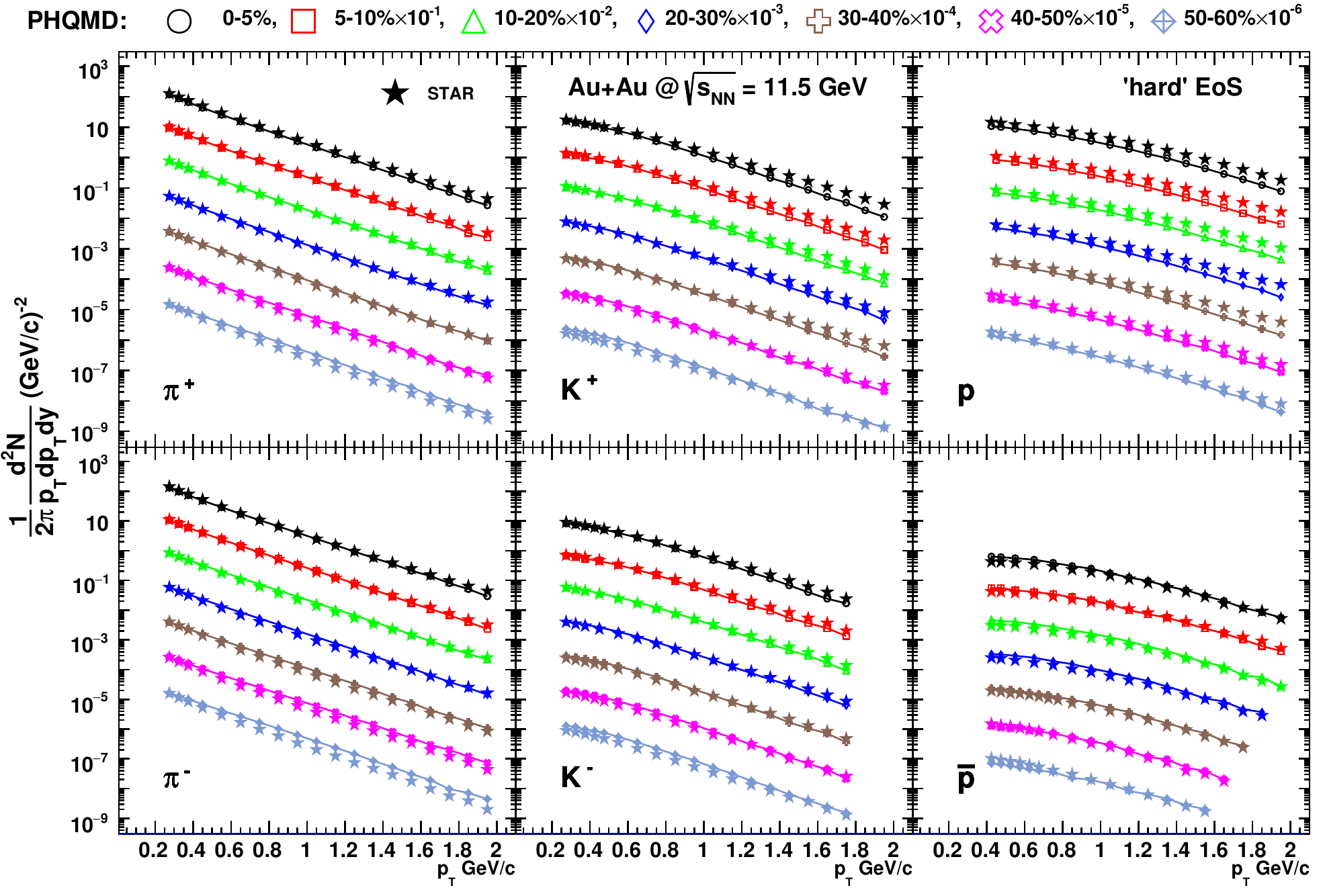}
\caption{\label{BES} (Color online) 
The midrapidity $p_T$- spectra of 
$\pi^\pm, K^\pm $, $p$ and $\bar p$ at midrapidity for Au+Au at $\sqrt{s} = 11.5$  GeV  
from PHQMD with hard EoS in comparison to the STAR
experimental data from Ref. \cite{Adamczyk:2017iwn} for different centrality classes. 
The spectra for different centralities are multiplied by corresponding factors for
better visibility: 0-5\%$\times 1$;   5-10\%$\times 10^{-1}$;
10-20\%$\times 10^{-2}$; 20-30\%$\times 10^{-3}$; 30-40\%$\times 10^{-4}$;
40-50\%$\times 10^{-4}$; 50-60\%$\times 10^{-6}$.}
\end{figure*}

Fig. \ref{BES} shows the transverse momentum spectra of produced mesons 
$\pi^\pm, K^\pm$, protons  and anti-protons at midrapidity for different centrality classes, measured 
by the STAR collaboration for Au+Au at $\sqrt{s} = 11.5$ GeV \cite{Adamczyk:2017iwn}. 
The PHQMD calculations correspond to the hard EoS.
We find that also the centrality dependence of the spectra of newly produced particles is 
well described in the PHQMD approach while the proton slope is slightly underestimated
at large $p_T$. A similar tendency has been observed for protons
at the SPS energies - cf. Fig. \ref{SPS}.

\subsection{Top RHIC energy }

\begin{figure}[htp]%
\centering
\includegraphics[scale=.5]{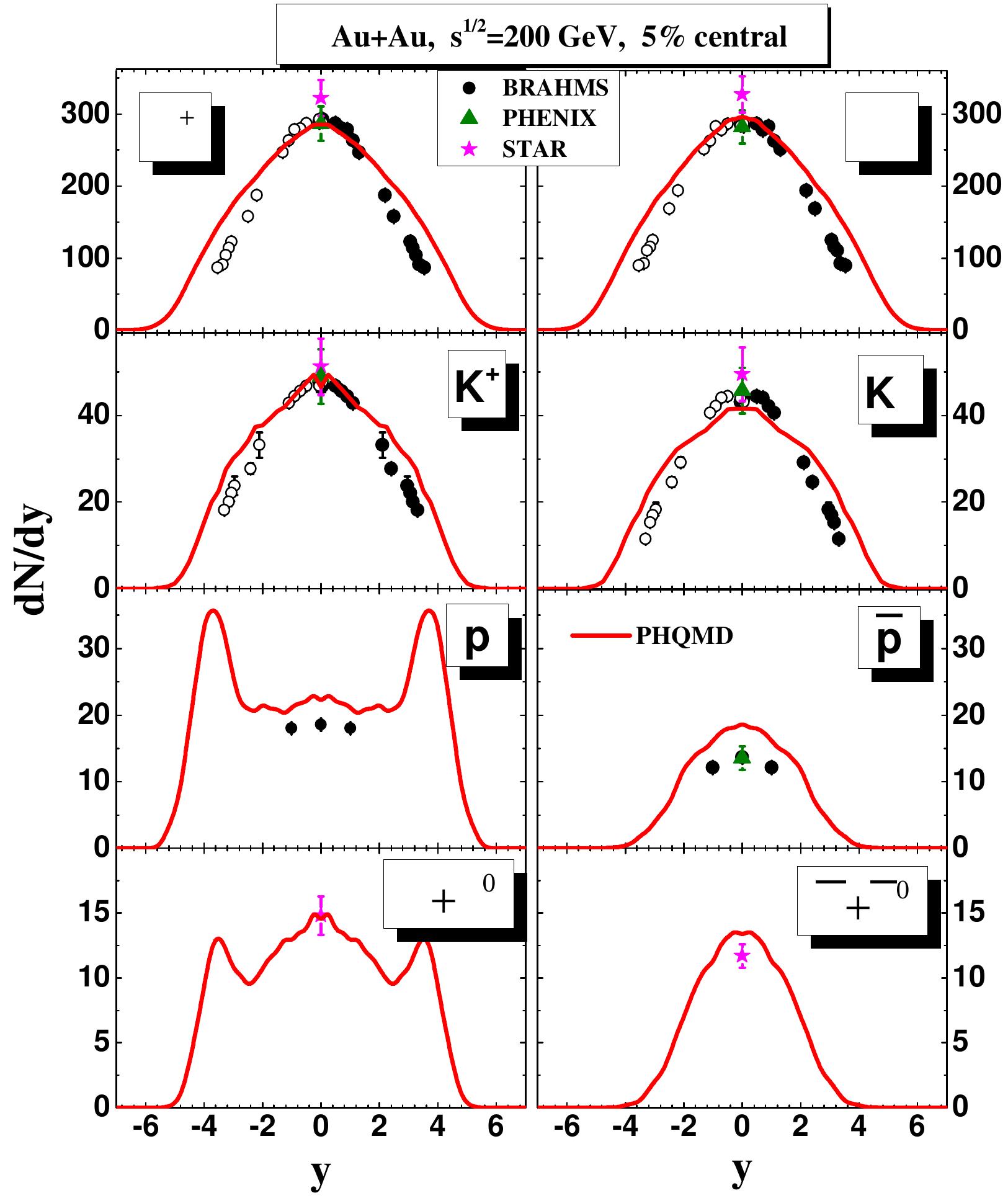}%
\caption{\label{y_RHIC} (Color online) 
The rapidity distributions of $\pi^+,  K^+, p$
 and $\Lambda+\Sigma^0$, left, ant their antiparticles$ \pi^-,  K^-, \bar p$
 and $\bar\Lambda+\bar\Sigma^0$, right,  for 5\% central Au+Au collisions 
 at $\sqrt{s}= 200$ GeV  in comparison 
to the experimental data from the BRAHMS \cite{Bearden:2004yx,Arsene:2005mr}, PHENIX \cite{Adler:2003cb} and STAR \cite{Agakishiev:2011ar} collaborations. }
\end{figure}

\begin{figure}[htp]%
\centering
\includegraphics[scale=.5]{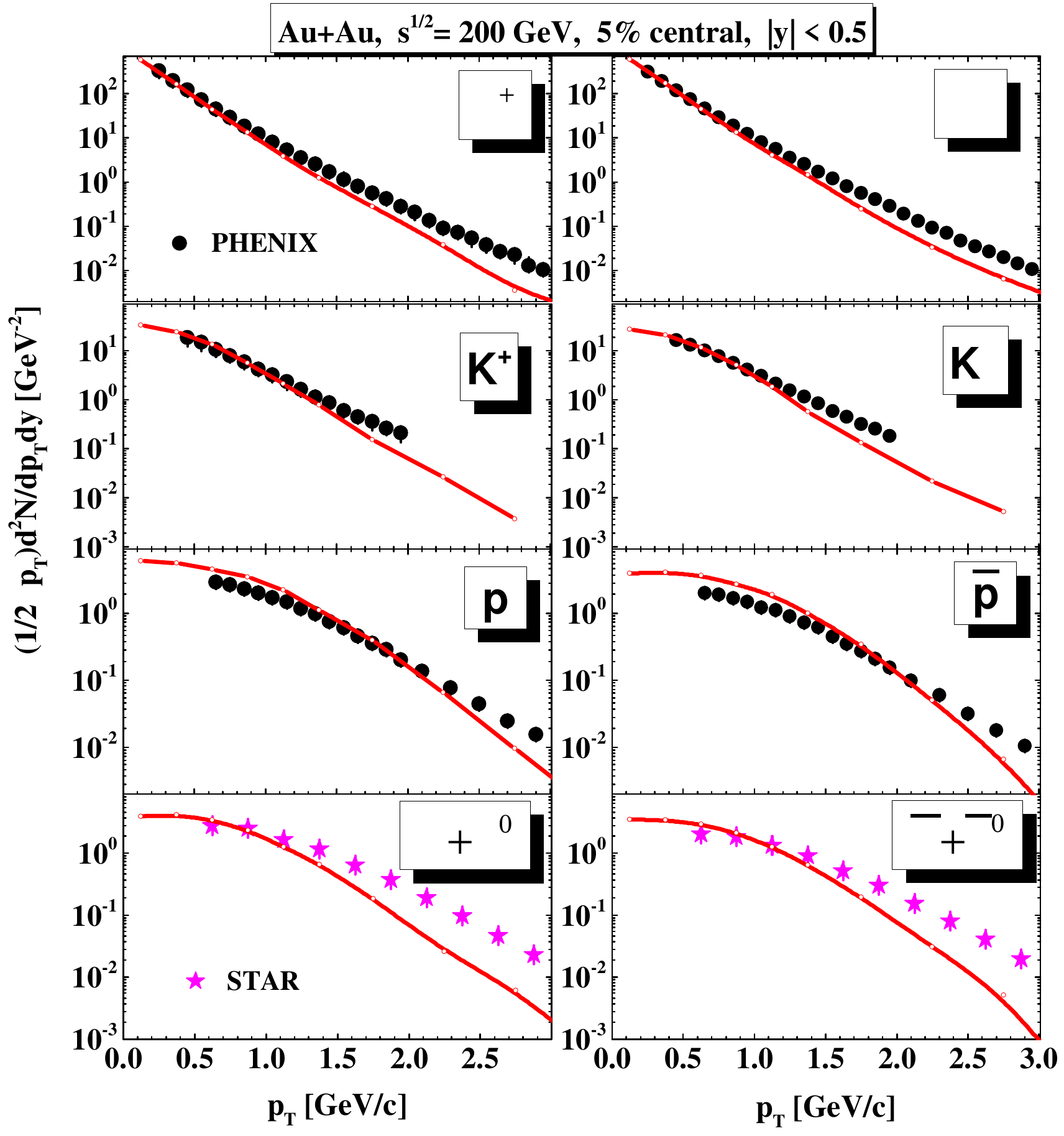}%
\caption{\label{mtRHIC} (Color online) 
The transverse momentum $p_T$- spectra of $\pi^+,  K^+, p$
 and $\Lambda+\Sigma^0$, left, and their antiparticles$ \pi^-,  K^-, \bar p$
 and $\bar\Lambda+\bar\Sigma^0$, right,  for 5\% central Au+Au collisions 
 at $\sqrt{s}= 200$ GeV  in comparison 
 to the experimental data from the PHENIX \cite{Adler:2003cb} and STAR \cite{Agakishiev:2011ar} collaborations. }
\end{figure}

This good agreement between the PHQMD results for the single particle rapidity and transverse momentum spectra and the experimental data continues for higher beam energies. 
In Figs. \ref{y_RHIC} and \ref{mtRHIC} we show the calculated rapidity distributions
and transverse momentum $p_T$ spectra 
of hadrons ($\pi^\pm, K^\pm, p, \bar p, \Lambda+\Sigma^0, \bar\Lambda+\bar\Sigma^0$)
for 5\% central Au+Au collisions at $\sqrt{s} = 200$ GeV in comparison to the experimental 
data from the BRAHMS \cite{Bearden:2004yx,Arsene:2005mr}, PHENIX \cite{Adler:2003cb} and STAR \cite{Agakishiev:2011ar} collaborations.

We note again that at RHIC energies we show only the PHQMD calculations since 
the PHSD and PHQMD give very similar results. At such ultra-relativistic energies the influence of the nucleon potential is negligible and the shape of the spectra (even for protons) is mainly defined by the partonic interactions.  
We note that at the highest energy,  PHQMD (as well as the PHSD) under-predicts the spectra 
at high $p_T$.  That can be attributed to the fact that some parts of the 
initial 'hard' processes is partially smeared out in the present realization of the 
PHSD by the melting of 'pre-hadrons' from the strings to massive dressed quasi-partons
in line with the DQPM model. By that procedure some mini-jets, present in the LUND
strings, can be melted to the QGP, too. This issue requires further investigation
which we leave for future studies.

\subsection{SIS energies }

\begin{figure}[htp]%
\centering
\includegraphics[scale=.4]{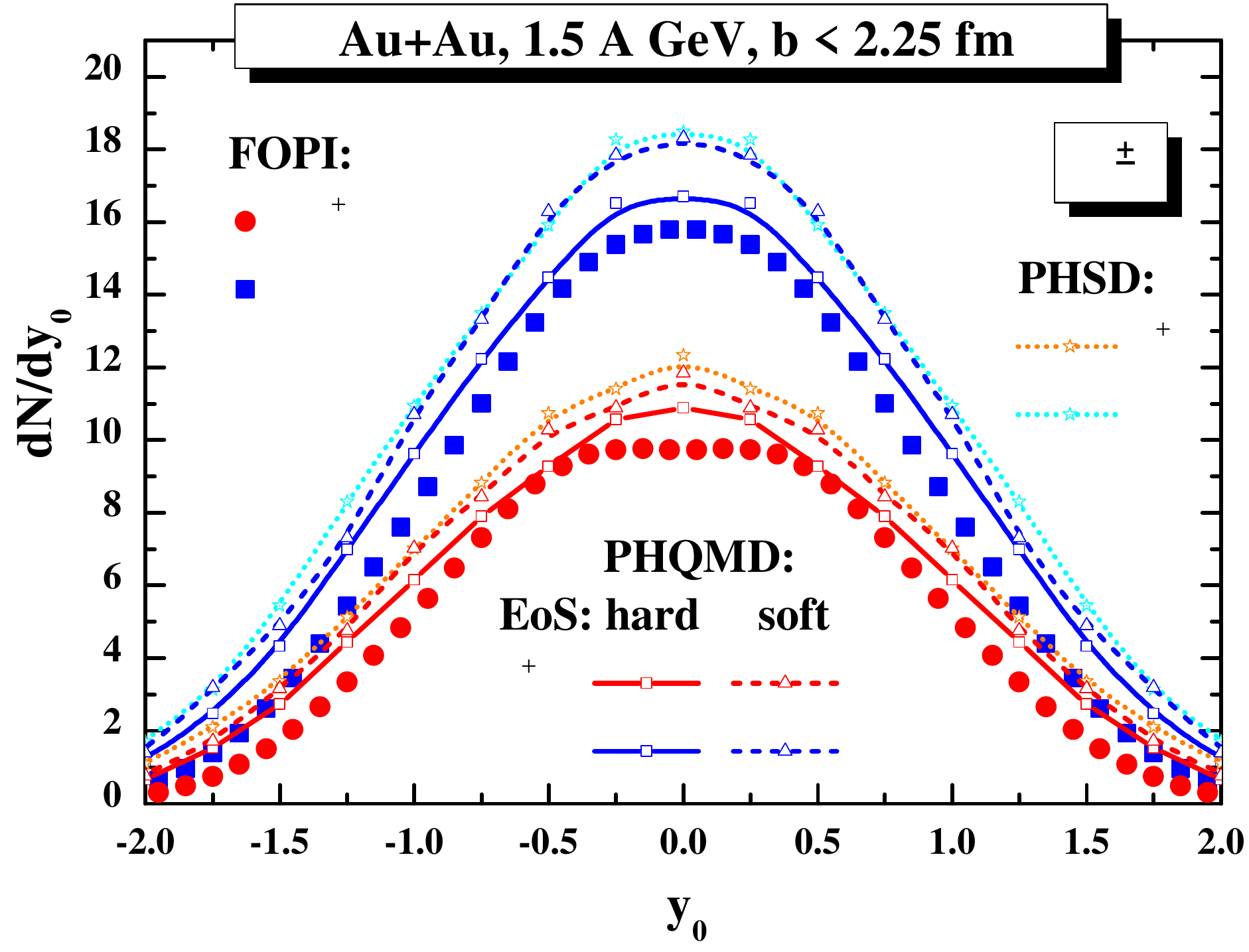}%
\caption{ (Color online) 
Scaled experimental rapidity distribution, $y_0=y/y_{proj}$, of $\pi^+$ and $\pi^-$ observed in central Au+Au reactions at 1.5 $A$GeV  \cite{Reisdorf:2006ie} in comparison with
 PHQMD calculations with a hard (solid lines with squares) 
and a soft EoS (dashed lines with triangles) as well as with the PHSD result 
(dotted lines with stars).}
\label{FOPI1} \end{figure}

We close this Section by going down in energy  to SIS energies which allows to show
the sensitivity  of newly produced particle spectra to the QMD and MF dynamics
as well as to the different EoS. 
We start with the pion spectra since - as discussed in the introduction -
the proton spectra can be compared to the data only after the subtraction
of the protons bound in the clusters. We will see in the next Section  
that the fraction of such bound protons is rather high at low energies 
since the cluster production grows with decreasing bombarding energy.

At $E_{beam}=1.5$ $A$GeV the pion rapidity spectra as a function of $y_0=y/y_{proj}$ 
in central Au+Au reactions  have been measured by the FOPI collaboration \cite{Reisdorf:2006ie}. 
In Fig. \ref{FOPI1} we compare the FOPI data with PHQMD calculations employing a hard
(solid lines with squares) and a soft EoS (dashed lines with triangles) as well as with 
the PHSD results (dotted lines with stars). As seen from Fig. \ref{FOPI1}, the 
pion rapidity distribution is sensitive to the  EoS:  
the experimental data are best in agreement with the PHQMD results for 
a hard EoS. The softening of the EoS leads to a small enhancement of the pion yield
as seen for the PHQMD results with a soft EoS as well as for the PHSD results, 
where the EoS is also soft.

Finally, we can conclude from this comparison that the rapidity as well as the $m_T$ spectra of produced particles, as well as of protons, are well reproduced in the PHQMD approach. 
This means also that the basic features like energy loss and elementary cross sections are under control. 
These findings allow us to proceed to investigate the
cluster production  based on the SACA and MST algorithms which we 
present in the next Section.


\section{Results for Clusters}

\subsection{Light Clusters}

\begin{figure}[htp]%
\centering
\includegraphics[scale=0.52]{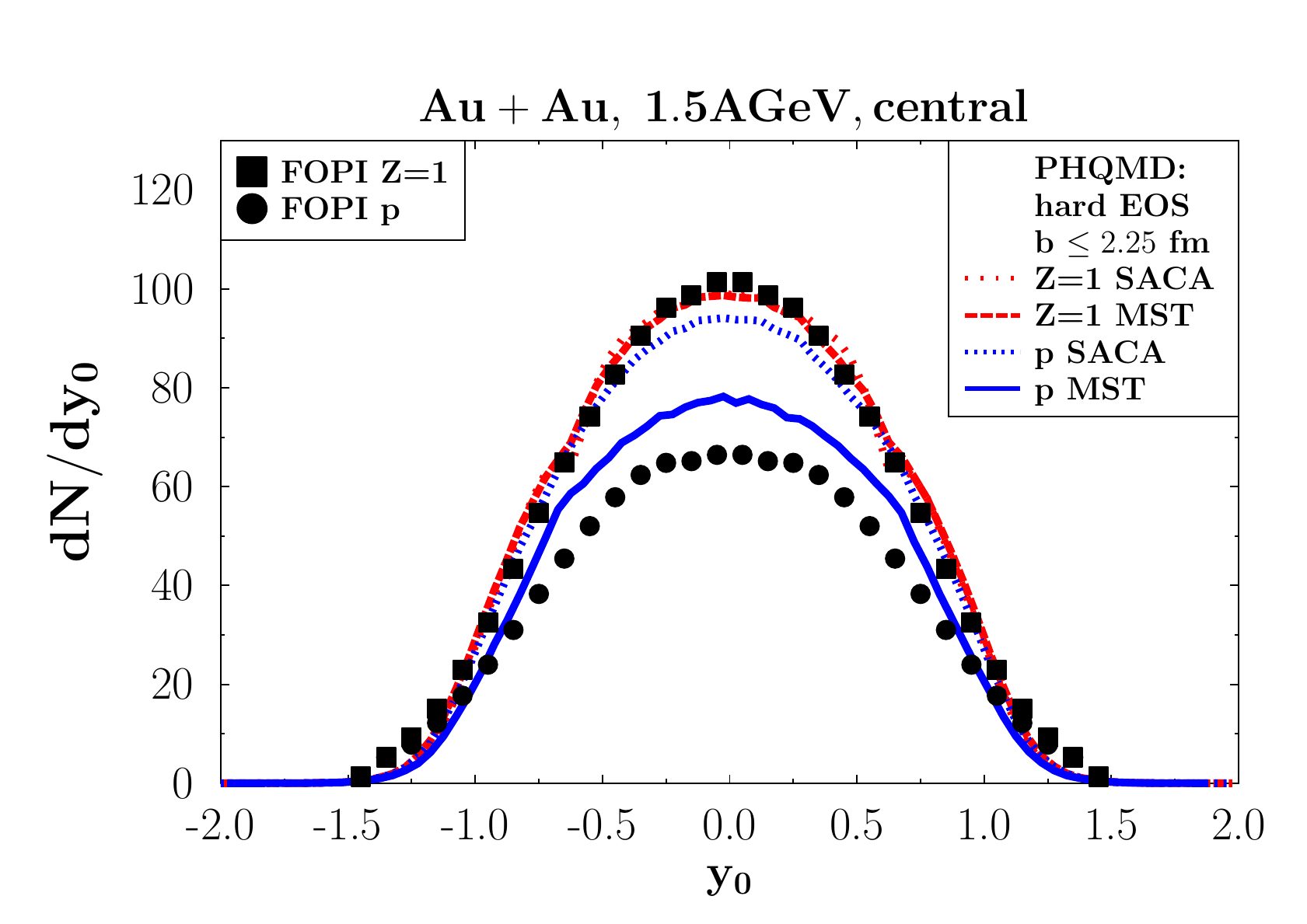}
\caption{\label{FOPI2} (Color online) 
Scaled experimental rapidity distribution, $y_0=y/y_{proj}$, of 
all bound and unbound protons ($Z=1$)\ -- solid squares, and 
free (unbound) protons - solid dots, observed by the FOPI collaboration 
in central Au+Au collisions at 1.5 $A$GeV  \cite{Reisdorf:2006ie} 
in comparison to the PHQMD results:
the rapidity distribution of all bound and unbound protons ($Z=1$) after the clusters have been identified by MST (red dotted line) or by  SACA (red dashed line);
the rapidity distributions of free protons after subtracting the protons bound 
in clusters identified by MST (blue solid line) or by SACA (blue short dotted line).
}
\end{figure}
At lower beam energies cluster production becomes important. 
According to the measurements by the FOPI Collaboration  \cite{Reisdorf:2006ie}
in central Au+Au collisions at 1.5 $A$GeV about of 111 free protons are found 
and 60 protons are bound mostly in $Z=1, 2$ clusters.
In Fig. \ref{FOPI2} we compare the PHQMD results for the scaled rapidity distributions  ($y_0= y/y_{proj}$) with $y_{proj}$ being the beam rapidity in the center-of-mass frame)
of the $Z=1$ 'clusters' (which includes unbound protons as well as light clusters as deuterons and tritons) and the (unbound) protons  with FOPI experimental data  for central Au+Au collisions at 1.5 $A$GeV \cite{Reisdorf:2006ie}. 
Here we present the results for clusters identified by MST (red dotted line) or by  SACA (red dashed line).
Since the integrated yield of the $Z=1$ clusters gives almost the total number of charges
(there are on the average only 6.8 clusters with $Z=2$), it is rather trivial that
the integrated PHQMD $Z=1$ yield agrees with data. In addition, also  the scaled rapidity
distribution of $Z=1$ 'clusters', which reflects the stopping, is well reproduced. 
This  explains that also the rapidity distributions of the produced particles, like that 
of $\pi^+$ and $\pi^-$, agree with experiments (cf. Fig. \ref{FOPI1}).
In Fig. \ref{FOPI2} we show also the rapidity distribution of free protons (blue lines).
The difference between the rapidity distribution of $Z=1$ (red lines) and protons (blue
lines) in Fig. \ref{FOPI2} is due to those protons which are bound in $Z=1$ clusters. 

As discussed already in Section III,  SACA with Skyrme type interactions only - as presently implemented in the PHQMD - (blue long dashed line)  is not very efficient 
to describe the light clusters at midrapidity and, correspondingly, underestimates the
number of nucleons which are bound in clusters. The MST algorithm - which is not accounting for the binding energy of clusters as SACA and, thus, less sensitive to the potential interaction of nucleons - (blue short dashed line in Fig. \ref{FOPI2}) 
comes much closer to the data, in spite of lucking the quantum nature of light clusters.
Moreover, as demonstrated in Fig. \ref{FOPI3} and discussed in Section III.B,
the MST algorithm provides rather stable yield of light clusters over time.
Therefore, for further analysis of light clusters at midrapidity in this Section
we employ  the MST algorithm.  


\begin{figure}[htp]%
\centering
\includegraphics[scale=0.3]{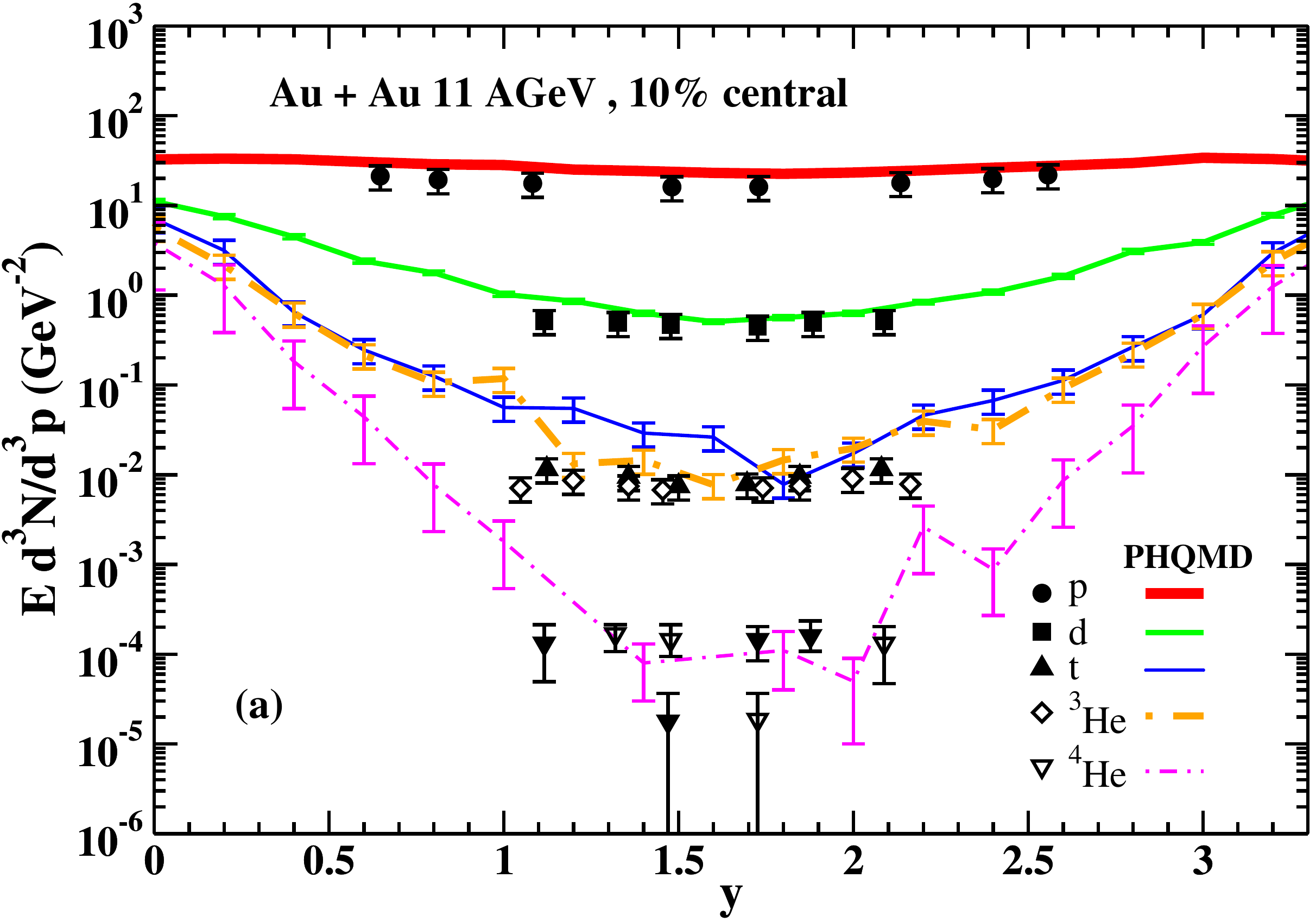}\\
\vspace{0.5cm}
\includegraphics[scale=0.3]{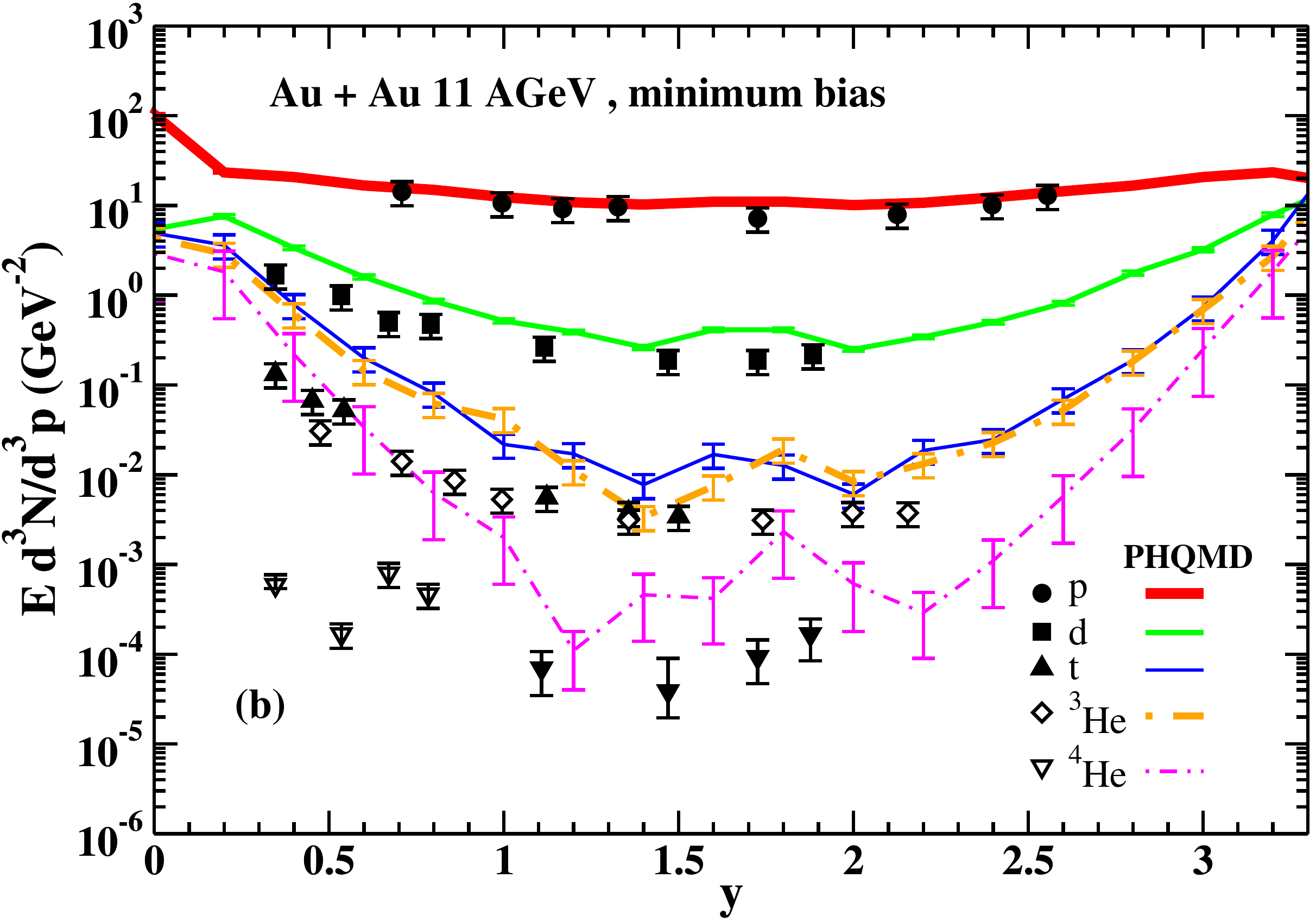}
\caption{\label{AGSclusters} 
(Color online) The invariant multiplicities for $p, d, t, ^{3}\!He, ^{4}\!He$ at $p_T \le 0.1$ GeV versus rapidity for $10 \%$ central (upper plot) and minimum-bias (lower plot) $Au+Au$ collisions at $E_{beam}=11 \, A$GeV. Experimental data from E886 and E878 Collaborations are taken from Refs.~ \cite{Saito:1994tg,Bennett:1998be}. The solid (dash-dotted) lines with different thickness correspond to the PHQMD calculations with hard EoS for charge value $Z=1$ ($Z=2$). Clusters are identified by the MST algorithm.}
\end{figure}

We step to high energies and confront expectations from PHQMD for light clusters with the available experimental data at AGS energies. In Fig.~\ref{AGSclusters} the PHQMD results with hard EoS of the invariant multiplicities for $p, d, t, ^{3}\!He, ^{4}\!He$ at $p_T \le 0.1$ GeV as function of rapidity $y$ at $10\%$ central (upper plot) and minimum bias (lower plot) Au+Au collisions at $E_{beam}=11 \, A$GeV  are compared to the experimental data from the E878~\cite{Bennett:1998be} and  E886~\cite{Saito:1994tg} collaborations, labeled by different symbols for the various species. For $ ^{4}\!He$ we represent separately the measurements from E886 (empty triangles) and E878 (filled triangles). 
The clusters are identified by the MST method and later selected through the physical isospin and charge combinations.
The colored lines in Fig.~\ref{AGSclusters} are the PHQMD results which we provide also with uncertainties which resemble the statistical fluctuations of the binned distributions.
 As one can see they are in line with the measured experimental data. 
We point out that in the final stage of heavy-ion reactions the MST algorithm finds approximately the same number of light clusters which are recognized in a rather stable and time-independent way by the SACA method.

Finalizing this section, we stress that the PHQMD is a 
consistent microscopic transport approach applicable to relativistic energies 
in which clusters are produced dynamically by the same potential interaction 
which governs the time evolution of the nucleons up to the end of the reaction.
Even more, the cluster finding algorithm (MST) applied  at different time finds a similar
cluster pattern.
No assumptions about a coalescence time or coalescence radii are necessary
in order to obtain 
these clusters. They are naturally produced by the interactions among the nucleons 
during the entire heavy-ion reaction.

Generally, the existence of light clusters at midrapidity of heavy-ion collisions 
is a amazing phenomena. There the participating  nucleons form a fireball 
which can well be described in thermal approaches assuming a temperature of the order of 100 MeV \cite{Cleymans:2005xv}. Also the transverse energy spectra show an inverse slope parameter of this order which is, however, composed of a radial flow and a thermal contribution.
This observation has triggered the suggestion that in high energy heavy-ion reactions a hot
thermal system is formed. On the other hand, the light clusters which are formed, have
binding energies of a couple of MeV and they cannot survive in such a hot environment. In addition, any collision of a cluster with hadrons from the fireball would destroy these clusters. It is, therefore, an open question how these midrapidity clusters, which can be observed up to the highest LHC beam energies, are formed and how they can survive in this hot fireball. Static models like the coalescence model or the statistical model cannot answer this question. The PHQMD results obtained with the MST cluster identification method show that clusters can be formed in such an environment but the MST method does not allow for a detailed investigation of why and when clusters are formed since this method can
identify clusters only at the end of the reaction. In order to overcome this limitation, 
a further development of the SACA algorithm for the light cluster finding is required 
which will help to shed light on the dynamical formation of the light clusters.
 
\subsection{Heavy Clusters}

In the past QMD approaches have been very successfully applied to 
describe many details of the cluster formation at energies below $E_{kin} = 200\ A$MeV \cite{LeFevre:2009er,Fevre:2007pr,Zbiri:2006ts,Nebauer:1998fy}. They could reproduce charge yields, cluster multiplicities, cluster spectra and complex phenomena like bimodality. At these energies the fragmentation of spectator matter is the dominate mechanism for cluster production and cluster identification methods like the minimum spanning tree  or the SACA method could identify the produced cluster \cite{Puri:1996qv,Puri:1998te}. 

Within the PHQMD we extend our research to a bit higher energies and confront first
the PHQMD results to the experimental data of the ALADIN collaboration which has measured the cluster formation at beam energies between  600 $A$MeV and 1000 $A$MeV \cite{Schuttauf:1996ci,Sfienti:2006zb}. This is presently the highest beam energy for which experimental data on heavy clusters are completely analyzed. For this investigation we use a hard EoS and employ the SACA algorithm. One of the key results of the ALADIN collaboration is the "rise and fall" of the multiplicity of intermediate mass clusters 
$3\le Z \le 30$ emitted in forward direction. This multiplicity is presented as 
a function of the sum of all forward emitted bound charges, $Z_\mathrm{bound\ 2}$ which can be expressed with help of the $\Theta$ function:
$$Z_\mathrm{bound\ 2}= \sum_i Z_i \ \Theta(Z_i-(1+\epsilon)),$$ 
with ($\epsilon <1$).
One obtains a distribution which is for Au projectiles almost independent of the beam energy in the interval $ 600\ A$MeV $\le E_{beam} \le 1000\ A$MeV and also independent of the target size. 
We note, that in the original publication \cite{Schuttauf:1996ci} the intermediate mass cluster multiplicity has been overestimated due to misidentified, mostly  $Z=3$, clusters which were in reality two $\alpha$ particles. Later, with an improved apparatus, this has been realized for smaller systems. A re-measurement  for the Au+Au system has shown that the multiplicity of intermediate mass clusters is about 15 \% lower than published in  \cite{Schuttauf:1996ci} . The corrected rise and fall curve for Au+Au reactions has been published in \cite{LeFevre:2017ygd} and will be used for the comparison
in our study.
 
\begin{figure}[htp]%
\centering
\includegraphics[scale=.5]{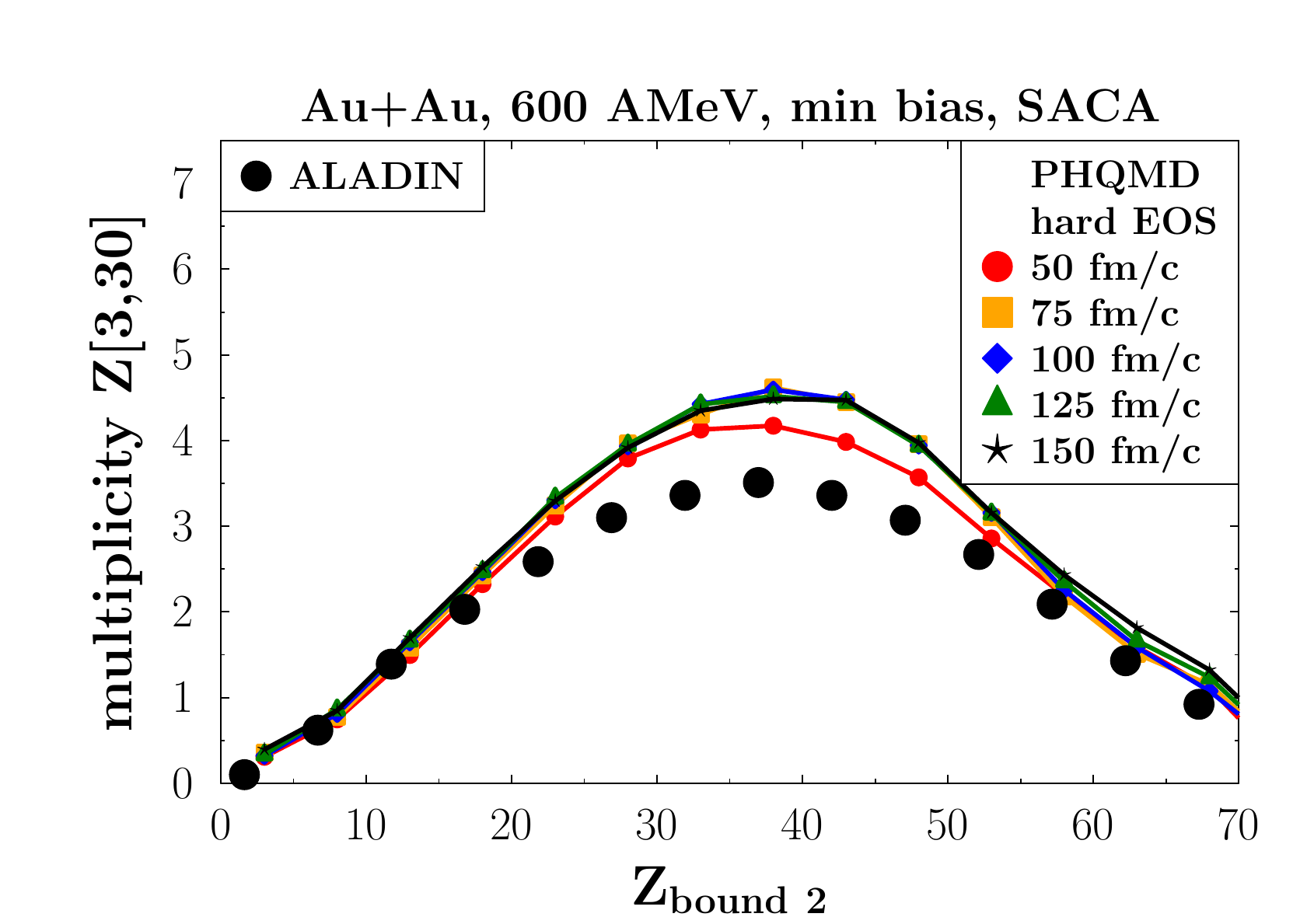}
\caption{\label{rafz} (Color online) 
'Rise and fall' of the multiplicity of clusters with $Z \in [3,30]$ as a function of the total bound charge  $Z_\mathrm{bound\ 2}$.
Both quantities are measured for forward emitted clusters. The experimental data of the ALADIN Collaborations  are from Ref. \cite{LeFevre:2017ygd, FRIGA2019}. 
The plot shows the PHQMD results with hard EoS using cluster identification 
by SACA for 600 $A$GeV at different times -- 
50 (red line), 75(orange line), 100 (blue line),  125 (green line), 150 (black line) fm/c.}
\end{figure} 
\begin{figure}[htp]%
\centering
\includegraphics[scale=.5]{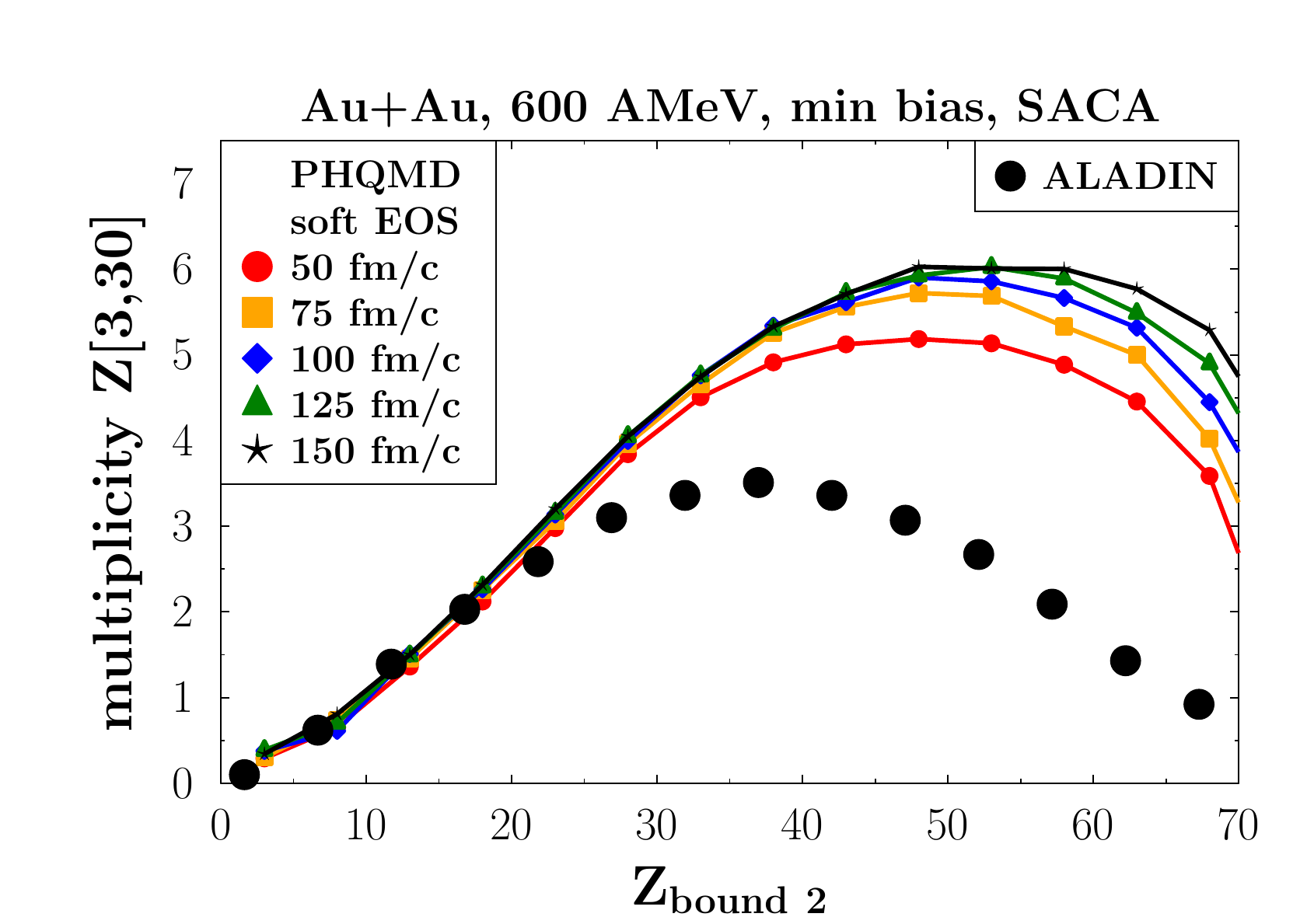}
\caption{\label{rafzs} (Color online) 
Same as Fig. \ref{rafz}, but for a soft EoS.}
\end{figure}

In Fig. \ref{rafz} we display our results for Au+Au at 600 $A$MeV calculated with a hard EoS 
in comparison with minimum bias ALADIN data  \cite{LeFevre:2017ygd}. The clusters identified by SACA are  stable
for time larger than 50 fm/c as shown in Fig. \ref{rafz} .
One can see clearly that PHQMD with a hard EoS reproduces quite nicely the experimentally observed 'rise and fall'.  

The rise and fall of the intermediate mass cluster multiplicity depends strongly on the nuclear equation-of-state. In Fig. \ref{rafzs} we show the rise and fall for a soft EoS.
There in semi-peripheral and peripheral collisions, where $Z_{bound\ 2}$ is large, the spectator matter is much less stable and fragments into a much larger number of intermediate mass clusters  as compared to a hard EoS (Fig. \ref{rafz}). The fragment pattern in semi-peripheral reactions can therefore serve as an additional observable to determine the hadronic EoS experimentally. 

\begin{figure}[htp]%
\centering
\includegraphics[scale=.5]{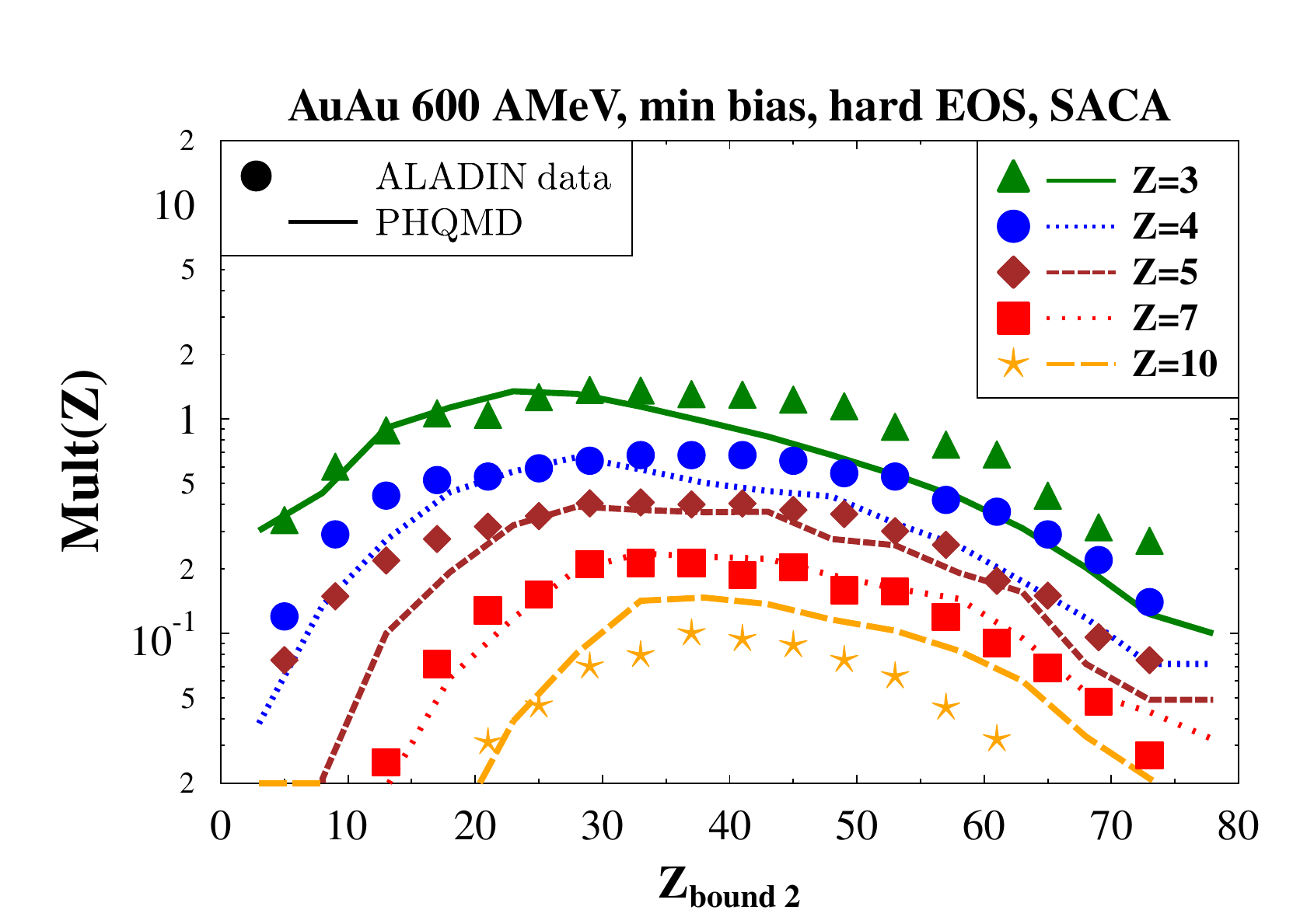}
\caption{\label{raf} (Color online) 
'Rise and fall' of the multiplicity of intermediate mass clusters of a given charge $Z$ ($Z=3,4,5,7,10$) as a function of the total bound charge $Z_\mathrm{bound\ 2}$. Both quantities are measured for forward emitted clusters. The results of PHQMD with cluster 
identification by SACA (lines) are compared to the ALADIN experimental data \cite{Schuttauf:1996ci} (symbols). The Z=3 data are corrected by 15\% , see text.}
\end{figure} 

The ALADIN collaboration has also measured the multiplicity of clusters of a given charge $Z$ 
($Z=3,4,5,7,10$) as a function of $Z_\mathrm{bound\ 2}$. The PHQMD result are compared with the experimental finding in Fig. \ref{raf}. Due to the arguments presented above we have multiplied the multiplicity of $Z=3$ clusters, published in  \cite{Schuttauf:1996ci}, by 0.85 assuming that the misidentified clusters have been exclusively $Z=3$ clusters. We observe a quite good agreement of the PHQMD results with experimental
data.

\begin{figure}[htp]%
\centering
\includegraphics[scale=.5]{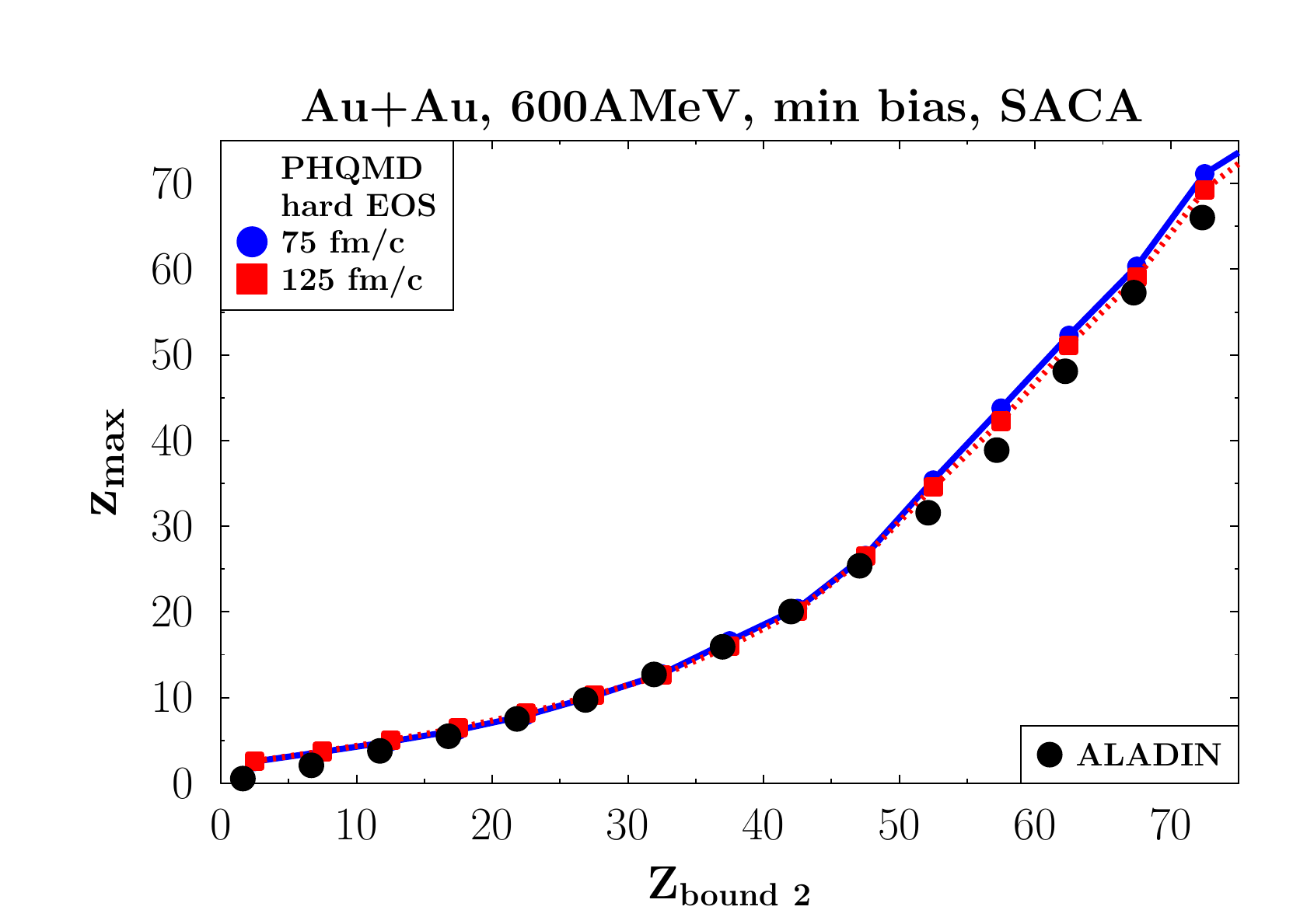}
\caption{\label{Aladin2} (Color online) 
The average charge of the largest cluster as a function of the total bound charge. Both quantities are measured for forward emitted clusters. 
The PHQMD results with cluster identification by SACA are presented for two times 
of 75 and 125 fm/c and compared to the ALADIN experimental data  \cite{Schuttauf:1996ci}.}
\end{figure} 

\begin{figure}[htp]%
\centering
\includegraphics[scale=.5]{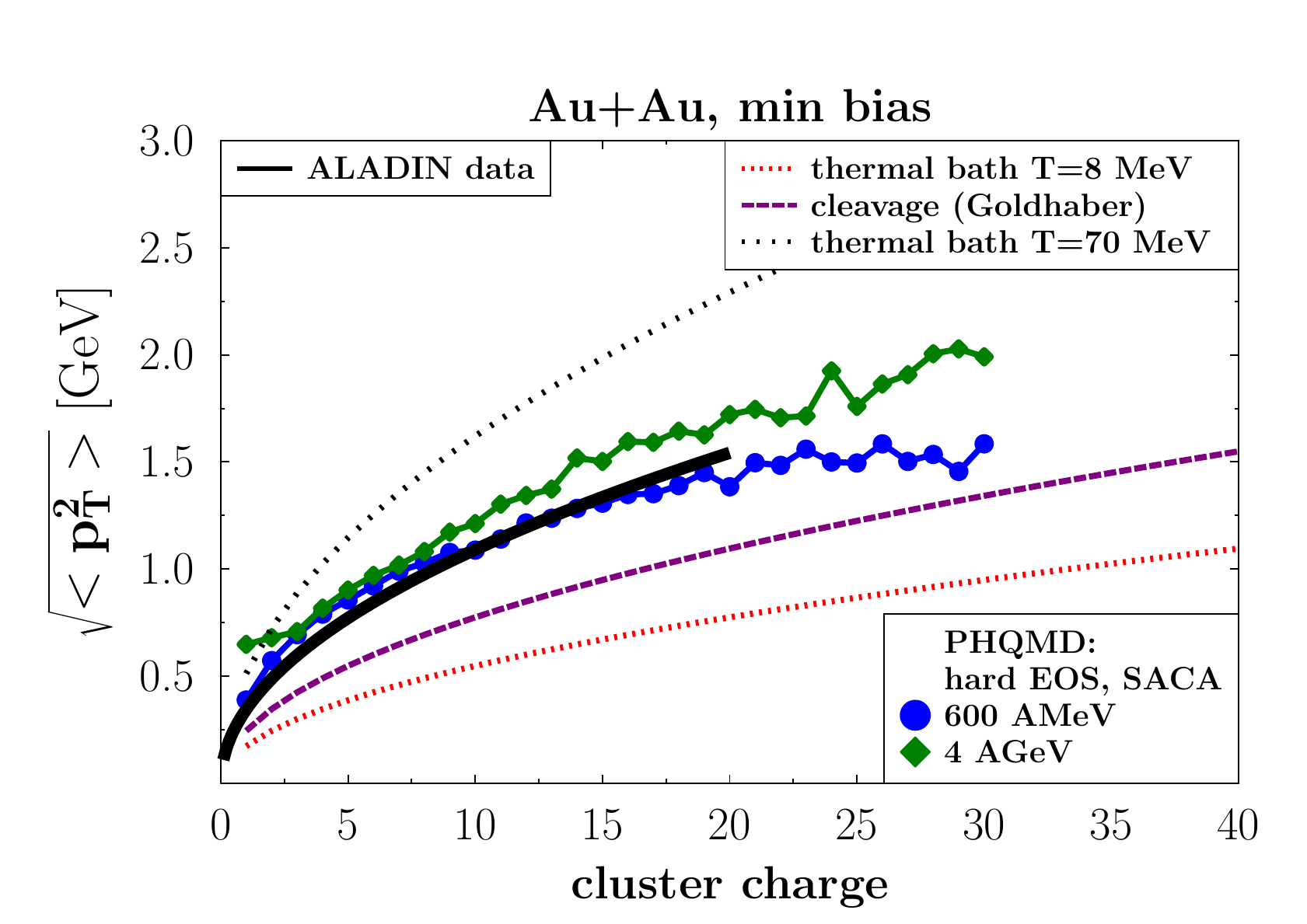}
\caption{\label{ALADIN3} (Color online) 
 $\sqrt{<p_T^2>}$ as a function of the cluster charge $Z$
for minimal bias Au+Au collisions at 600 $A$MeV and 4 $A$GeV. 
The black solid line is an interpolation of the experimental data \cite{Schuttauf:1996ci}, 
the blue line with dots and green line with diamonds are the result of PHQMD calculations. 
The red short-dotted line (black dotted line) presents the result of the thermal model for $T=8 (70)$ MeV.
The blue dashed line is the result of the cleavage model of Goldhaber (see text).}
\end{figure} 

Fig. \ref{Aladin2} shows the charge of the largest cluster as a function of $Z_{bound}$ for forward emitted clusters in Au+Au collisions at 600 $A$MeV. In central collisions, where $Z_{bound}$ is small, we see also no large clusters whereas in very peripheral reactions $Z_{bound\  2}$ approaches  the charge of the projectile. The PHQMD calculations with the SACA algorithm for cluster identification
reproduce the experimental data. Even more important, the result does not depend on the time 
when we apply the SACA algorithm because the cluster pattern changes only little with time.

From Figs. \ref{raf} and \ref{Aladin2} we can conclude that PHQMD describes the size and the multiplicity of clusters $Z \ge 2$ from very central to peripheral Au+Au reactions at  600 $A$MeV 
if the SACA algorithm is employed. Beyond $E_{beam} = 1$ $A$GeV (where the cluster distribution is very similar to the more extensively analyzed 600 $A$MeV data), there are no measurements of  heavy clusters, only that of light midrapidity clusters.

Another observable, measured by the ALADIN collaboration \cite{Schuttauf:1996ci},  is the  {\it rms} of the transverse momentum distribution, $\sqrt{<p_T^2(Z)>}$, as a function of the cluster charge. 
In Fig. \ref{ALADIN3}  we show these data in terms of an interpolation line provided by the ALADIN collaboration  \cite{Schuttauf:1996ci}. 
Additionally to the PHQMD results  for the 600 $A$MeV and 4 $A$GeV,
we also show the expectations from three different theoretical models: a  thermal model
for temperatures of 8 and  70 MeV and the 'cleavage' model of Goldhaber \cite{Goldhaber:1974qy}.
All three models predict that $\sqrt{p_T^2(Z) }\propto \sqrt{Z}$. 
The dotted  lines are the expected {\it rms} momenta if the clusters were in thermal equilibrium 
with a heat bath of a temperature of $T=8$ MeV and $T =$ 70 MeV, respectively. 
Since the binding energy per nucleon of a cluster is around 8 MeV, a temperature considerably higher than 8 MeV would not allow for the existence of clusters. We see that the experimental {\it rms} momenta are higher than expected for a heat bath of $T =$ 8 MeV indicated as the red short dotted line
in Fig. \ref{ALADIN3}. 
This questions the assumption that clusters are emitted by a thermal source
as assumed in statistical models. 

On the other hand, the apparent inverse slope of the transverse energy spectra of protons at midrapidity 
for Au+Au at 600 $A$MeV is about 100 MeV. It is a superposition of a thermal contribution and the contribution from the radial flow.
70 MeV is a reasonable value for the termal part.  If clusters are formed from the nucleons 
of the expanding fireball at the end of the expansion by momentum space coalescence, one would expect that
the {\it rms} of the transverse momenta distribution of the clusters is of the same order as the black dotted line. 
Since this scenario is substantially overestimating the experimental data, 
one would conclude that the late clusterization by coalescence is also not
supported by the ALADIN data, even not for light clusters.

The dashed line shows the result expected from the 'cleavage' model of Goldhaber 
which assumes that the spectator matter is cleaved instantaneously into clusters 
by penetrating participant nucleons and that the {\it rms} momenta of the clusters 
are  reminiscent of the Fermi motion of the nucleons 
\cite{Goldhaber:1974qy,Aichelin:1984asp,Aichelin:1984xb}.  
The difference to the prediction of the Goldhaber model comes mainly from the Coulomb repulsion among
the clusters and protons which is not taken into account in the Goldhaber model.
The PHQMD calculations agree with data and show the same $\sqrt{p_T^2(Z)} \propto \sqrt{Z}$ dependence  as the data.

\subsection{Hyper-clusters}

The production of the hypernuclei in heavy-ion collisions is one of the challenging 
experimental and theoretical topics nowadays.
Hyperons ($\Lambda$'s and $\Sigma$'s) are produced in heavy-ion collisions already
at the SIS energies above 1.6 $A$GeV (which corresponds to the $NN$ threshold).
For details of the strangeness production at low energy we refer the reader to a review \cite{Hartnack:2011cn}. 
In heavy-ion collisions at lower energies  the hyperons are almost exclusively produced in the overlapping fireball,
however, they may penetrate into the spectator matter and form  hyper-clusters  
with spectator nucleons or, during the expansion of the fireball, may find other nucleons 
with which they form light hyper-clusters at midrapidity. 
Thus, hyper-clusters in the projectile/target rapidity regime give information on 
how these hyperons penetrate the fast moving spectator matter 
and get accelerated in order to form clusters with spectator nucleons. 
Hyper-nuclei around midrapidity are sensitive to the time evolution of the high density 
zone in the center of the reaction where the hyperons are produced.  
The study of hyper-clusters is one of the research priorities of the upcoming NICA facility and 
for the CMB experiment at FAIR. 
Statistical model calculations \cite{Andronic:2010qu} predict that hyper-clusters  are produced copiously in the energy regime accessible with these facilities.

\begin{figure}[htp]%
\includegraphics[scale=.3]{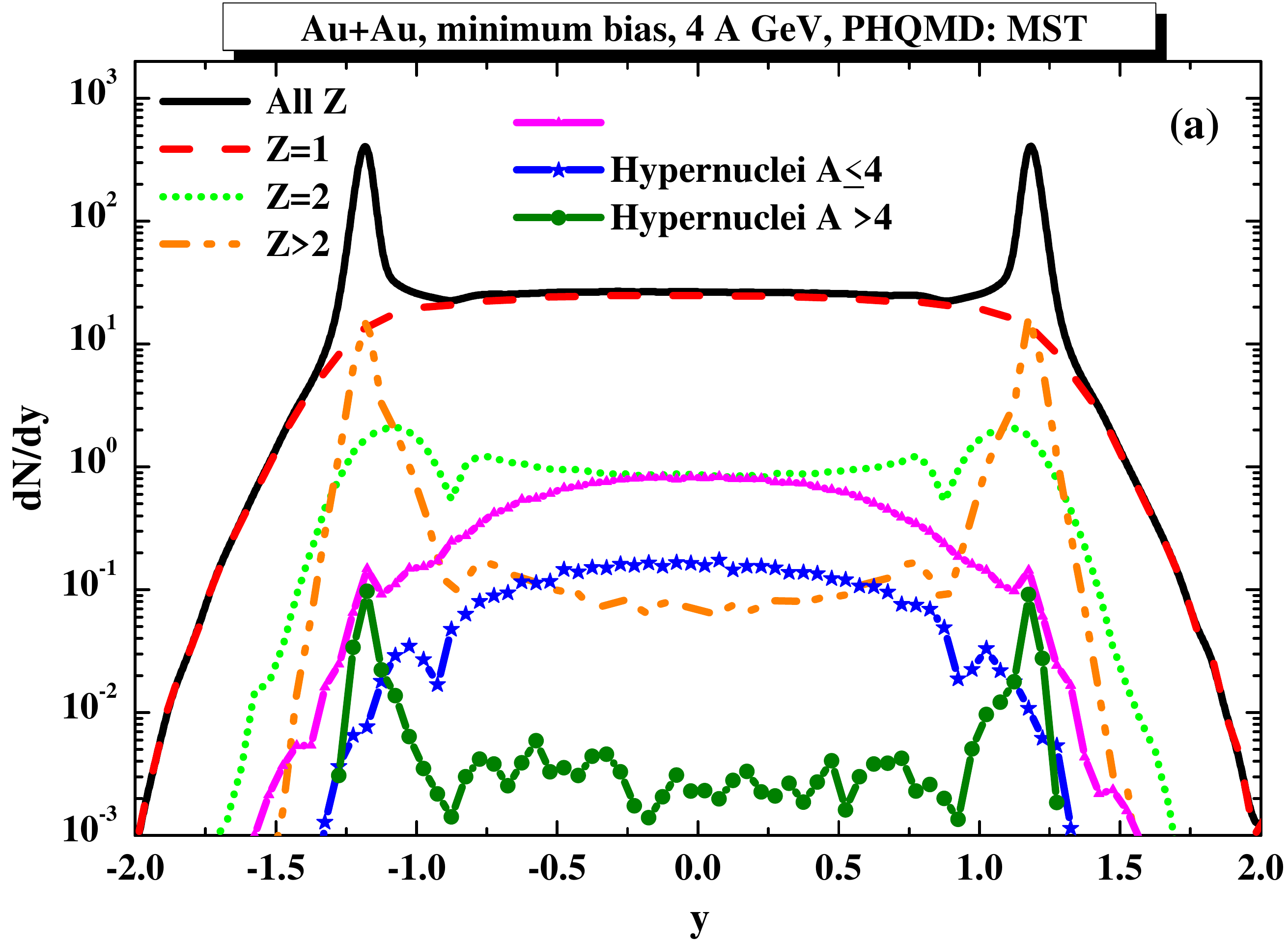}\\[1mm]
\includegraphics[scale=.3]{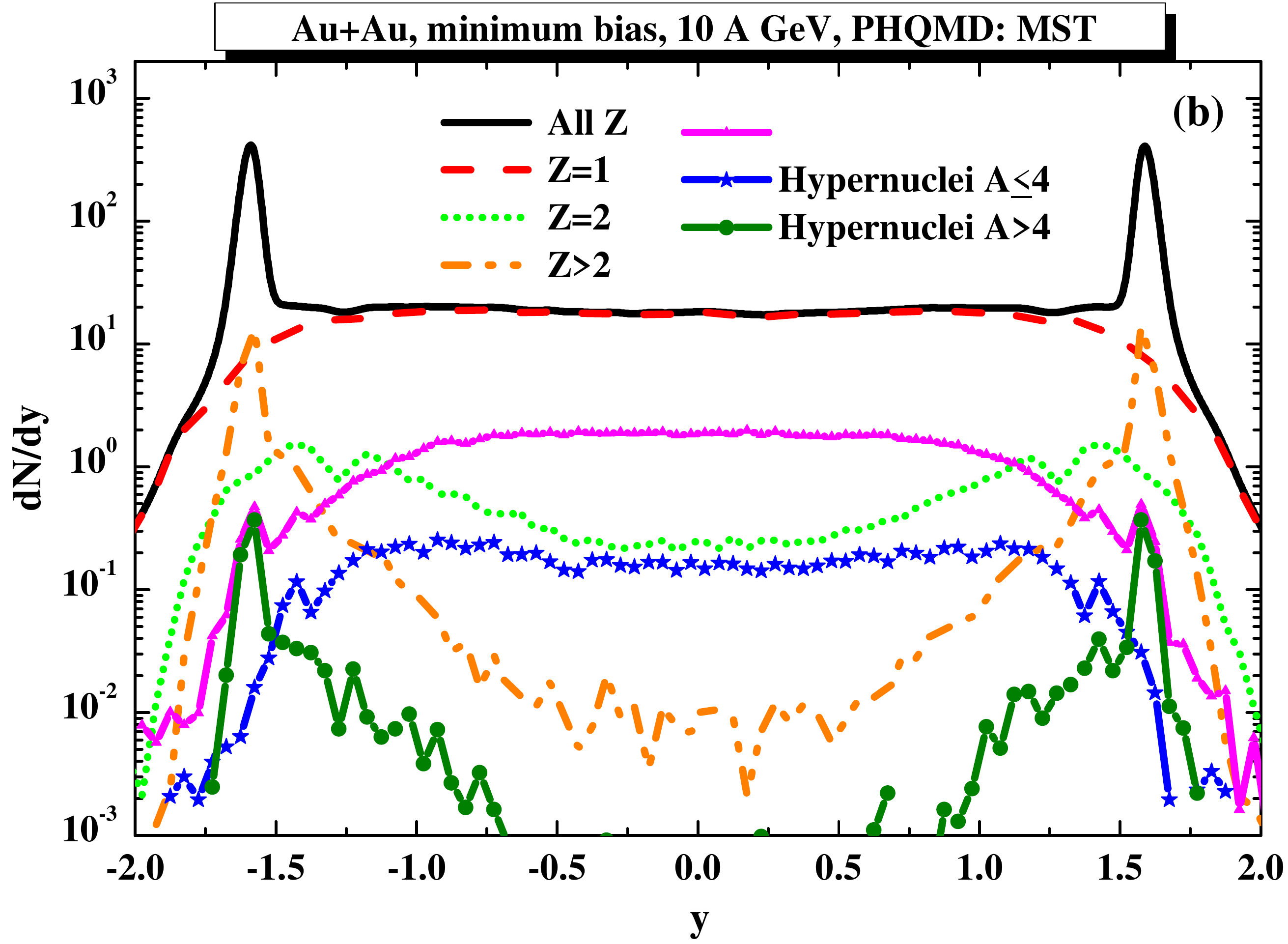}
\caption{\label{fig:BMN} (Color online) 
The PHQMD results (with a hard EoS and the MST algorithm) 
for the rapidity distributions of all charges (black solid line),
$Z=1$ particles (red dashed line),
$Z=2$ clusters (green dotted line), $Z>2$ (orange dot-dot-dashed line), 
all (bound and unbound) $\Lambda$'s (magenta line with triangles) 
as well as  light hypernuclei with $A \le 4$ 
(blue line with stars) and heavy hypernuclei with $A>4$ (green line with dots)
as a function of the rapidity for central Au+Au collisions at 4 $A$GeV (upper plot (a))  
and at 10 $A$GeV (lower plot (b)). }
\end{figure}

In this Section we extended our study on cluster formation within the PHQMD 
to the hyper-clusters using the MST and SACA cluster finding algorithms.
When calculating the hyper-nuclei with the SACA algorithm, we assume that the strength of the
hyperon-nucleon potential is 2/3 of that of nucleon-nucleon potential.
We note, the PHQMD describes the hyperon production
rather well as demonstrated in Section IV for AGS, SPS and RHIC energies.
This gives us a solid basis to study the hyper-cluster production within PHQMD.

Fig. \ref{fig:BMN} shows the distribution of $Z=1$, $Z=2$ particles, 
heavier clusters ($Z>2$), all $\Lambda$'s (bound or unbound) as well as of light ($A \le 4$)
and heavy ($A>4$) hypernuclei identified by MST algorithm
as a function of the rapidity for Au+Au collisions at 4 $A$GeV (upper plot) 
and at 10 $A$GeV (lower plot). We see an enhancement of
the yields of $Z=1$ particles, $\Lambda$'s and  heavier clusters close to 
projectile and target rapidity and an almost constant distribution for $Z=1$ particles in between. The production of hyperons increases towards midrapidity. 
We note that in these calculations we did not make a selection of clusters
according to the realistic isospin contents.
At midrapidity only a small fraction of the hyperons end up in light hypernuclei, in contradistinction to the projectile/target rapidities  where many of the produced hyperons end up as part of a larger hyper-cluster. 

\begin{figure}[htp]%
\centering
\includegraphics[scale=.5]{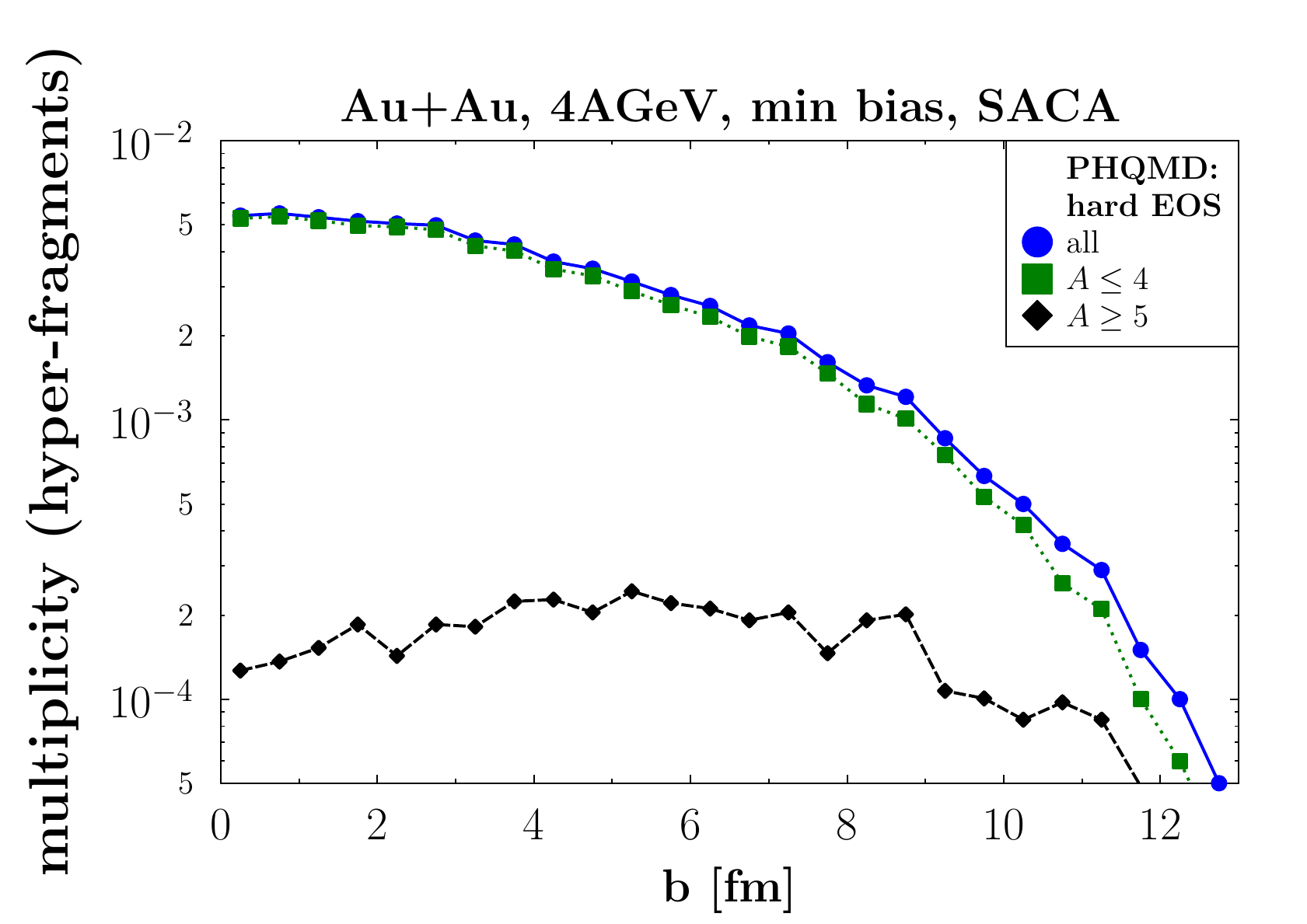}
\caption{\label{b3} (Color online)  
The multiplicity of light hyperclusters as a function of 
the impact parameter for Au+Au collisions at 4 $A$GeV
calculated with the PHQMD using the SACA cluster recognition algorithm.
The blue dots show the multiplicity of all hyper-nuclei, 
while the green squares and black rhombus stand for $A \le 4$ and $A\ge 5$, respectively.}  
\end{figure}  

In Fig. \ref{b3} we show the multiplicity of light and heavy hypercluster as a function of the 
impact parameter for Au+Au collisions at 4 $A$GeV. 
As seen from this figure, the yield of light hyper-clusters decreases  
with the impact parameter, mainly because the overlap region between projectile and target gets smaller and hence less hyperons are produced. In central collisions,  mainly light hypernuclei ($A \le 4$) are formed while mid-central collisions  
are better suited for a study of heavier hypernuclei ($A \ge 5$). 
Hypernuclei with $A \ge 5$ are dominantly produced by hyperons which enter the spectator matter and get caught there. Therefore, for heavy hyper-nuclei production there is a competition between the hyperon production which decreases with impact parameter and the spectator matter whose size increases with impact parameter.

\section{Conclusion}

We have presented a novel microscopic transport approach -- PHQMD, to study the dynamics of 
heavy-ion collisions, cluster and hypernuclei formation at beam energies from a couple of 
hundred $A$MeV to ultra-relativistic energies.
The PHQMD approach extends, on the one side, the study of cluster formation within the QMD model 
at lower beam energies and, on the other side, the particle production from SIS to LHC 
energies within the PHSD approach. 
The PHQMD  adopts the hadronic and partonic collisional interactions 
from the PHSD approach via the same collision integral. However, it extends the PHSD approach 
by replacing the mean-field dynamics for the baryon propagation by a n-body quantum molecular 
dynamics based on density dependent 2-body interactions between all baryons in the system.
This allows to propagate all baryonic correlations and fluctuations what is necessary  to
study the dynamical cluster formation in heavy-ion reactions. This implies 
that clusters are produced dynamically during the whole heavy-ion collisions by the same potential interaction among nucleons which drives their interaction during the heavy-ion collision. Consequently, there is no need to switch to other assumptions 
for modeling the cluster formation as it is done in some other transport approaches by introducing, for example, a coalescence model or a statistical fragmentation model.
  
For the cluster finding we use the MST and SACA algorithms. The MST finds clusters based 
on spacial correlations at the end of the reaction while the SACA algorithm, which is based
on the finding of the most bound configuration, allows to identify clusters during the early 
heavy-ion dynamics when clusters still overlap in coordinate space. 
Moreover, the availability of the mean-field and QMD propagation in one numerical code  
PHQMD allows to explore the differences in the dynamical description of HIC and their influences 
on cluster formation.

First of all, we have validated the PHQMD approach by comparing the "bulk" hadronic
observables as rapidity distributions and $m_T$ or $p_T$  spectra of baryons 
($p, \bar p, \Lambda, \bar\Lambda$) and mesons ($\pi^\pm, K^\pm$) from 
low SIS to top RHIC energies. We find a reasonable good
agreement between the PHQMD and experimental data. For the QMD dynamics we explore two EoS:
"hard" and "soft", realized by static potentials. 
We find that 

i) for the protons the PHQMD results with a soft EoS agree very well with PHSD results.
The QMD with a hard EoS shows slightly harder spectra of protons at AGS energies which is 
favored by experimental data. However, we give a note of caution that in order to draw
robust conclusions about the 'softening/hardening' of the EoS, one needs to include the 
momentum dependence of the nuclear potential. This work is under way.

ii) for the newly produced hadrons the sensitivity to the EoS is minor in the QMD dynamics.
At relativistic energies and at midrapidity the dynamics is driven by 
hadronic/partonic collisions. The results are thus  less sensitive to the baryonic potentials during propagation,
and, consequently, the PHSD and PHQMD results are similar.
Secondly, within the PHQMD approach we have studied the cluster (including hypernuclei)  production which are identified with the MST and SACA models.

iii) We have demonstrated that the QMD dynamics allows to form clusters at midrapity as well
as at target/projectile rapidity and to keep them stable over time.
When using the mean-field propagation, the clusters are note stable 
and disintegrate with time. This demonstrates the importance of nucleon correlations 
for the cluster dynamics which are smeared out in the mean-field propagation.

iv) We have validated the PHQMD approach by reproducing the complex cluster 
pattern observed by the ALADIN collaboration at the highest energies where experimental 
data for heavy clusters are available (i.e. the beam energies of 600-1000 $A$MeV). 
We observed that these heavy clusters are produced  close to target and projectile rapidity 
and with increasing energies also hyper-clusters can be formed in this kinematic region. 
We find a good description of the ALADIN data for the  "rise and fall" of the multiplicity 
of intermediate mass clusters $3\le Z \le 30$ emitted in forward direction as 
a function of the sum of all forward emitted bound charges, $Z_\mathrm{bound\ 2}$.
Moreover, the PHQMD calculations with the SACA algorithm show a stability of the clusters versus time.
We compared also $Z_{max}$
versus $Z_\mathrm{bound\ 2}$ as well as $\sqrt{<p_T^2(Z)>}$ as a function of the cluster charge.
The latter agrees well with the prediction for an instantaneous break up of the nucleus 
and disagrees with the assumptions that clusters are created in a thermal heat bath 
of a temperature around the binding energy.

v) We have studied also the light cluster production at midrapidity within the PHQMD approach.
The identification of light clusters is important for the understanding of the proton spectra
at low energies. 
As has been found by the FOPI collaboration, in central Au+Au reactions at 
1.5 $A$GeV around 40\% of all nucleons are bound  in clusters. 
The PHQMD calculations show a good agreement with the FOPI proton data only
when subtracting the protons which are bound in clusters. 
With increasing  beam energy up to relativistic energies, the fraction of nucleons bound 
in clusters decreases, however, at beam energies below 5 $A$GeV the 
identification of clusters is an important issue also for proton observables \cite{Reisdorf:2010aa}.
We also reproduce the rapidity distribution of light clusters observed at AGS energies,
for central as well as for minimum bias data. 

vi) We made predictions for the production of clusters and hypernuclei at higher beam energies (4-10 $A$GeV) 
relevant for the FAIR and NICA experiments. In particular, we presented the rapidity distribution and 
centrality dependence of hypernuclei production. We investigated also the collective flow of clusters
in terms of the $v_1$ coefficient.

We note, that the microscopic origin of the cluster and hypernucleus formation
at midrapidity at relativistic energies is one of the intriguing problems of present heavy-ion physics. 
The measured hadronic transverse energy spectra at midrapidity show an inverse slope parameter 
in between 100 and 150 MeV, to a large part due to thermal movement of the particles,
even if the radial flow contributes as well. 
Additionally a thermal model fit of the particle ratios 
at RHIC and LHC energies  yield a  temperature of the same order. 
On the other hand clusters are weakly bound objects (with a binding energy of a couple of $A$MeV) 
and with a large distance between the cluster nucleons. Consequently, they are not stable in an environment 
of a temperature of around 100 $MeV$ and collisions with other hadrons can easily destroy them. 
One may talk about pieces of "ice in a fire". Therefore it is all but evident how 
these clusters are created and  survive the expansion of the system.
In this respect the PHQMD approach provides the basis of a more detailed study of their origin
since it is based on a microscopic description of the interaction  and can be applied early during the collision.
The MST method applied in this study for the identification of midrapidity clusters at 
high energies can identify clusters only at the end of the expansion and is presently 'charge blind'. 
To study the cluster formation process in more detail we have to develop further the SACA approach 
to a method which can deal with strange baryons and with the quantum 
features which determine the binding energy of light clusters. Such a development is also 
necessary to study quantitatively the production of hyper-nuclei which PHQMD produces copiously. 
First step in this direction are under way \cite{Fevre:2015fua,FRIGA2019}.

\section*{Acknowledgements}
The authors acknowledge inspiring discussions with C. Blume, W. Cassing, C. Hartnack,  
P. Moreau, L. Oliva, T. Song, O. Soloveva, Io. Vassiliev and M. Winn. 
Furthermore, we acknowledge support by the Deutsche Forschungsgemeinschaft 
(DFG, German Research Foundation): grant BR 4000/7-1,  by the Russian Science Foundation grant 19-42-04101
and   by the GSI-IN2P3 agreement under contract number 13-70.
Also we thank the COST Action THOR, CA15213.
This project has also received funding from the European Union’s Horizon 2020 research and innovation programme under grant agreement No 824093.
The computational resources have been provided by the LOEWE-Center for Scientific Computing
and the "Green Cube" at GSI.



\appendix

\section*{Appendix}
\label{AppendixA}
\section{Dynamical Quasiparticle Model (DQPM)}

The Dynamical Quasiparticle Model (DQPM) has been introduced in Refs. 
\cite{Peshier:2005pp,Cassing:2007nb,Cassing:2007yg} for the effective description
of the properties of the QGP in terms of strongly interacting quarks and gluons
with properties and interactions which 
are adjusted to reproduce lQCD results
on the thermodynamics of the equilibrated QGP at finite temperature $T$ and 
baryon (or quark) chemical potential $\mu_q$.
In the DQPM the quasiparticles are characterized by single-particle Green's functions
(in propagator representation) with complex self-energies. 
The real part of the self-energies is related to the mean-field properties, whereas the imaginary part provides information about the lifetime and/or reaction rates of the particles. This described by a Lorentzian spectral function \cite{Linnyk:2015rco}
of quasiparticles:
\begin{eqnarray}
\rho_{j}(\omega,{\bf p}) && = \frac{\gamma_{j}}{\tilde{E}_j}
	\left(\frac{1}{(\omega-\tilde{E}_j)^2+\gamma^{2}_{j}}
	-\frac{1}{(\omega+\tilde{E}_j)^2+\gamma^{2}_{j}}\right) \nonumber \\
&& \equiv \frac{4\omega\gamma_j}{\left( \omega^2 - \mathbf{p}^2 - M^2_j \right)^2 + 4\gamma^2_j \omega^2}
	\label{spectral_function}
\end{eqnarray}	
separately for quarks, antiquarks, and gluons ($j = q,\bar q,g$).
Here, $\tilde{E}_{j}^2({\bf p})={\bf p}^2+M_{j}^{2}-\gamma_{j}^{2}$;  
the widths $\gamma_{j}$ and the masses $M_{j}$ from the DQPM are functions of the temperature $T$ and the chemical potential $\mu_q$. 

Since the DQPM is an effective model, one has to assume the actual form of the 
$(T,\mu_q)$-dependences of the dynamical masses and widths of quasiparticles
as well as the coupling. 
By fixing the quasiparticle properties, one can evaluate the entropy density 
$s(T,\mu_B)$ and number density in the propagator representation from Baym \cite{Vanderheyden:1998,Blaizot:2000fc} and then, by comparison to the corresponding lQCD data, one can fix the few parameters of the DQPM. After that the DQPM provides a consistent
description of the QGP thermodynamics \cite{Cassing:2007nb,Cassing:2007yg} 
and has a predictive power, additionally.

The effective masses are assumed to be given in line with the HTL thermal mass in the asymptotic high-momentum regime, i.e., for gluons by \cite{Linnyk:2015rco}
\begin{equation}
M^2_{g}(T,\mu_q)=\frac{g^2(T,\mu_q)}{6}\left(\left(N_{c}+\frac{1}{2}N_{f}\right)T^2
+\frac{N_c}{2}\sum_{q}\frac{\mu^{2}_{q}}{\pi^2}\right)\ ,
\label{Mg9}
\end{equation}
and for quarks (antiquarks) by
\begin{equation}
M^2_{q(\bar q)}(T,\mu_q)=\frac{N^{2}_{c}-1}{8N_{c}}g^2(T,\mu_q)\left(T^2+\frac{\mu^{2}_{q}}{\pi^2}\right)\ ,
\label{Mq9}
\end{equation}
~\\
where $N_{c}=3$ stands for the number of colors while $N_{f}\ (=3)$ denotes the number of flavors. Furthermore, the effective quarks, antiquarks, and gluons in the DQPM have finite widths $\gamma$, which are adopted in the form \cite{Linnyk:2015rco}

\begin{equation}
\label{widthg}
\gamma_{g}(T,\mu_q) = \frac{1}{3}N_{c}\frac{g^2(T,\mu_q)T}{8\pi}\ln\left(\frac{2c}{g^2(T,\mu_q)}+1\right),
\end{equation}
\begin{equation}
\label{widthq}
\gamma_{q(\bar
	q)}(T,\mu_q)=\frac{1}{3}\frac{N^{2}_{c}-1}{2N_{c}}\frac{g^2(T,\mu_q)T}{8\pi}
\ln\left(\frac{2c}{g^2(T,\mu_q)}+1\right),
\end{equation}
~\\
where $c=14.4$  is related to a magnetic cutoff, which is a parameter of the DQPM. Furthermore, we assume that the width of the strange quarks is the same as that for the light ($u,d$) quarks. With the choice of Eq. (\ref{spectral_function}), the complex self-energies for gluons $\Pi = M_g^2-2i \omega \gamma_g$ and for (anti)quarks $\Sigma_{q} = M_{q}^2 - 2 i \omega \gamma_{q}$ are fully defined via Eqs. (\ref{Mg9}), (\ref{Mq9}), (\ref{widthg}), and (\ref{widthq}).

The coupling $g^2$, which defines the strength of the interaction in the DQPM, is extracted from lQCD thermodynamics. 
There are few realizations of the DQPM for the evaluation of the $g^2$:
i) its temperature dependence at vanishing chemical potential can either be obtained by using an ansatz with a few parameters  adjusted to results of lQCD thermodynamics \cite{Berrehrah:2013mua,Berrehrah:2015ywa}, or ii) $g^2$ can directly be obtained by a parametrization of the entropy density from lQCD as  in Ref. \cite{Moreau:2019vhw}. 
We indicate that for the present version of the PHQMD we adopted the DQPM model
in the first realization - as used in the PHSD v4.0 \cite{Cassing:2008sv,Cassing:2008nn,Cassing:2009vt,Bratkovskaya:2011wp,Linnyk:2015rco}. 

The extension of the DQPM to finite baryon chemical potential, $\mu_B$, is performed by using a scaling ansatz which works up to $\mu_B \approx 450$ MeV \cite{Steinert:2018bma}, and which assumes that $g^2$ is a function of the ratio of the effective temperature $T^* = \sqrt{T^2+\mu^2_q/\pi^2}$ and the $\mu_B$-dependent critical temperature $T_c(\mu_B)$ as \cite{Cassing:2007nb}
\begin{equation}
g^2(T/T_c,\mu_B) = g^2\left(\frac{T^*}{T_c(\mu_B)},\mu_B =0 \right)
\label{coupling}
\end{equation}
with $\mu_B=3\mu_q$ and $T_c(\mu_B) = T_c \sqrt{1-\alpha \mu_B^2}$, where $T_c$ is
the critical temperature at vanishing chemical potential ($\approx 0.158$ GeV) and $\alpha= 0.974\ \text{GeV}^{-2}$.
By employing the quasiparticle properties and dressed propagators as given by the DQPM, one can deduce the differential partonic scattering cross sections as well as the interaction rates of light and charm quarks in the QGP as a function of the temperature and the chemical potential \cite{Berrehrah:2013mua,Moreau:2019vhw} by calculating the scattering diagrams of the corresponding processes in leading order. This extended version of the DQPM
has been employed recently in the PHSD 5.0 \cite{Moreau:2019vhw} and will
be adopted by the PHQMD also in future.


\label{AppendixB}
\section{Hadronization}

The hadronization, i.e. the transition from partonic to hadronic
degrees-of-freedom, is described in PHQMD (as well as in PHSD) by local covariant 
transition rates as introduced in Ref.~\cite{Cassing:2008sv}.
For $q+\bar{q}$ fusion to an off-shell meson $m$ of four-momentum 
$p= (\omega, {\bf p})$ at space-time point $x=(t,{\bf x})$ it is:
\begin{eqnarray}
&&\phantom{a}\hspace*{-5mm} \frac{d N_m(x,p)}{d^4x d^4p}= Tr_q
Tr_{\bar q} \
  \delta^4(p-p_q-p_{\bar q}) \
  \delta^4\left(\frac{x_q+x_{\bar q}}{2}-x\right) \nonumber\\
&& \times \omega_q \ \rho_{q}(p_q)
   \  \omega_{\bar q} \ \rho_{{\bar q}}(p_{\bar q})
   \ |v_{q\bar{q}}|^2 \ W_m(x_q-x_{\bar q},(p_q-p_{\bar q})/2) \nonumber \\
&& \times N_q(x_q, p_q) \
  N_{\bar q}(x_{\bar q},p_{\bar q}) \ \delta({\rm flavor},\, {\rm color}).
\label{trans}
\end{eqnarray}
In Eq.~(\ref{trans}) the shorthand notation is introduced:
\begin{equation}
Tr_j = \sum_j \int d^4x_j \int \frac{d^4p_j}{(2\pi)^4} \ ,
\end{equation}
where $\sum_j$ denotes a summation over discrete quantum numbers
(spin, flavor, color); $N_j(x,p)$ is the phase-space density of
parton $j$ at space-time position $x$ and four-momentum $p$.  In
Eq.~(\ref{trans}) $\delta({\rm flavor},\, {\rm color})$ stands
symbolically for the conservation of flavor quantum numbers as
well as color neutrality of the formed meson $m$. 
Furthermore, $v_{q{\bar q}}(\rho_p)$ is the effective quark-antiquark interaction  
from the DQPM  (displayed in Fig. 10 of Ref.~\cite{Cassing:2007nb}) as a
function of the local parton ($q + \bar{q} +g$) density $\rho_p$
(or energy density). Furthermore, $W_m(x,p)$ is the dimensionless phase-space
distribution of the formed off-shell meson, i.e.
\begin{equation} \label{Dover} W_m(\xi,p_\xi) =
\exp\left( \frac{\xi^2}{2 b^2} \right)\ \exp\left( 2 b^2 (p_\xi^2- (M_q-M_{\bar
q})^2/4) \right)
\end{equation} 
with $\xi = x_1-x_2 = x_q - x_{\bar q}$ and $p_\xi = (p_1-p_2)/2 =
(p_q - p_{\bar q})/2$. The width parameter $b$ is fixed by
$\sqrt{\langle r^2 \rangle} = b$ = 0.66 fm (in the rest frame) which
corresponds to an average rms radius of mesons. We note that the
expression~(\ref{Dover}) corresponds to the limit of independent
harmonic oscillator states and that the final hadron-formation rates
are approximately independent of the parameter $b$ within reasonable
variations. By construction the quantity~(\ref{Dover}) is Lorentz
invariant; in the limit of instantaneous 'hadron formation',
i.e. $\xi^0=0$, it provides a Gaussian dropping in the relative
distance squared $({\bf r}_1 - {\bf r}_2)^2$. The four-momentum
dependence reads explicitly 
\begin{equation} 
(E_1 - E_2)^2 - ({\bf p}_1 - {\bf p}_2)^2 -
(M_1-M_2)^2 \leq 0
\end{equation} 
and leads to a negative argument of the second
exponential in (\ref{Dover}) favoring the fusion of partons with
low relative momenta $p_q - p_{\bar q}= p_1-p_2$.

Related transition rates (to Eq.~(\ref{trans})) have been defined 
in Ref.~\cite{Cassing:2009vt} also for the fusion of three off-shell 
quarks ($q_1+q_2+q_3 \leftrightarrow B$) to color neutral baryonic 
($B$ or $\bar{B}$) resonances of finite width (or strings) fulfilling 
energy and momentum conservation as well as flavor current conservation 
using Jacobi coordinates.



\begin{thebibliography}{999}

\bibitem{QM2017} 
The Proceedings of the 26th International Conference on Ultra-relativistic 
Nucleus-Nucleus Collisions: Quark Matter 2017,
Edited by Ulrich Heinz, Olga Evdokimov, Peter Jacobs;
Nucl. Phys. A {\bf 961},1 (2017).


\bibitem{Borsanyi:2013bia} 
  S.~Borsanyi, Z.~Fodor, C.~Hoelbling, S.~D.~Katz, S.~Krieg and K.~K.~Szabo,
  Phys.\ Lett.\ B {\bf 730}, 99 (2014).
  
\bibitem{Bazavov:2014pvz} 
  A.~Bazavov {\it et al.} [HotQCD Collaboration],
  Phys.\ Rev.\ D {\bf 90}, 094503 (2014).
  
\bibitem{ref1storder1}
M. Asakawa and K. Yazaki, Nucl. Phys. A {\bf 504}, 668 (1989).

\bibitem{ref1storder2}
M.A. Stephanov, Prog. Theor. Phys. Suppl. {\bf 153}, 139 (2004). 


\bibitem{Schuttauf:1996ci}
  A.~Schuttauf {\it et al.},
  Nucl.\ Phys.\ A {\bf 607}, 457 (1996).

\bibitem{Sfienti:2006zb}
  C.~Sfienti {\it et al.} [ALADiN2000 Collaboration],
  Nucl.\ Phys.\ A {\bf 787}, 627 (2007).
 
\bibitem{Nebauer:1998fy}
  R.~Nebauer {\it et al.} [INDRA Collaboration],
  Nucl.\ Phys.\ A {\bf 658}, 67 (1999).
  
\bibitem{Reisdorf:2010aa}
  W.~Reisdorf {\it et al.} [FOPI Collaboration],
  Nucl.\ Phys.\ A {\bf 848}, 366 (2010).

\bibitem{Rappold:2015una} 
  C.~Rappold {\it et al.} [HypHi Collaboration],
  Phys.\ Lett.\ B {\bf 747}, 129 (2015).
  
\bibitem{Anticic:2016ckv} 
  T.~Anticic {\it et al.} [NA49 Collaboration],
  Phys.\ Rev.\ C {\bf 94}, 044906 (2016).
  
\bibitem{Abelev:2010rv} 
  B.~I.~Abelev {\it et al.} [STAR Collaboration],
  Science {\bf 328}, 58 (2010).

\bibitem{Agakishiev:2011ib} 
  H.~Agakishiev {\it et al.} [STAR Collaboration],
  Nature {\bf 473}, 353 (2011);
  Erratum: [Nature {\bf 475}, 412 (2011)].
  

\bibitem{Adam:2015yta} 
  J.~Adam {\it et al.} [ALICE Collaboration],
  Phys.\ Lett.\ B {\bf 754}, 360 (2016).
  
\bibitem{Adam:2015vda}
  J.~Adam {\it et al.} [ALICE Collaboration],
  Phys.\ Rev.\ C {\bf 93}, 024917 (2016).
  
\bibitem{Acharya:2017bso} 
  S.~Acharya {\it et al.} [ALICE Collaboration],
  Nucl.\ Phys.\ A {\bf 971}, 1 (2018).
\bibitem{Shuryak:2018lgd} 
  E.~Shuryak and J.~M.~Torres-Rincon,
  Phys.\ Rev.\ C {\bf 100}, 024903 (2019).

  
\bibitem{Wakai88}
  M. Wakai, H. Bando and M. Sano, Phys. Rev. {\bf C 38},  748 (1988).
\bibitem{Rudy95}
  Z. Rudy, T. Demski, L. Jarczyk, B. Kamys, P. Kulessa et al., Z. Phys. {\bf A 351}, 217 (1995).
\bibitem{Gaitanos09}
  T. Gaitanos, H. Lenske and U. Mosel, Phys. Lett. {\bf B 675},  297 (2009).
\bibitem{Topor10}
  V. Topor Pop and S. Das Gupta, Phys. Rev. {\bf C 81}, 054911 (2010).
\bibitem{Botvina}
  A. S. Botvina, K. K. Gudima, J. Steinheimer, M. Bleicher and I. N. Mishustin,  
       Phys. Rev. {\bf C 84}, 064904 (2011);
       A. S. Botvina, K. K. Gudima, J. Steinheimer, M. Bleicher and J. Pochodzalla,  
       Phys. Rev. {\bf C 95}, 014902 (2017).
\bibitem{Ko15}
  L. Zhu, C. M. Ko and X. Yin, Phys. Rev. {\bf C 92}, 064911 (2015).
  
\bibitem{Andronic:2010qu}
  A.~Andronic, P.~Braun-Munzinger, J.~Stachel and H.~Stocker,
  Phys.\ Lett.\ B {\bf 697}, 203 (2011).

\bibitem{Ritman:1995tn}
  J.~L.~Ritman {\it et al.} [FOPI Collaboration],
  Z.\ Phys.\ A {\bf 352}, 355 (1995).



\bibitem{David:1998qu}
  C.~David, C.~Hartnack and J.~Aichelin,
  Nucl.\ Phys.\ A {\bf 650}, 358 (1999).


\bibitem{Aichelin:1991xy}
  J.~Aichelin,
  Phys.\ Rept.\  {\bf 202}, 233 (1991).

\bibitem{Aichelin:1987ti}
  J.~Aichelin, A.~Rosenhauer, G.~Peilert, H.~Stoecker and W.~Greiner,
  Phys.\ Rev.\ Lett.\  {\bf 58}, 1926 (1987).

\bibitem{Aichelin:1988me}
  J.~Aichelin, A.~Bohnet, G.~Peilert, H.~Stoecker, W.~Greiner and A.~Rosenhauer,
  Phys.\ Rev.\ C {\bf 37}, 2451 (1988).


\bibitem{Hartnack:1997ez} 
  C.~Hartnack, R.~K.~Puri, J.~Aichelin, J.~Konopka, S.~A.~Bass, H.~Stoecker and W.~Greiner,
  Eur.\ Phys.\ J.\ A {\bf 1}, 151 (1998).


\bibitem{Bass:1998ca}
 S.~A.~Bass {\it et al.},
  Prog.\ Part.\ Nucl.\ Phys.\  {\bf 41}, 255 (1998).

\bibitem{Bleicher:1999xi}
 M.~Bleicher {\it et al.},
  J.\ Phys.\  {\bf G 25}, 1859 (1999).
  
\bibitem{Kruse:1985pg}
  H.~Kruse, B.~V.~Jacak, J.~J.~Molitoris, G.~D.~Westfall and H.~Stoecker,
  Phys.\ Rev.\ C {\bf 31}, 1770 (1985).

\bibitem{Aichelin:1985zz}
  J.~Aichelin and G.~Bertsch,
  Phys.\ Rev.\ C {\bf 31}, 1730 (1985).
 
\bibitem{BUU}
  P.~Danielewicz and G.~F.~Bertsch,
  Nucl.\ Phys.\ A {\bf 533}, 712 (1991).
   
\bibitem{AMPT} 
  Z.~W.~Lin, C.~M.~Ko, B.~A.~Li, B.~Zhang and S.~Pal,
  Phys.\ Rev.\ C {\bf 72}, 064901 (2005).
  
\bibitem{Ehehalt:1996uq}
  W.~Ehehalt and W.~Cassing,
  Nucl.\ Phys.\ A {\bf 602}, 449 (1996).
  
\bibitem{Cassing:1999es}
  W.~Cassing and E.~L.~Bratkovskaya,
  Phys.\ Rept.\  {\bf 308}, 65 (1999).

\bibitem{GiBUU}
  O.~Buss {\it et al.},
  Phys.\ Rept.\  {\bf 512}, 1 (2012).

\bibitem{Weil:2016zrk}
  J.~Weil {\it et al.},
  Phys.\ Rev.\ C {\bf 94}, 054905 (2016).


\bibitem{Cassing:2008sv} 
  W.~Cassing and E.~L.~Bratkovskaya,
  Phys.\ Rev.\ C {\bf 78}, 034919 (2008).
  
\bibitem{QGSM}  
V.D. Toneev and K.K. Gudima, Nucl. Phys. {\bf A 400}, 173c (1983);
V.D. Toneev, N.S. Amelin, K.K. Gudima, and S.Yu. Sivoklokov, 
 Nucl. Phys. {\bf A 519}, 463c (1990);
N.S. Amelin, E.F. Staubo, L.S. Csernai et al., Phys. Rev. {\bf C 44}, 1541 (1991).

\bibitem{Gossiaux:1994jq}
  P.~B.~Gossiaux, D.~Keane, S.~Wang and J.~Aichelin,
  Phys.\ Rev.\ C {\bf 51}, 3357 (1995).

\bibitem{Zhu:2015voa} 
  L.~Zhu, C.~M.~Ko and X.~Yin,
  Phys.\ Rev.\ C {\bf 92}, 064911 (2015).

\bibitem{Feckova:2016kjx}
  Z.~Fecková, J.~Steinheimer, B.~Tomášik and M.~Bleicher,
  Phys.\ Rev.\ C {\bf 93}, 054906 (2016).
 
 
\bibitem{Botvina_HSD}
  A. S. Botvina, J. Steinheimer, E. Bratkovskaya, M. Bleicher and J. Pochodzalla,
  Phys. Lett. {\bf B 742}, 7 (2015).

\bibitem{Danielewicz:1991dh}
  P.~Danielewicz and G.~F.~Bertsch,
  Nucl.\ Phys.\ A {\bf 533}, 712 (1991).

\bibitem{Oliinychenko:2018ugs} 
  D.~Oliinychenko, L.~G.~Pang, H.~Elfner and V.~Koch,
  Phys. Rev. {\bf C 99}, 044907 (2019).
 
\bibitem{Cassing:2008nn}
   W.~Cassing,
  Eur.\ Phys.\ J.\ ST {\bf 168}, 3 (2009).

  
\bibitem{Cassing:2009vt}
  W.~Cassing and E.~L.~Bratkovskaya,
  Nucl.\ Phys.\ A {\bf 831}, 215 (2009).
  
\bibitem{Bratkovskaya:2011wp}
  E.~L.~Bratkovskaya, W.~Cassing, V.~P.~Konchakovski and O.~Linnyk,
  Nucl.\ Phys.\ A {\bf 856}, 162 (2011).
    
\bibitem{Linnyk:2015rco} 
  O. Linnyk, E. L.  Bratkovskaya,  W. Cassing,
  Prog. Part. Nucl. Phys. {\bf 87}, 50 (2016).    
  
\bibitem{Ozvenchuk:2012fn} 
  V.~Ozvenchuk, O.~Linnyk, M.~I.~Gorenstein, E.~L.~Bratkovskaya and W.~Cassing,
  Phys.\ Rev.\ C {\bf 87},  024901 (2013).
  
\bibitem{Cassing:2001ds} 
  W.~Cassing,
  Nucl.\ Phys.\ A {\bf 700}, 618 (2002).
  
\bibitem{Seifert:2017oyb} 
  E.~Seifert and W.~Cassing,
  Phys.\ Rev.\ C {\bf 97}, 024913 (2018); 
  Phys.\ Rev.\ C {\bf 97}, 044907 (2018).

\bibitem{Bratkovskaya:2007jk}
  E.~L.~Bratkovskaya and W.~Cassing,
  Nucl.\ Phys.\ A {\bf 807}, 214 (2008).
  
\bibitem{Ilner:2016xqr} 
  A.~Ilner, D.~Cabrera, C.~Markert and E.~Bratkovskaya,
  Phys.\ Rev.\ C {\bf 95},  014903 (2017); 
  A.~Ilner, J.~Blair, D.~Cabrera, C.~Markert and E.~Bratkovskaya,
  Phys.\ Rev.\ C {\bf 99}, 024914 (2019).
 
\bibitem{HSDK} 
  W. Cassing, L. Tol\'os, E. L. Bratkovskaya, A. Ramos,
  Nucl. Phys. {\bf A 727}, 59 (2003). 
  
\bibitem{Marty:2012vs} 
 R.~Marty and J.~Aichelin,
  Phys.\ Rev.\  {\bf C 87}, 034912 (2013).


\bibitem{Marty:2014zka} 
  [40] R.~Marty, E.~Bratkovskaya, W.~Cassing and J.~Aichelin,
  Phys.\ Rev.\ {\bf C 92}, 015201 (2015).

\bibitem{Puri:1996qv}
  R.~K.~Puri, C.~Hartnack and J.~Aichelin,
  Phys.\ Rev.\ C {\bf 54}, R28 (1996).
\bibitem{Puri:1998te}
  R.~K.~Puri and J.~Aichelin,
  J.\ Comput.\ Phys.\  {\bf 162}, 245 (2000).
  
\bibitem{FRIGA2019}
A.~Le F\`evre, J.~Aichelin, C.~Hartnack and Y.~Leifels,
arXiv:1906.06162.


\bibitem{Dorso:1992ch}
  C.~Dorso and J.~Randrup,
  Phys.\ Lett.\ B {\bf 301}, 328 (1993).
  

\bibitem{LeFevre:2009er} 
  A.~Le F\`evre {\it et al.},
  Phys.\ Rev.\ {\bf C 80}, 044615 (2009).

\bibitem{Gossiaux:1997hp} 
  P.~B.~Gossiaux, R.~Puri, C.~Hartnack and J.~Aichelin,
  Nucl.\ Phys.\ {\bf A 619}, 379 (1997).

\bibitem{Fevre:2007pr} 
   A.~L.~F\`evre and J.~Aichelin,
  Phys.\ Rev.\ Lett.\  {\bf 100}, 042701 (2008).


 
\bibitem{Fevre:2015fua} 
   A.~Le F\`evre, Y.~Leifels, J.~Aichelin, C.~Hartnack, V.~Kireyev and E.~Bratkovskaya,
  J.\ Phys.\ Conf.\ Ser.\  {\bf 668}, 012021 (2016).
  
\bibitem{Kadanoff1962}
  L. P. Kadanoff and G. Baym, {\it Quantum Statistical Mechanics},
  Benjamin, New York, 1962.
  
\bibitem{Juchem:2004cs}
  S.~Juchem, W.~Cassing and C.~Greiner,
  Nucl.\ Phys.\ A {\bf 743}, 92 (2004).

\bibitem{Peshier:2005pp} 
  A.~Peshier and W.~Cassing,
  Phys.\ Rev.\ Lett.\  {\bf 94}, 172301 (2005).
  
\bibitem{Cassing:2007nb}
  W.~Cassing,
  Nucl.\ Phys.\ {\bf A 795}, 70 (2007).
  
\bibitem{Cassing:2007yg}
  W.~Cassing,
  Nucl.\ Phys.\ {\bf A 791}, 365 (2007).

 \bibitem{LUND} 
  B. Nilsson-Almqvist and E. Stenlund, Comp. Phys. Comm.
  {\bf 43}, 387 (1987);
 B. Andersson, G. Gustafson, and H. Pi, Z. Phys. {\bf C 57}, 485 (1993).
 
\bibitem{FRITIOF}
B. Nilsson-Almqvist and E. Stenlund, {\em Comp. Phys. Comm.} {\bf 43}, 387
(1987); B. Andersson, G. Gustafson, and H. Pi, {\em Z. Phys. C} {\bf 57},
485 (1993).

\bibitem{Sjostrand:2006za}
  T.~Sjostrand, S.~Mrenna and P.~Z.~Skands,
  {\em JHEP} {\bf 0605}, 026 (2006).
  
\bibitem{Schwinger}
    J. Schwinger, {\em Phys. Rev.} {\bf 83}, 664 (1951).

\bibitem{PHSD_CSR}
W. Cassing, A. Palmese, P. Moreau and E.L. Bratkovskaya, {\em Phys.
Rev. C} {\bf 93}, 014902 (2016).

\bibitem{Alessia}  A. Palmese, W. Cassing, E. Seifert, T. Steinert, P. Moreau, and E.L. Bratkovskaya
{\em Phys. Rev. C} {\bf 94}, 044912 (2016).

    

  
\bibitem{Aoki:2009sc}
  Y.~Aoki, S.~Borsanyi, S.~Durr, Z.~Fodor, S.~D.~Katz, S.~Krieg and K.~K.~Szabo,
  JHEP {\bf 0906}, 088 (2009).

\bibitem{Cheng:2007jq}
  M.~Cheng {\it et al.},
  Phys.\ Rev.\ {\bf D 77}, 014511 (2008).


\bibitem{Borsanyi:2015waa}
  S.~Borsanyi {\it et al.},
  Phys.\ Rev.\  {\bf D 92}, 014505 (2015).
  

  

\bibitem{Feldmeier:1989st}
  H.~Feldmeier,
  Nucl.\ Phys.\ A {\bf 515}, 147 (1990).

\bibitem{Ono:1992uy}
  A.~Ono, H.~Horiuchi, T.~Maruyama and A.~Ohnishi,
  Prog.\ Theor.\ Phys.\  {\bf 87}, 1185 (1992).



\bibitem{Hartnack:2011cn}
  C.~Hartnack, H.~Oeschler, Y.~Leifels, E.~L.~Bratkovskaya and J.~Aichelin,
  Phys.\ Rept.\  {\bf 510}, 119 (2012).

\bibitem{Konchakovski:2012yg} 
  V.~P.~Konchakovski, E.~L.~Bratkovskaya, W.~Cassing, V.~D.~Toneev, S.~A.~Voloshin and V.~Voronyuk,
  Phys.\ Rev.\ C {\bf 85}, 044922 (2012).
  
\bibitem{Moreau:2019vhw} 
  P.~Moreau, O.~Soloveva, L.~Oliva, T.~Song, W.~Cassing and E.~Bratkovskaya,
   Phys.\ Rev.\ C {\bf 100},  014911 (2019).
  
\bibitem{Zbiri:2006ts}
  K.~Zbiri {\it et al.},
  Phys.\ Rev.\ C {\bf 75}, 034612 (2007).


\bibitem{FOPI:2011aa}
  W.~Reisdorf {\it et al.} [FOPI Collaboration],
  Nucl.\ Phys.\ A {\bf 876}, 1 (2012).

\bibitem{E866E917} 
    L. Ahle et al. [E866 and E917 Collaborations],
    Phys. Lett. B {\bf 476}, 1 (2000);
    Phys. Lett. B {\bf 490}, 53 (2000).
\bibitem{E917K} 
     W.-C. Chang et al. [E917 Collaboration], nucl-ex/99040110;
      Proceedings of the 15Th Winter Workshop on Nuclear Dynamics,
      ParkCity, UT, Janyary 1999.
\bibitem{E866pi11} 
    Y. Akiba et al. [E866 Collaboration],
    Nucl. Phys. A {\bf 610}, 139c (1996).
\bibitem{E877pi11} 
      L. Ahle et al. [E877 Collaboration], Phys. Rev. C {\bf 57}, R466 (1998);
      J. Barrette et al. [E877 Collaboration], Phys. Rev. C {\bf 63}, 014902 (2001).
\bibitem{E891Lam} 
      S. Ahmad et al. [E891 Collaboration], Phys. Lett. B {\bf 382}, 35 (1996);
      C. Pinkenburg et al. [E866 Collaboration],
    Nucl. Phys. A {\bf 698}, 495c (2002).
\bibitem{E896Lam} 
       S. Albergo et al., [E896 Collaboration],
	Phys. Rev. Lett. {\bf 88}, 062301 (2002).
\bibitem{E917p02} 
    B. Holzman et al. [E917 Collaboration], Nucl. Phys. A {\bf 698}, 643 (2002).
    
\bibitem{NA49pold}
    H. Appelsh\"auser {\it et al.} [NA49 Collaboration],
    Phys. Rev. Lett. {\bf 82}, 2471 (1999).
  C.~Blume {\it et al.},
  J.\ Phys.\ G {\bf 34}, S951 (2007);
 C. Alt {\it et al.}, [NA49 Collaboration], Phys. Rev. {\bf C73}, 044910 (2006);
  T. Anticic {\it et al.}, Phys. Rev. {\bf C77}, 024903 (2008).      
\bibitem{NA49_T}
    V. Friese {\it et al.} [NA49 Collaboration],
    J. Phys. G {\bf 30}, 119 (2004);       
  S.V. Afanasiev {\it et al.}, Phys. Rev. {\bf C66}, 054902 (2002);
     C. Alt {\it et al.}, Phys. Rev. {\bf C77}, 024903 (2008).  
\bibitem{NA49_Lam} 
    A.~Mischke {\it et al.} [NA49 Collaboration],
    J. Phys. G. {\bf 28}, 1761 (2002);
     Nucl. Phys. A {\bf 715}, 453 (2993);
     C. Alt {\it et al.}, Phys. Rev. {\bf C78}, 034918 (2008).


\bibitem{Adamczyk:2017iwn}
  L.~Adamczyk {\it et al.} [STAR Collaboration],
  Phys.\ Rev.\ C {\bf 96}, 044904 (2017).
 
\bibitem{Bearden:2004yx} 
  I.~G.~Bearden {\it et al.} [BRAHMS Collaboration],
  Phys.\ Rev.\ Lett.\  {\bf 94}, 162301 (2005).
  
\bibitem{Arsene:2005mr} 
  I.~Arsene {\it et al.} [BRAHMS Collaboration],
  Phys.\ Rev.\ C {\bf 72}, 014908 (2005).

\bibitem{Adler:2003cb} 
  S.~S.~Adler {\it et al.} [PHENIX Collaboration],
  Phys.\ Rev.\ C {\bf 69}, 034909 (2004).
 
\bibitem{Agakishiev:2011ar} 
  G.~Agakishiev {\it et al.} [STAR Collaboration],
  Phys.\ Rev.\ Lett.\  {\bf 108}, 072301 (2012).
\bibitem{Reisdorf:2006ie}
  W.~Reisdorf {\it et al.} [FOPI Collaboration],
  Nucl.\ Phys.\ A {\bf 781}, 459 (2007).
\bibitem{Cleymans:2005xv}
  J.~Cleymans, H.~Oeschler, K.~Redlich and S.~Wheaton,
  Phys.\ Rev.\ C {\bf 73}, 034905 (2006).


 
\bibitem{LeFevre:2017ygd}
  A.~Le F\`evre, Y.~Leifels, J.~Aichelin, C.~Hartnack, V.~Kireyev and E.~Bratkovskaya,
  Nuovo Cim.\ C {\bf 39}, 399 (2017).

 
\bibitem{Goldhaber:1974qy}
  A.~S.~Goldhaber,
  Phys.\ Lett.\  {\bf 53B}, 306 (1974).


\bibitem{Aichelin:1984asp}
  J.~Aichelin and J.~Huefner,
  Phys.\ Lett.\  {\bf B136}, 15 (1984).


\bibitem{Aichelin:1984xb}
  J.~Aichelin, J.~Hufner and R.~Ibarra,
  Phys.\ Rev.\ C {\bf 30}, 107 (1984).
 
\bibitem{Bennett:1998be} 
  M.~J.~Bennett {\it et al.} [E878 Collaboration],
  Phys.\ Rev.\ C {\bf 58}, 1155 (1998).
  
\bibitem{Saito:1994tg} 
  N.~Saito {\it et al.} [E886 Collaboration],
  Phys.\ Rev.\ C {\bf 49}, 3211 (1994).
  
\bibitem{Berrehrah:2013mua}
H.~Berrehrah, E.~Bratkovskaya, W.~Cassing, P.~B.~Gossiaux, J.~Aichelin and M.~Bleicher,
Phys.\ Rev.\ C {\bf 89},  054901 (2014).

\bibitem{Berrehrah:2015ywa}
H.~Berrehrah, E.~Bratkovskaya, W.~Cassing, P.~B.~Gossiaux and J.~Aichelin,
Phys.\ Rev.\ C {\bf 91},  054902 (2015).

\bibitem{Steinert:2018bma}
T.~Steinert and W.~Cassing,
J.\ Phys.\ Conf.\ Ser.\  {\bf 1024},  012029 (2018).

\bibitem{Vanderheyden:1998}
B. Vanderheyden and G. Baym, J. Stat. Phys. {\bf 93}, 843 (1998).

\bibitem{Blaizot:2000fc}
J.~P.~Blaizot, E.~Iancu and A.~Rebhan,Phys. Rev. D {\bf 63}, 065003
(2001).
\end{thebibliography}
\end{document}